\documentclass[10pt,a4]{article}
\usepackage{mathrsfs,amsmath,amsthm,latexsym,amssymb,amsfonts,epsfig,txfonts,undertilde,cancel}
\newcommand{\makeSymbol}[1]{\mathord{\vcenter{\hbox{#1}}}}
\usepackage{hyperref}
\hypersetup{bookmarks=true, bookmarksnumbered=true,
bookmarksopen=true, colorlinks, linkcolor=red, citecolor=magenta,
plainpages=false, pdfstartview=FitH}
\numberwithin{equation}{section}
\usepackage{appendix}
\allowdisplaybreaks
\begin{document}

\title{Graphical method in loop quantum gravity: I. Derivation of the closed formula for the matrix element of the volume operator}

\author{Jinsong Yang$^{1,2}$\thanks{yangksong@gmail.com}, Yongge Ma$^3$\thanks{Corresponding author, mayg@bnu.edu.cn} \vspace{0.5em}\\
$^1${\small Department of Physics, Guizhou University, Guiyang 550025, China} \\
$^2${\small Institute of Physics, Academia Sinica, Taiwan}\\
$^3${\small Department of Physics, Beijing Normal University, Beijing 100875, China}}
\date{}

\maketitle

\begin{abstract}
To adopt a practical method to calculate the action of geometrical operators on quantum states is a crucial task in loop quantum gravity. In the series of papers, we will introduce a graphical method, developed by Yutsis and Brink, to loop quantum gravity along the line of previous works. The graphical method provides a very powerful technique for simplifying complicated calculations. In this first paper, the closed formula of volume operator is derived via the graphical method. By employing suitable and non-ambiguous graphs to represent the acting of operators as well as the spin network states, we use the simple rules of transforming graphs  to yield the resulting formula. Comparing with the complicated algebraic derivation in some literatures, our procedure is more concise, intuitive and visual. The resulting matrix elements of volume operator is compact and uniform, fitting for both gauge-invariant and gauge-variant spin network states.
\end{abstract}

{\hspace{1em}PACS numbers: 04.60.Pp, 04.60.Ds}

\section{Introduction}

As a non-perturbative approach to quantum gravity, loop quantum gravity (LQG) has made considerable achievements (see \cite{Ashtekar:2004eh,Han:2005km} for review arcticles, and \cite{Rovelli:2004tv,Thiemann:2007bk} for books). This theory rigorously enforces the lesson of general relativity and is built on a strict mathematical fundament. In LQG, the quantum kinematical Hilbert space was successfully constructed with the spin network states as its orthonormal basis. The elementary operators are the holonomy and flux operators. By suitable regularization schemes, quantum geometric operators, such as the length, area, and volume operators corresponding to their classical quantities, were well defined on the kinematical Hilbert space ${\cal H}_{\rm kin}$ \cite{Thiemann:1996at,Bianchi:2008es,Ma:2010fy,Rovelli:1994ge,Ashtekar:1996eg,Ashtekar:1997fb,Thiemann:1996au}. The area operator plays an important role in computing the entropy of black hole. The volume operator is a cornerstone on which some physical interesting operators, for instance, the Hamiltonian constraint operator determining the quantum dynamics of LQG can be constructed.

There are two versions of volume operator in the literature. The first one, based on `external' regularization scheme, was introduced by Rovelli and Smollin in loop representation \cite{Rovelli:1994ge}, and re-obtained in the connection representation \cite{Ashtekar:1997fb}. The second one, based on the `internal' regularization, was firstly defined by Ashtekar and Lewandowski \cite{Ashtekar:1997fb}. In \cite{Thiemann:1996au}, Thiemann presented a rather short and compact regularization procedure to re-derive the second version of volume operator. Playing a crucial role in LQG, the spectra of the volume operator is pursued. Certain matrix elements of volume operator were calculated in the framework of loop representation by using graphical tangle-theoretic Temperley-Lieb formulation in \cite{DePietri:1996pja}. Then they were also derived in connection representation by a rigorous but tedious algebraic method in \cite{Thiemann:1996au,Brunnemann:2004xi}, and their special case was re-derived using generalized Wigner-Echart theory \cite{Dass:2006sp}. Although those components of the volume operator are rigorously defined, the computation of their actions on the spin network states are  difficult. The main reason is the following. The volume element operator at a vertex $v$ of a graph $\gamma$ reads $V_v=\sqrt{|\hat{q}_v|}$. Although the matrix elements of $\hat{q}_v$ can be calculated using recoupling theory, the matrix has no obvious symmetries and hence is difficult to be diagonalized analytically for the case that the dimension of the matrix is bigger than nine. On the one hand, the derivation of closed formula in \cite{Brunnemann:2004xi} is rigorous. But there is no universal formula with so tedious and abstract method. Hence, it is desirable to develop certain rigorous, but concise and intuitive method to calculate the matrix elements of volume operator, as well as other geometric operators.

Graphic calculus has been introduced in LQG in a few papers (see e.g., \cite{DePietri:1996pja,Borissov:1997ji,Alesci:2007tx,Alesci:2010gb,Alesci:2011ia,Alesci:2013kpa}). But a concise, non-ambiguous and complete scheme is still desirable. In this first paper of the series, we will introduce a graphical method to LQG, which is based on the Brink's graphical method \cite{brink1968angular} and its suitable extension \footnote{A similar scheme is introduced independently at almost the same time by E. Alesci el. in \cite{Alesci:2015wla}.}. This method consists of two ingredients, graphical representation and graphical calculation. We will represent the algebraic formula by corresponding graphical formula in an unique and unambiguous way. Then the graphical calculation will be performed following the simple rules of transforming graphs, corresponding uniquely to the algebraic manipulation of the formula. The basic components of the graphical representation and the simple rules of transforming graphs are presented in Appendix A. A central goal of this paper is to derivate the closed formula for the matrix element of the volume operator, which involves only the flux operator, based on our rigorous graphical method. Comparing to the algebraic method, our derivation is obviously more compact and simple. Our analysis shows that the formula of the matrix elements for certain case in \cite{Brunnemann:2004xi} is also valid for other cases and hence can be regarded as a general expression.
In the second paper of the series \cite{graph-II} , we will consider the actions of the gravitational Hamiltonian constraint operator and the inverse volume operator, depending also on the honolomies, on spin network states in the graphical method. In order to obtain the on-shell anomaly-free quantum constraint algebra, one has to employ degenerate triangulation at the co-planar vertices of spin networks in the regularization procedure of Thiemann's Hamiltonian. This problem has been overcome by a new proposed Hamiltonian constrain operator in \cite{Yang:2015zda}. 

 This paper organized as follows. Section \ref{section-II} is devoted to review briefly the elements of LQG. In section \ref{section-III}, we will introduce the graphical method to LQG. In section \ref{section-IV}, we will derive the closed formula for the matrix element of the volume operator by the graphical method. It is shown how the simple rules of transforming graphs tremendously simplify our calculation. The results are discussed in section \ref{sec-summary}. In \ref{appendix-A}, we will review the representation theory of $SU(2)$ group, including the notation of intertwiners and basic components of Brink's graphical representation and some rules of transforming graphs. The detail proof of some identities used in main text are presented in \ref{appendix-B} and \ref{appendix-C} separately.

\section{Preliminaries}\label{section-II}
In this section, we briefly summarize the elements of LQG to establish our notations and conventions. The classical starting point of LQG is the Hamiltonian formalism of GR, formulated on a 3-dimensional manifold $\Sigma$ of arbitrary topology. With Ashtekar-Barbero variables \cite{Ashtekar:1987gu,Barbero:1994ap}, GR can be cast as a dynamical theory of connection with $SU(2)$ gauge group. We denote spatial indices by $a,b,c,...$ and internal indices by $i,j,k,...=1,2,3$.  The phase space consists of canonical pairs $(A^i_a, \tilde{E}^a_i)$ of fields on $\Sigma$, where $A^i_a$ is a connection 1-form which takes values in the Lie algebra $su(2)$, and $\tilde{E}^a_i$ is a vector density of weight 1. The densitized triad $\tilde{E}^a_i$ is related to the cotriad $e_a^i$ by $\tilde{E}^a_i=\frac{1}{2}\,\tilde{\epsilon}^{abc}\epsilon_{ijk}e^j_be^k_c{\rm sgn}
(\det(e^l_d))$, wherein $\tilde{\epsilon}^{abc}$ is the Levi-Civit\`a tensor density of weight 1, and ${\rm sgn}
(\det(e^l_d))$ denotes the sign of $\det(e^l_d)$. The 3-metric on $\Sigma$ is expressed in terms of cotriads through $q_{ab}=e^i_ae^j_b\delta_{ij}$. The only nontrivial Poisson bracket reads
\begin{align}
\{A^i_a(x),\tilde{E}^b_j(y)\}=\kappa\beta\delta^b_a\delta^i_j\delta^3(x,y)\,,
\end{align}
where $\kappa=8\pi G$, and $\beta$ is the Barbero-Immirzi parameter. The behavior of connection under finite gauge transformations is
\begin{align}\label{connection-gauge}
A\mapsto A^g&=-({\rm d}g)g^{-1}+gAg^{-1}\,.
\end{align}
The fundamental variables in LQG are the holonomy of the connection along a curve and the flux of densitized triad through a 2-surface.
Given an edge $e:[0,1]\rightarrow\Sigma$, the holonomy $h_e(A)$ of connection $A^i_a$ along the edge $e$ is
\begin{align}\label{holonomy-def}
 h_e(A)={\cal P}{\rm
 exp}\left(\int_eA\right)={\mathrm I}_2+\sum_{n=1}^{\infty}\int_0^1{\rm d}t_1
 \int_{t_1}^1{\rm d}t_2\cdots\int_{t_n}^1{\rm d}t_n\;A(e(t_1))\cdots
 A(e(t_{n}))\,,
\end{align}
wherein $A(e(t))\equiv\dot{e}^a(t)A^i_a(e(t))\tau_i$, with $\dot{e}^a(t)$ being the tangent vector of $e$, and $\tau_i:=-i\sigma_i/2$ (with $\sigma_i$ being the Pauli matrices), ${\cal P}$ denotes the path ordering which orders the smallest path parameter to the left. Define a combination $\circ$ of two edges $e_1,e_2$ satisfying $e_1(1)=e_2(0)$ as
\begin{align}
[e_1\circ e_2](t):=\left\{
\begin{array}{cc}
e_1(2t), & t\in [0,\frac12]\\
e_2(2t-1), & t\in [\frac12,1]
\end{array}
\right.\,,
\end{align}
and the inversion of an edge as
\begin{align}
e^{-1}(t):=e(1-2t)\,.
\end{align}
The holonomy \eqref{holonomy-def} has two key properties
\begin{align}\label{hol-prop}
h_{e_1\circ e_2}(A)=h_{e_1}(A)h_{e_2}(A), \qquad h_{e^{-1}}(A)=h_e(A)^{-1}\,.
\end{align}
The transformation behavior \eqref{connection-gauge} of connection $A$ under gauge transformation leads to the corresponding transformation behavior of holonomy as
\begin{align}\label{holonomy-gauge}
h_e(A^g)=g(b(e))h_e(A)g(f(e))^{-1}\,,
\end{align}
where $b(e), f(e)$ denote the beginning and final points of $e$, respectively.

Consider a finite piecewise analytic graph $\gamma$ in $\Sigma$, which consists of analytic edges $e$ incident at vertices $v$. We insert a pseudo-vertex $\tilde{v}$ into each edge $e$ and split $e$ into two segments $s_e$ and $l_e$ such that $e=s_e\circ l_e^{-1}$ and the orientations of $s_e$ and $l_e$ are all outgoing from the two endpoints of $e$. We call the new graph {\em the standard graph} obtained from the original graph by splitting edges and adding pseudo-vertices. Denote the standard graph by $\gamma$, the set of its edges by $E(\gamma)$, and the set of vertices, containing the true vertices $v$ and pseudo vertices $\tilde{v}$, by $V(\gamma)$. Our following discussion is based on the standard graphs.

To construct quantum kinematics, one has to extend the configuration space ${\cal A}$ of smooth connections to the space $\bar{\cal A}$ of distributional connections. A function $f$ on $\bar{\cal A}$ is said to be cylindrical with respect to a graph $\gamma$ if and only if it can be written as $f=f_\gamma\circ p_{\gamma}$, wherein $p_\gamma(A)=(h_{e_1}(A),..,h_{e_n}(A))$ and $e_1,..,e_n$ are the edges of $\gamma$. Here $h_e(A)$ is the holonomy along $e$ evaluated at $A\in\bar{\cal A}$ and $f_\gamma$ is a complex-valued function on $SU(2)^n$. Since a function cylindrical with respect to a graph $\gamma$ is automatically cylindrical with respect to any graph bigger than $\gamma$, a cylindrical function is actually given by a whole equivalence class of functions $f_\gamma$. We will henceforth not distinguish the functions in one equivalence class. The set of cylindrical functions is denoted by ${\rm Cyl}(\bar{\cal A})$. The space ${\rm Cyl}(\bar{\cal A})$ can be completed as a Hilbert space, {\em the kinematical Hilbert space} ${\cal H}_{\rm kin}$.

Given a standard graph $\gamma$, we assign each edge $e$ with an unitary irreducible representation with spin $j_e$, two edges incoming to a pseudo-vertex $\tilde{v}$ with the same representation, each true vertex $v$ with an intertwiner $i_v$, and pseudo vertex $\tilde{v}$ with the conjugate intertwiner $i^*_{\tilde{v}}$. Then {\em the spin network state} reads \footnote{The spin network state can be normalized by multiplying $\sqrt{2j_e+1}$ for all $e\in E(\gamma)$.}
\begin{align}\label{spin-network-state}
T_{\gamma,\vec{j},\vec{i}}(A):=\bigotimes_{v\in V(\gamma)} i_v\cdot \bigotimes_{e\in E(\gamma)}\;\pi_{j_e}(h_{e}(A))\cdot\bigotimes_{\tilde{v}\in V(\gamma)}i_{\tilde{v}}^*\,,
\end{align}
where $\cdot$ stands for contracting the upper (or former) indices of representation matrices $\pi_{j_e}(h_{e}(A))$ with indices of intertwines $i_v$ at true vertices $v$, the lower (or later) indices of $\pi_{j_e}(h_{e}(A))$ with indices of conjugate intertwiners $i_{\tilde{v}}^*$ at pseudo vertices $\tilde{v}$. Given $n$ edges with $n$ spins $j_1,\cdots, j_n$ incident at a true $v$, matrix elements of the intertwiner $i_v$ associated to $v$ take the complex conjugate of (generalized) Clebsch-Gordan coefficients (CGCs) up to a constant factor (see \ref{appendix-A} for detailed explanation), i.e.,
\begin{align}\label{intertw-true}
{\left(i^{\,J;\,\vec{a}}_v\right)_{m_1m_2\cdots m_n}}^M\equiv{\left(i^{\,J;\,\vec{a}}_{j_1\cdots j_n}\right)_{\,m_1m_2\cdots m_n}}^M&:=(-1)^{j_1-\sum_{i=2}^nj_i-J}\langle JM;\vec{a}\;|\,j_1m_1j_2m_2\cdots j_nm_n\rangle\notag\\
&=\/(-1)^{j_1-\sum_{i=2}^nj_i-J}\sum_{k_2,\cdots,k_{n-1}}\langle a_2k_2|j_1m_1j_2m_2\rangle\langle a_3k_3|a_2k_2j_3m_3\rangle\cdots\langle JM|a_{n-1}k_{n-1}j_nm_n\rangle\,,
\end{align}
where $\langle JM;\vec{a}\;|\,j_1m_1j_2m_2\cdots j_nm_n\rangle$ is the complex conjugate of generalized CGCs, which describe coupling $n$ angular momenta $j_1,\cdots,j_n$ to a total angular momentum $J$ in the standard coupling scheme (that is, we first couple $j_1$ to $j_2$ to give a resultant $a_2$, and then couple $a_2$ to $j_3$ to give $a_3$, and so on), and $\vec{a}\equiv\{a_2,\cdots,a_{n-1}\}$ denotes the set of the angular momenta appeared in the intermediate coupling. Notice that the intertwiner presented in Eq. \eqref{intertw-true}, differing the factor $(-1)^{j_1-\sum_{i=2}^nj_i-J}$ from CGCs, is more convenient to be simply represented in graphical formula. Matrix elements of the conjugate intertwiner $i_{\tilde{v}}^*$ associated to pseudo vertex $\tilde{v}$ at which two edges with the same spin $j$ incoming are given by
\begin{align}
\left({i^0_{\tilde{v}}}^*\right)^{n_1n_2}\equiv{\left({i^0_{\tilde{v}}}^*\right)_0}^{n_1n_2}:=\langle jn_1jn_2|00\rangle\,.
\end{align}
The assignment of intertwiners to the true vertices and conjugate intertwiners to the pseudo vertices is compatible with the transformation behavior \eqref{holonomy-gauge} of holonomy.
The CGCs are usually chosen to be real so that
\begin{align}
\langle j_1m_1j_2m_2|JM\rangle=\langle JM|j_1m_1j_2m_2\rangle\,.
\end{align}
It is, therefore, not necessary to sedulously distinguish intertwiner from its conjugate when we do calculation.

The {\em gauge-invariant} spin network states, which keep invariant under gauge transformations, correspond to the states whose intertwiners in \eqref{intertw-true} associated to true vertices are specially chosen such that the resulting angular momenta $J=0$. The spin network states which are variant under gauge transformations are called {\em gauge-variant} spin network states. The normalized gauge-invariant/variant states consist of the orthonormal basis of the gauge-invariant/variant Hilbert space \cite{Ashtekar:1995zh}.

Two elementary operators in LQG are the holonomy and the flux operators. The holonomy operator acts as a multiplication operator, while flux, essentially the self-adjoint right-invariant operator $J^i$, acts as a derivative operator,
\begin{align}
{[\hat{h}_{e_I}(A)]^{B}}_{C}\cdot f_\gamma(h_{e_1}(A),\cdots,h_{e_n}(A))&:={\left[\pi_{1/2}(h_{e_I}(A))\right]^{B}}_{C}f_\gamma(h_{e_1}(A),\cdots,h_{e_n}(A))\,,\label{holonomy-operator-def}\\
 J^i_{e_I}\cdot\,f_\gamma(h_{e_1}(A),\cdots,h_{e_I}(A),\cdots,h_{e_n}(A))&:=-i\left.\frac{{\rm d}}{{\rm d}t}\right|_{t=0}f_\gamma(h_{e_1}(A),\cdots,e^{t\tau_i}h_{e_I}(A),\cdots,h_{e_n}(A))\,.\label{right-inv-operator-def}
\end{align}

\section{Graphical method for LQG}\label{section-III}
\subsection{Algebraic formula}
In LQG, under different physical considerations, one needs to construct operators, e.g., the geometric operators and the Hamiltonian operators, corresponding to their classical quantities based on the two elementary operators $\hat{h}_e(A)$ and $J^i_e$. The action of those operators on a given spin network state will involve the actions of the two elementary operators. The action of $\hat{h}_e(A)$ on the spin network states involves essentially the decomposition of the tensor product representation of $SU(2)$, which is well known as the Clebsch-Gordan series
\begin{align}\label{reps-couple-main}
{[\pi_{j_1}(g)]^{m_1}}_{\,n_1}{[\pi_{j_2}(g)]^{m_2}}_{n_2}&=\sum_{J,M,N}{\left((i^{\,J}_{j_1j_2})^{-1}\right)_M}^{\;m_1m_2}{[\pi_J(g)]^M}_{\,N}{\left(i^{\,J}_{j_1j_2}\right)_{n_1n_2}}^N\,,
\end{align}
where ${\left(i^{\,J}_{j_1j_2}\right)_{n_1n_2}}^N\equiv(-1)^{j_1-j_2-J}\langle JN|j_1n_1j_2n_2\rangle$.
The fact that the operator $i^{\,J}_{j_1j_2}$ is unitary and its matrix elements take real numbers results in
\begin{align}
{\left((i^{\,J}_{j_1j_2})^{-1}\right)_M}^{\;m_1m_2}={\left(i^{\,J}_{j_1j_2}\right)_{m_1m_2}}^M\,.
\end{align}
 Given a spin network state $T_{\gamma,\vec{j},\vec{i}}(A)$ on a graph $\gamma$, we consider a true vertex $v\in V(\gamma)$ at which $n$ edges $e_1,\cdots e_n$ incident and denote $T^v_{\gamma,\vec{j},\vec{i}}(A)$ the terms, in $T_{\gamma,\vec{j},\vec{i}}(A)$ , directly associated to $v$.  Then the action of the holonomy operator ${[\hat{h}_{e_I}(A)]^B}_C$ on $T^v_{\gamma,\vec{j},\vec{i}}(A)$ reads
 \begin{align}\label{acton-h-alg}
 {[\hat{h}_{e_I}(A)]^B}_C\cdot T^v_{\gamma,\vec{j},\vec{i}}(A)&={\left(i^{\,J;\,\vec{a}}_v\right)_{m_1\cdots m_I\cdots  m_n}}^M\,{[\pi_{j_1}(h_{e_1})]^{m_1}}_{\,n_1}\cdots{[\pi_{j_I}(h_{e_I})]^{\,m_I}}_{n_I}{\left[\pi_{1/2}(h_{e_I})\right]^B}_C\cdots{[\pi_{j_n}(h_{e_n})]^{m_n}}_{\,n_n}\notag\\
 &={\left(i^{\,J;\,\vec{a}}_v\right)_{m_1\cdots m_I\cdots  m_n}}^M\,{[\pi_{j_1}(h_{e_1})]^{m_1}}_{\,n_1}\cdots\sum_{j'_I,m'_I,n'_I}{\left((i^{j'_I}_{j_I1/2})^{-1}\right)_{m'_I}}^{\;m_IB}{[\pi_{j'_I}(h_{e_I})]^{\,m'_I}}_{n'_I}{\left(i^{j'_I}_{j_I1/2}\right)_{n_IC}}^{n'_I}\cdots{[\pi_{j_n}(h_{e_n})]^{m_n}}_{\,n_n}\,.
 \end{align}
 where ${[\pi_{j_I}(h_{e_I})]^{\,m_I}}_{n_I}\equiv {[\pi_{j_I}(h_{e_I}(A))]^{\,m_I}}_{n_I}$.
 On the other hand,  $J^i_e$ is a derivate operator acting on a given spin network state. The action of $J^i_{e_I}$ defined in \eqref{right-inv-operator-def} on $T^v_{\gamma,\vec{j},\vec{i}}(A)$ reads
\begin{align}\label{J-i-snf-vertex}
J^i_{e_I}\cdot T^v_{\gamma,\vec{j},\vec{i}}(A)&={\left(i^{\,J;\,\vec{a}}_v\right)_{m_1\cdots m_I\cdots  m_n}}^M\,{[\pi_{j_1}(h_{e_1})]^{m_1}}_{\,n_1}\cdots\left(-i\left.\frac{{\rm d}}{{\rm d}t}\right|_{t=0}{[\pi_{j_I}(e^{t\tau_i}h_{e_I})]^{\,m_I}}_{n_I}\right)\cdots{[\pi_{j_n}(h_{e_n})]^{m_n}}_{\,n_n}\notag\\
&={\left(i^{\,J;\,\vec{a}}_v\right)_{m_1\cdots m_I\cdots  m_n}}^M\,{[\pi_{j_1}(h_{e_1})]^{m_1}}_{\,n_1}\cdots \left(-i\,{[\pi_{j_I}\left(\tau_i\right)]^{m_I}}_{\,m'_I}{[\pi_{j_I}\left(h_{e_I}\right)]^{m'_I}}_{\,n_I}\right)\cdots{[\pi_{j_n}(h_{e_n})]^{m_n}}_{\,n_n}
\notag\\
&=\left[{\left(i^{\,J;\,\vec{a}}_v\right)_{m_1\cdots m'_I\cdots  m_n}}^M\left(-i\,{[\pi_{j_I}(\tau_i)]^{m'_I}}_{\,m_I}\right)\right]{[\pi_{j_1}(h_{e_1})]^{m_1}}_{\,n_1}\cdots{[\pi_{j_I}(h_{e_I})]^{m_I}}_{\,n_I} \cdots{[\pi_{j_n}(h_{e_n})]^{m_n}}_{\,n_n}\,,
\end{align}
which indicates that $J^i_{e_I}$ leaves $\gamma$ and $\vec{j}$ invariant, but does change the intertwiner associated to $v$ by contracting matrix elements of the $i$-th $\tau$ with the intertwiner in the following way,
\begin{align}\label{action-inva}
J^i_{e_I}\cdot \;{\left(i^{\,J;\,\vec{a}}_v\right)_{m_1\cdots m_I\cdots  m_n}}^M&={\left(i^{\,J;\,\vec{a}}_v\right)_{m_1\cdots m'_I\cdots  m_n}}^M\left(-i\,{[\pi_{j_I}(\tau_i)]^{m'_I}}_{\,m_I}\right)\,.
\end{align}
However, in practical calculation, it is not convenient to directly compute the contraction of matrix elements of $\tau_i$ with an intertwiner. One usually introduces the irreducible tensor operators \cite{Edmonds}, or the spherical tensors of $\tau_i$, to replace the original $\tau_i$ (for a reason that will become clear in a moment). The spherical tensors $\tau_\mu$ ($\mu=0,\pm1$), corresponding to $\tau_i$ ($i=1,2,3$), are defined by
\begin{align}\label{tau-mu}
\tau_0:=\tau_3,\qquad \tau_{\pm1}:=\mp\frac{1}{\sqrt{2}}\left(\tau_1\pm i \tau_2\right)\,.
\end{align}
Then the contraction of matrix elements of $\tau_i$ with an intertwiner is transformed to that of their tensor operators with the intertwiner. The matrix elements ${[\pi_j(\tau_\mu)]^{m'}}_{\,m}$ can be related to the $3j$-symbols (or CGCs) by (see Appendix \ref{appendix-B-1} for proof)
\begin{align}\label{spher-rep}
{[\pi_j(\tau_\mu)]^{m'}}_{\,m}&=\frac{i}{2}\sqrt{2j(2j+1)(2j+2)}\,
\begin{pmatrix}
  1 & j & j \\
 \mu & m''  &  m \\
\end{pmatrix}C^{m''m'}_{(j)}\,,
\end{align}
where $C^{m''m'}_{(j)}:=(-1)^{j+m'}\delta_{m',-m''}$ is the contravariant ``metric'' tensor on the irreducible representation space ${\cal H}_j$ of $SU(2)$ with spin $j$ (see Appendix \ref{appendix-A-1} for a detailed explaination for the $C^{m''m'}_{(j)}$) \cite{Wigner-bk}.
The spherical tensor $\tau_\mu$ generates the self-adjoint right-invariant operator $J^\mu_{e_I}$ defined by
\begin{align}
 J^\mu_{e_I}\cdot\,f_\gamma(h_{e_1}(A),\cdots,h_{e_I}(A),\cdots,h_{e_n}(A))&:=-i\left.\frac{{\rm d}}{{\rm d}t}\right|_{t=0}f_\gamma(h_{e_1}(A),\cdots,e^{t\tau_\mu}h_{e_I}(A),\cdots,h_{e_n}(A))\,.\label{mu-right-inv-operator-def}
\end{align}
The action of $J^\mu_{e_I}$ on ${\left(i^{\,J;\,\vec{a}}_v\right)_{m_1\cdots m_I\cdots  m_n}}^M$ reads
\begin{align}\label{action-inva-mu}
J^\mu_{e_I}\cdot \;{\left(i^{\,J;\,\vec{a}}_v\right)_{m_1\cdots m_I\cdots  m_n}}^M&={\left(i^{\,J;\,\vec{a}}_v\right)_{m_1\cdots m'_I\cdots  m_n}}^M\left(-i\,{[\pi_{j_I}(\tau_\mu)]^{m'_I}}_{\,m_I}\right)\,.
\end{align}
Any gauge-invariant operator, e.g., the volume operator considered in this paper, defined by $J^i$s can be expressed in terms of the corresponding $J^\mu$s. Hence its action on the spin-network states essentially is equivalent to contracting $3j$-symbols (or CGCs) with corresponding intertwiners.

\subsection{Graphical representation and graphical calculation}\label{subsect-III-2}
The basic components of the graphical representation and the simple rules of transforming graphs are presented in Appendix \ref{appendix-A-1}. In graphical representation, the $3j$-symbol is represented by an oriented node with three lines, each of which represents a value of $j$, namely
\begin{align}
\begin{pmatrix}
j_1 & j_2 & j_3\\
m_1 & m_2 & m_3
\end{pmatrix}
&=\makeSymbol{
\includegraphics[width=2cm]{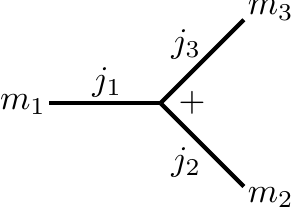}}=\makeSymbol{
\includegraphics[width=2cm]{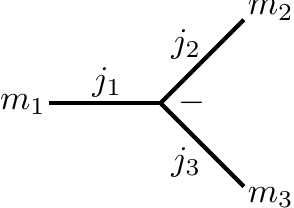}}\,,
\end{align}
where $-$ and $+$ denote a clockwise orientation and an anti-clockwise orientation respectively. A rotation of the diagram does not change the cyclic order of lines, and the angles between two lines as well as their lengths at a node have no significance.  The ``metric'' tensor $C^{(j)}_{m'm}$ in Eq. \eqref{metric-tensor} which occurs in the contraction of two $3j$-symbols with the same values of $j$ is denoted by a line with an arrow on it, i.e.,
\begin{align}
C^{(j)}_{m'm}=(-1)^{j-m}\delta_{m,-m'}=(-1)^{j+m'}\delta_{m,-m'}=\makeSymbol{
\includegraphics[width=2cm]{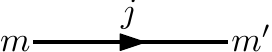}}\,,
\end{align}
and its inverse in Eq. \eqref{metric-tensor-inverse} can be expressed as
\begin{align}
C^{mm'}_{(j)}=(-1)^{j-m}\delta_{m,-m'}=(-1)^{j+m'}\delta_{m,-m'}=\makeSymbol{
\includegraphics[width=2cm]{graph/wigner-symbol/wigner-3j-symbol-4}}\,.
\end{align}
Summation over the magnetic quantum numbers $m$ is graphically represented by joining the free ends of the corresponding lines.
The contraction of a $3j$-symbol with a ``metric'' is represented by a node with one arrow, which provides us a way to representing the CGC, e.g.,
\begin{align}
\langle j_3m_3| j_1m_1j_2m_2\rangle&=(-1)^{j_1-j_2-j_3}\sqrt{2j_3+1}\begin{pmatrix}
j_1 & j_2 & j_3\\
m_1 & m_2 & m'_3
\end{pmatrix}C^{m'_3m_3}_{(j_3)}\notag\\
&=(-1)^{j_1-j_2-j_3}\sqrt{2j_3+1}\makeSymbol{
\includegraphics[width=2.4cm]{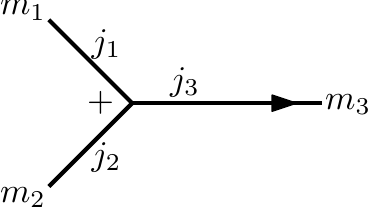}}=(-1)^{j_1-j_2-j_3}\sqrt{2j_3+1}\makeSymbol{
\includegraphics[width=2.6cm]{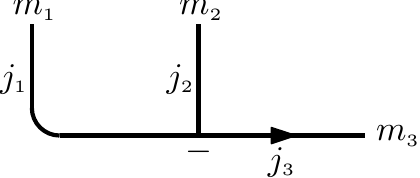}}\,.
\end{align}
This is the main motivation for Brink to modify the original Yutsis scheme \cite{Yutsis:1962bk,brink1968angular}.
Hence the intertwiner ${\left(i^{\,J;\,\vec{a}}_v\right)_{\,m_1m_2\cdots m_n}}^M$ in Eq. \eqref{intertw-true} associated to a true vertex $v$, from which $n$ edges are outgoing, is represented in graphical formula by Eq. \eqref{graph-intertwiner} as (see Appendix \ref{appendix-A-2} for a detailed interpretation)
\begin{align}\label{intertwined-form}
{\left(i^{\,J;\,\vec{a}}_v\right)_{\,m_1m_2\cdots m_n}}^M&\equiv{\left(i^{\,J;\,\vec{a}}_{j_1\cdots j_n}\right)_{\,m_1m_2\cdots m_n}}^M=\prod_{i=2}^{n-1}\sqrt{2a_i+1}\sqrt{2J+1}\;
\makeSymbol{\includegraphics[width=6.2cm]{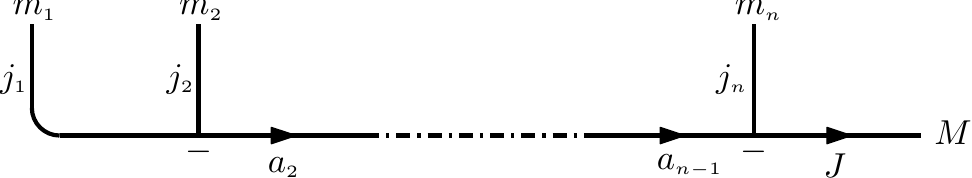}}\,.
\end{align}
Now we will extend Brink's representation and propose a graphical representation for the unitary irreducible representation $\pi_j$ of $SU(2)$. The matrix element ${[\pi_j(g)]^m}_n$ is denoted by a blue line with a hollow arrow (triangle) in it as
\begin{align}\label{rep-group-graph-main}
{[\pi_j(g)]^m}_n=:\;\makeSymbol{
\includegraphics[width=2.8cm]{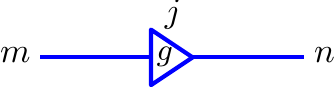}}\,.
\end{align}
The orientation of the arrow is from its row index $m$ to its column index $n$. The Clebsch-Gordan series in \eqref{reps-couple-main} can be represented by
\begin{align}\label{reps-couple-graph-main}
\makeSymbol{
\includegraphics[width=3cm]{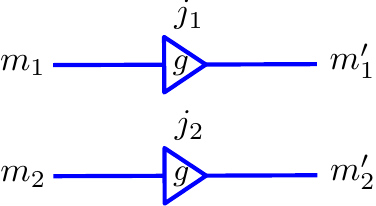}}&=\sum_{j_3}(2j_3+1)\makeSymbol{
\includegraphics[width=5cm]{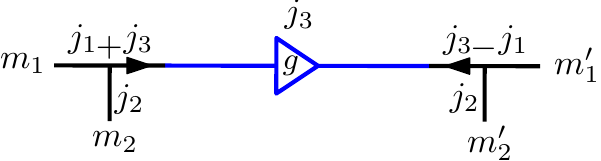}}
=\sum_{j_3}(2j_3+1)\makeSymbol{
\includegraphics[width=5cm]{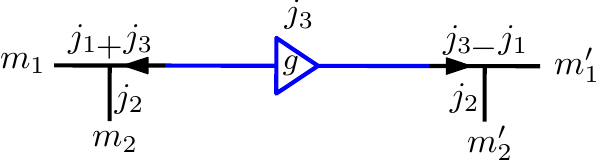}}\,,
\end{align}
where, in the second step, we used \eqref{arrow-flip} to flip the orientations of two arrows. The matrix element ${[\pi_j(g^{-1})]^n}_{\,m}$ can be presented in Eq. \eqref{rep-inverse-graph} by 
\begin{align}\label{rep-inverse-graph-main}
{[\pi_j(g^{-1})]^n}_{\,m}=C^{(j)}_{mm'}\,{[\pi_j(g)]^{m'}}_{n'}C^{n'n}_{(j)}=\makeSymbol{
\includegraphics[width=2.8cm]{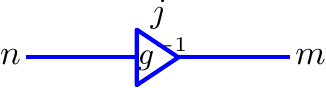}}=\makeSymbol{
\includegraphics[width=4cm]{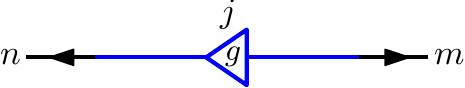}}\,.
\end{align}
Using Eqs. \eqref{rep-inverse-graph-main} and \eqref{reps-couple-graph-main}, we have
\begin{align}\label{reps-inverse-couple-graph-main}
\makeSymbol{
\includegraphics[width=3cm]{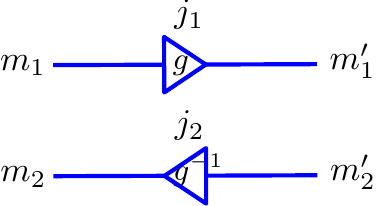}}&=\makeSymbol{
\includegraphics[width=4.1cm]{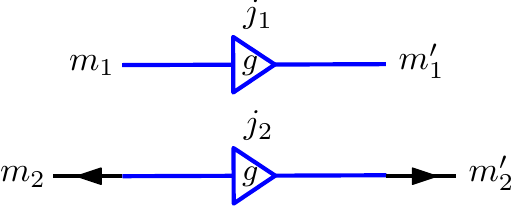}}=\sum_{j_3}(2j_3+1)\makeSymbol{
\includegraphics[width=4.8cm]{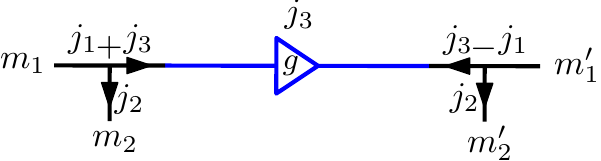}}\notag\\
&=\sum_{j_3}(2j_3+1)\makeSymbol{
\includegraphics[width=4.8cm]{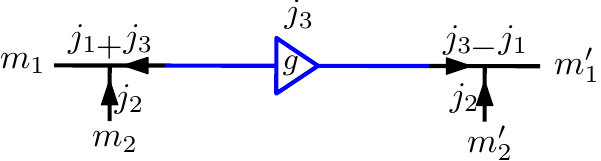}}\,.
\end{align}
For group elements as the holonomies $h_{e}\equiv h_e(A)$ of connection $A$ along an edge $e$, their matrix elements can be simply represented by
\begin{align}\label{holonomy-arrow}
{[\pi_j(h_e)]^m}_n&=\makeSymbol{
\includegraphics[width=2.8cm]{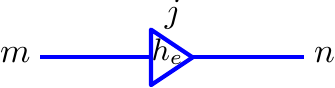}}\notag\\
&=:\makeSymbol{
\includegraphics[width=2.4cm]{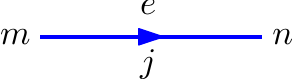}}\,.
\end{align}
By Eqs. \eqref{hol-prop}, \eqref{rep-inverse-graph-main} and \eqref{holonomy-arrow}, the matrix elements of the inverse of a holonomy can be represented by
\begin{align}\label{holonomy-inverse-arrow}
{[\pi_j(h_e^{-1})]^n}_{\,m}={[\pi_j(h_{e^{-1}})]^n}_{\,m}=\makeSymbol{
\includegraphics[width=2.8cm]{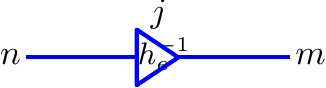}}&=\makeSymbol{
\includegraphics[width=2.8cm]{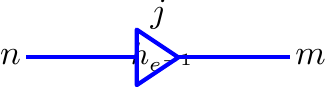}}=\makeSymbol{
\includegraphics[width=4cm]{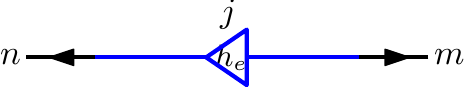}}\notag\\
&=:\makeSymbol{
\includegraphics[width=2.4cm]{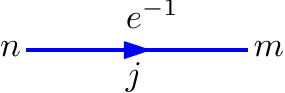}}\hspace{0.3cm}=\makeSymbol{
\includegraphics[width=3.6cm]{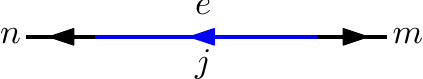}}\,.
\end{align}
Eqs. \eqref{reps-couple-graph-main} and \eqref{reps-inverse-couple-graph-main} yield the coupling rules of representations of holonomies as
\begin{align}\label{holonomy-reps-couple}
\makeSymbol{
\includegraphics[width=2.8cm]{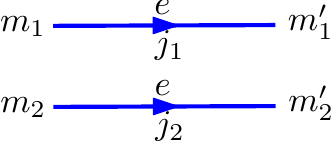}}&=\makeSymbol{
\includegraphics[width=4.8cm]{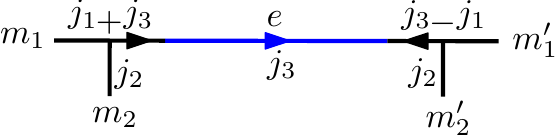}}=\sum_{j_3}(2j_3+1)\makeSymbol{
\includegraphics[width=4.8cm]{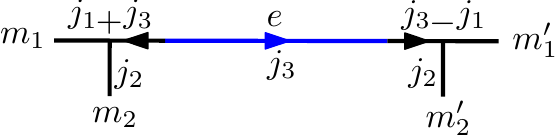}}
\end{align}
and
\begin{align}
\makeSymbol{
\includegraphics[width=2.8cm]{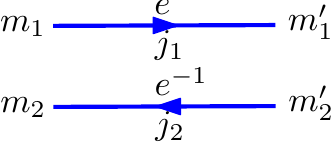}}&=\makeSymbol{
\includegraphics[width=4.8cm]{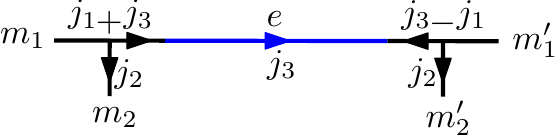}}=\sum_{j_3}(2j_3+1)\makeSymbol{
\includegraphics[width=4.8cm]{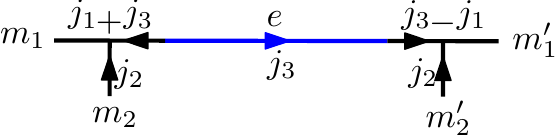}}\label{holonomy-reps-inverse-couple}\,.
\end{align}
Then the spin network states $T^v_{\gamma,\vec{j},\vec{i}}(A)$ associated to $v$ can be represented by
\begin{align}
T^v_{\gamma,\vec{j},\vec{i}}(A)&={\left(i^{\,J;\,\vec{a}}_v\right)_{m_1\cdots m_I\cdots  m_n}}^M\,{[\pi_{j_1}(h_{e_1})]^{m_1}}_{\,n_1}\cdots{[\pi_{j_I}(h_{e_I})]^{\,m_I}}_{n_I}\cdots{[\pi_{j_n}(h_{e_n})]^{m_n}}_{\,n_n}\notag\\
&=\prod_{i=2}^{n-1}\sqrt{2a_i+1}\sqrt{2J+1}\;
\makeSymbol{\includegraphics[width=10cm]{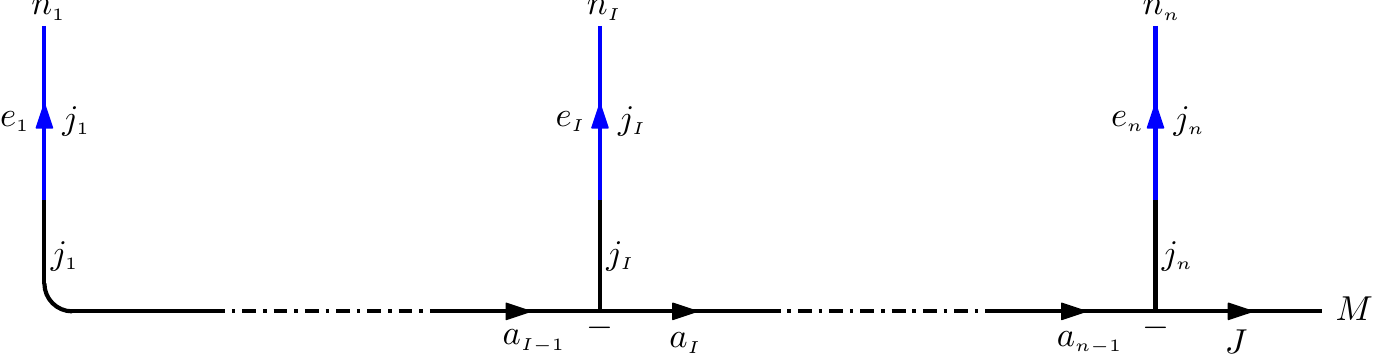}}\,.
\end{align}
The action of $ {[\hat{h}_{e_I}(A)]^B}_C$ on the spin network state given in \eqref{acton-h-alg} can be represented by
\begin{align}
 {[\hat{h}_{e_I}(A)]^B}_C\cdot T^v_{\gamma,\vec{j},\vec{i}}(A)&=\prod_{i=2}^{n-1}\sqrt{2a_i+1}\sqrt{2J+1}\;\makeSymbol{
\includegraphics[width=10cm]{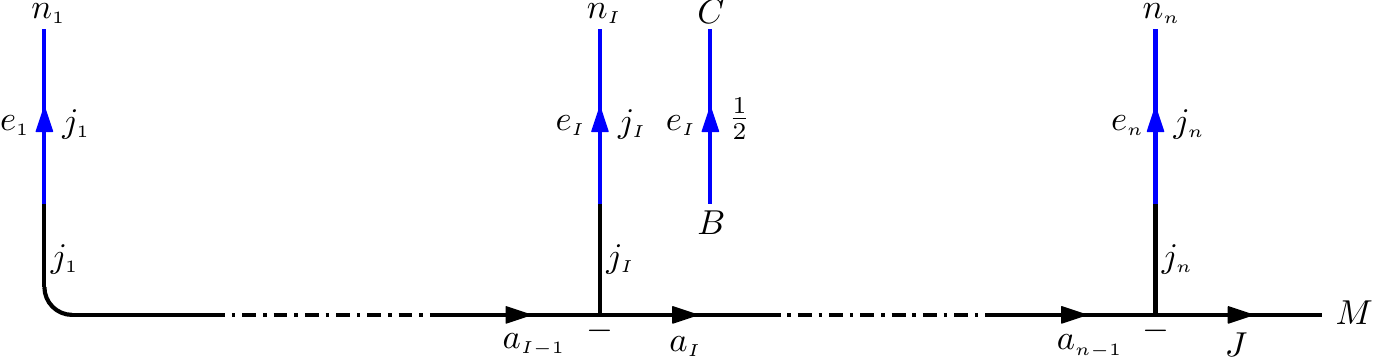}}\notag\\
&=\prod_{i=2}^{n-1}\sqrt{2a_i+1}\sqrt{2J+1}\sum_{j'_I}(2j'_I+1)\;\makeSymbol{
\includegraphics[width=10cm]{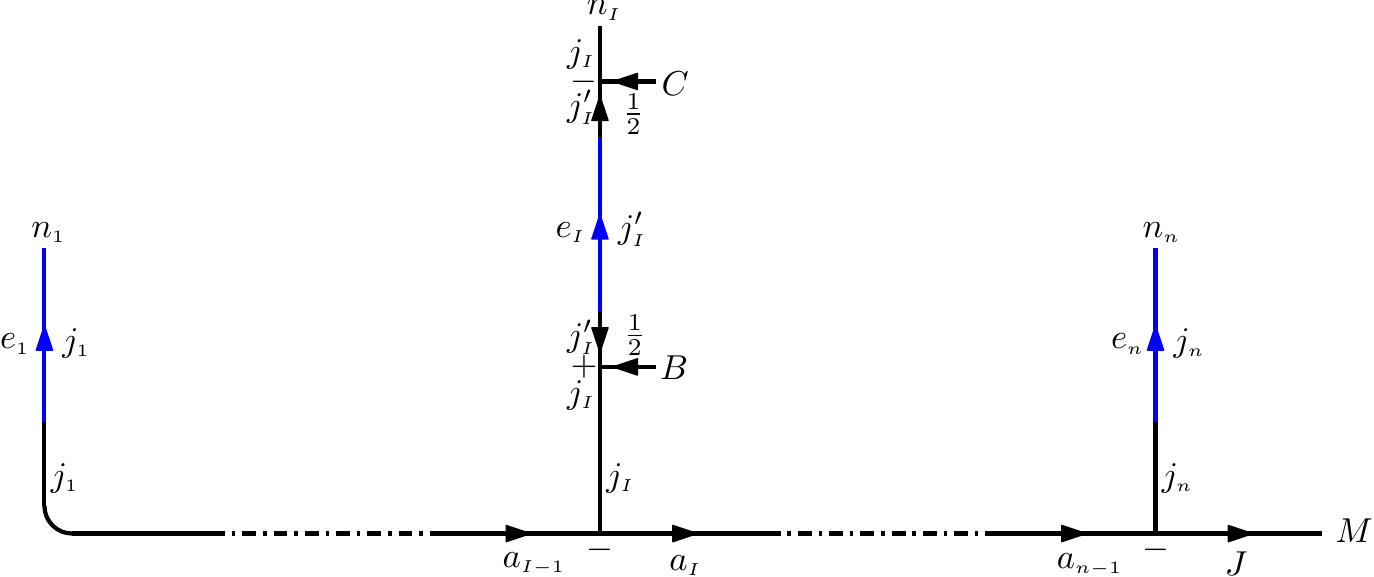}}\,.
\end{align}
The spherical tensor ${[\pi_j(\tau_\mu)]^{m'}}_m$ in Eq. \eqref{spher-rep} can be represented graphically by
\begin{align}\label{spher-rep-graph}
{[\pi_j(\tau_\mu)]^{m'}}_m&=\frac{i}{2}\sqrt{2j(2j+1)(2j+2)}\makeSymbol{
\includegraphics[width=1.5cm]{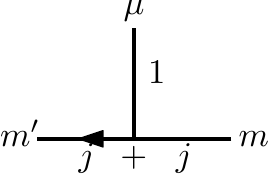}}=\frac{i}{2}\sqrt{2j(2j+1)(2j+2)}\makeSymbol{
\includegraphics[width=1.6cm]{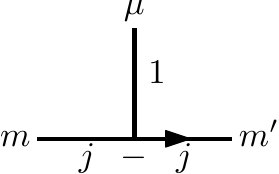}}\,.
\end{align}
Hence the action of $J^\mu_{e_I}$ on ${\left(i^{\,J;\,\vec{a}}_v\right)_{\,m_1m_2\cdots m_n}}^M$ in \eqref{action-inva-mu} can be represented by
\begin{align}
J^\mu_{e_I}\cdot{\left(i^{\,J;\,\vec{a}}_v\right)_{\,m_1m_2\cdots m_n}}^M&=\frac12\sqrt{2j(2j+1)(2j+2)}\prod_{i=2}^{n-1}\sqrt{2a_i+1}\sqrt{2J+1}\;\notag\\
&\hspace{3cm}\times\makeSymbol{
\includegraphics[width=10cm]{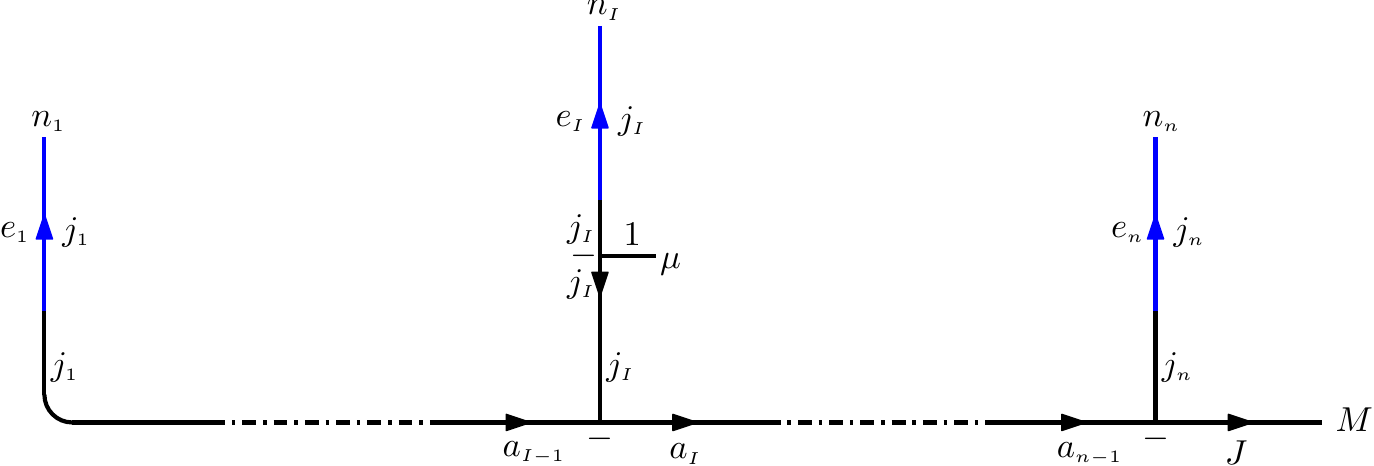}}\,.
\end{align}
The graphical calculation is to use the simple rules of transforming graphs (see \ref{appendix-A-2}) to derive the result.

The starting point of our scheme is the so-called standard graph $\gamma_{\rm std}$, which is obtained from its original graph $\gamma_{\rm org}$ by splitting edges and adding pseudo-vertices. We still need to show that the spin network function associated to the original graph $\gamma_{\rm org}$ is equivalent to the one associated to its corresponding standard graph $\gamma_{\rm std}$ when acted by an operator, e.g., the volume operator. Recall that the standard graph $\gamma_{\rm std}$ is obtained from $\gamma_{\rm org}$ by the following procedure. We insert a pseudo-vertex $\tilde{v}$ into each edge $e$ of $\gamma_{\rm org}$ and split $e$ into two segments $s_e$ and $l_e$, such that $e=s_e\circ l_e^{-1}$ and the orientations of $s_e$ and $l_e$ are all outgoing from the two endpoints of $e$. The standard graph $\gamma_{\rm std}$ consists of the new segments $s_e$ and $l_e$, the new adding pseudo-vertices $\tilde{v}$, and the (old) vertices of $\gamma_{\rm org}$. We can transform the spin networks based on the original graph into those on its standard graph by explicit transformation rules, and then find out their relation. Consider an edge $e$ with irrepresentaion $j$ in $\gamma_{\rm org}$ starting from $v$ and ending at $v'$, assigning the intertwiners $i_v$ and $i_{v'}$, respectively. We assume the edge $e$ in the original graph $\gamma_{\rm org}$ is regarded as the $k$-th edge and the $k'$-th edge in the set of edges which incident at $v$ and $v'$ respectively, i.e., $b(e)=b(e_k), f(e)=f(e'_{k'})$. The relevant ingredient of a spin network state associated to the edge $e$ takes the form (see \eqref{i-general-expression-graph} for the graphical representation of the intertwiner associated to $v$ at which there are coming and outgoing edges, \eqref{holonomy-arrow} and \eqref{holonomy-inverse-arrow} for the graphcal representation of the holonomy)
\begin{align}
\left.T_{\gamma_{\rm org},\vec{j},\vec{i}}\right|_{e,v,v'}&={(i^{\,J;\,\vec{a}}_v)_{\cdots m_k\cdots}}^{\,\cdots M}\,{[\pi_j(h_e)]^{m_k}}_{\,m'_{k'}}{(i^{\,J';\,\vec{a}'}_{v'})_{\cdots}}^{\,\cdots m'_{k'}\cdots M'}=\makeSymbol{\includegraphics[width=7cm]{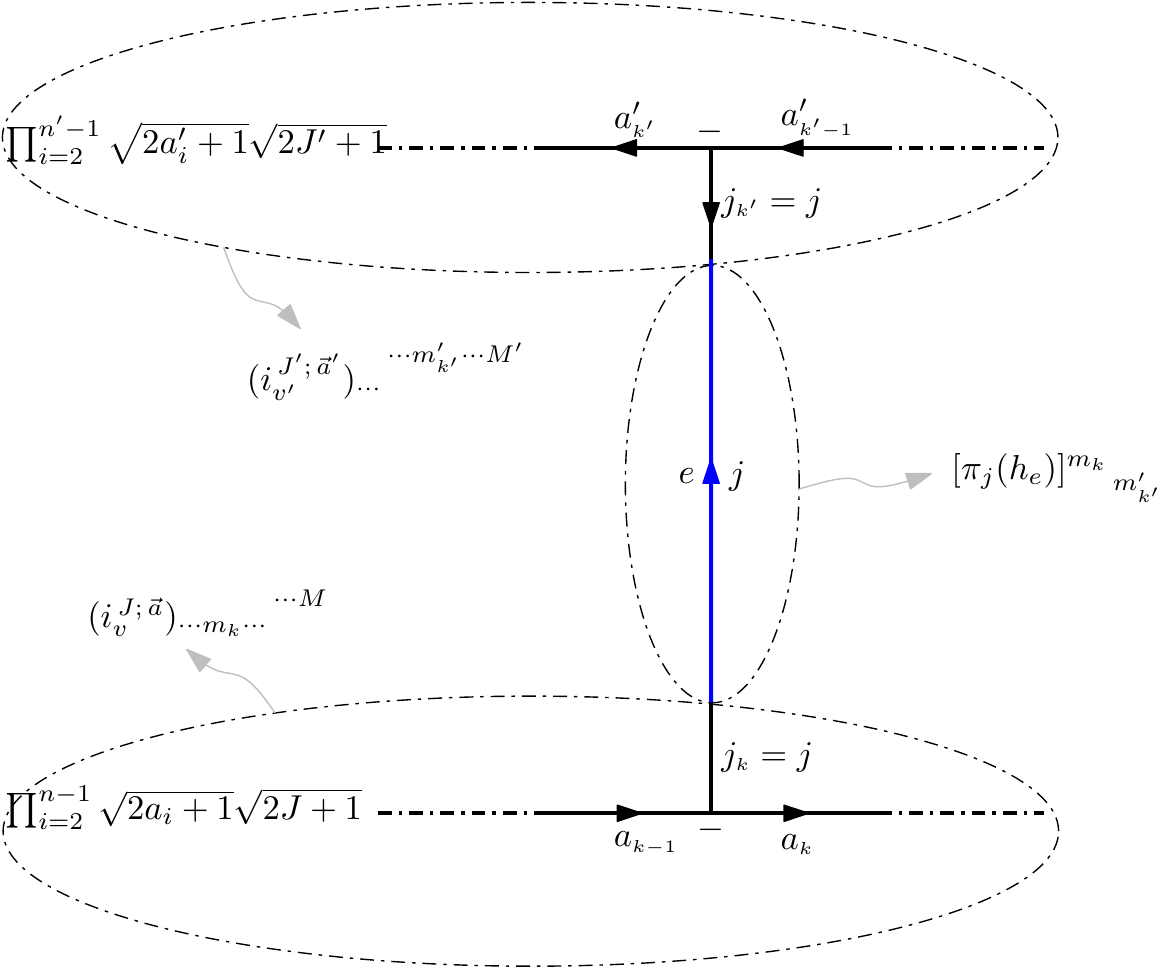}}\,.
\end{align}
The transformation from $\gamma_{\rm org}$ to $\gamma_{\rm std}$ induces the following transformation in the spin network state,
\begin{align}\label{standard-snf}
\makeSymbol{\includegraphics[width=3.8cm]{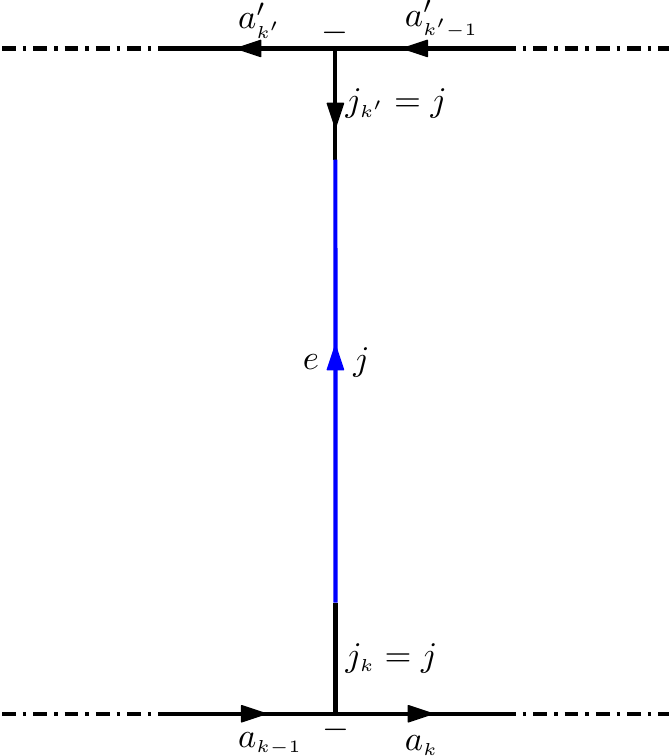}}&=\makeSymbol{\includegraphics[width=3.8cm]{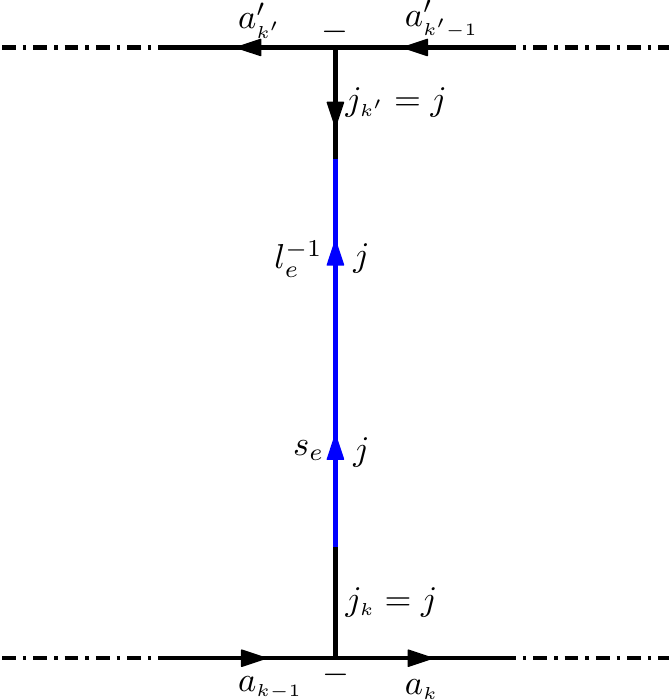}}=\makeSymbol{\includegraphics[width=3.8cm]{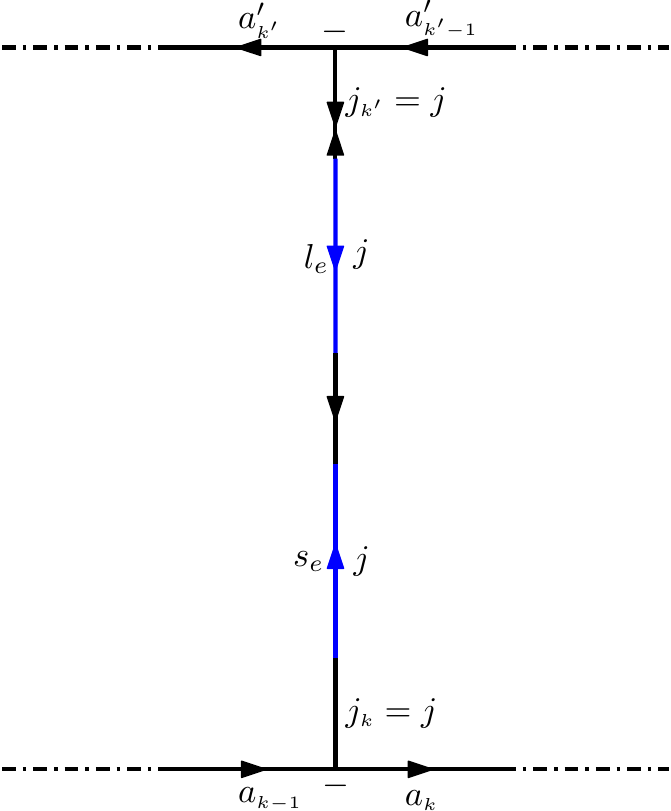}}=\makeSymbol{\includegraphics[width=3.8cm]{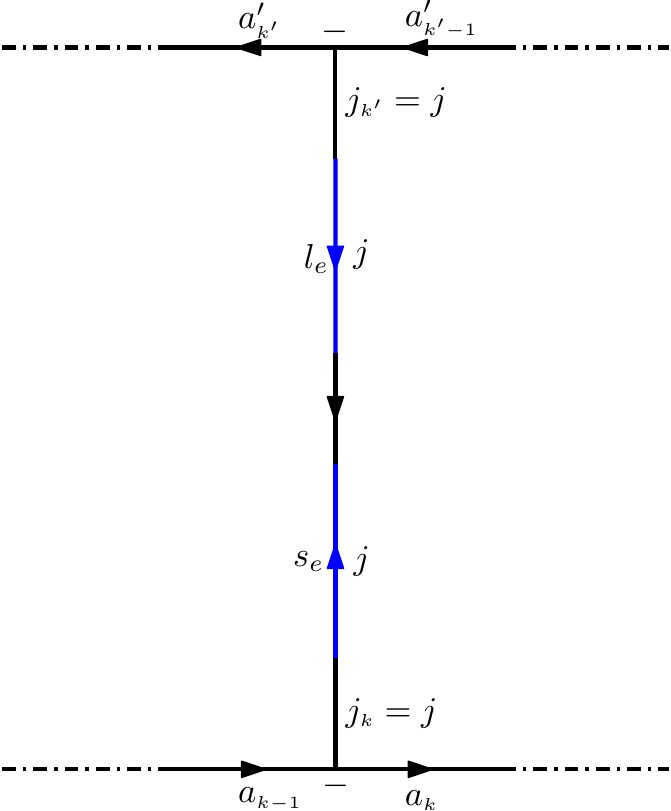}}\,,
\end{align}
where we have used \eqref{holonomy-inverse-arrow} in second step and the rule \eqref{two-arrow-cancel} to remove two arrows in the last step.
Repeating the above procedure, we can transform the spin network states associated to the origin graph into those corresponding to its standard graph. By this trick, we finally find out the corresponding relation of the spin network states between the original graph $\gamma_{\rm org}$ and its standard graph $\gamma_{\rm std}$.
The intertwiners associated to the vertices of the origin graph is replaced by its standard formula, and the adding pseudo vertex $\tilde{v}$ for each edge $e$ with a di-valent intertwiners is graphically represented by an arrow with orientation opposed to that of the original edge\footnote{The intertwiner, a line with an arrow, associated to the pseudo vertex $\tilde{v}$ is not normalizable since we adopt $\pi_{j_{e}}(h_e(A))$ rather that its normalized form $\sqrt{2j_e+1}\,\pi_{j_{e}}(h_e(A))$ in the spin network function (see Eq. \eqref{spin-network-state}). If the original spin network is normalized, the intertwiner associated to $\tilde{v}$ will automatically be normalized. Then it will be expressed as $\frac{1}{\sqrt{2j+1}}$ times a line with a spin-$j$ arrow in the graphical representation in Eq. \eqref{standard-snf}.}.

Let $Y^i_{v,e_I}$ be an operator assigned to vertex $v$  and edge $e_I$ intersecting $v$ by the following formula,
\begin{align}
Y^i_{v,e_I}\cdot  f_{\gamma_{\rm org}}(h_{e_1},\cdots,h_{e_I})=
\begin{cases}
J^i_{v,e_I;\,L}\cdot  f_{\gamma_{\rm org}}(h_{e_1},\cdots,h_{e_I}):=i\left.\frac{{\rm d}}{{\rm d}t}\right|_{t=0}f_{\gamma_{\rm org}}(h_{e_1},\cdots,h_{e_I}e^{t\tau_i},\cdots,h_{e_I}), \quad \text{when $e_I$ is incoming}\\
J^i_{v,e_I;\,R}\cdot  f_{\gamma_{\rm org}}(h_{e_1},\cdots,h_{e_I}):=-i\left.\frac{{\rm d}}{{\rm d}t}\right|_{t=0}f_{\gamma_{\rm org}}(h_{e_1},\cdots,e^{t\tau_i}h_{e_I},\cdots,h_{e_I}), \quad \text{when $e_I$ is outgoing}
\end{cases}\,.
\end{align}
Our task is to show that the action of $J^i_{v',e;\,L}$ on an arbitrary given spin network state of $\gamma_{\rm org}$ is equivalent to the action of $J^i_{v',l_e;\,R}$ on the spin network state of its corresponding standard graph $\gamma_{\rm std}$. The proof can be done by using their tensor operators $J^\mu$ ($\mu=0,\pm1$) in the graphical calculus. Notice that
\begin{align}
Y^\mu_{v',e}\cdot {[\pi_j(h_e)]^{m}}_{\,m'}=J^{\mu}_{v',e;\,L}\cdot {[\pi_j(h_e)]^{m}}_{\,m'}=i\left.\frac{{\rm d}}{{\rm d}t}\right|_{t=0}{[\pi_j(h_ee^{t\tau_\mu})]^{m}}_{\,m'}=i\left.\frac{{\rm d}}{{\rm d}t}\right|_{t=0}\left({[\pi_j(h_e)]^{m}}_{\,n}{[\pi_j(e^{t\tau_\mu})]^{n}}_{\,m'}\right)={[\pi_j(h_e)]^{m}}_{\,n}i{[\pi_j(\tau_\mu)]^{n}}_{\,m'}\,.
\end{align}
We obtain
\begin{align}
&J^{\mu}_{v',e;\,L}\cdot\makeSymbol{\includegraphics[width=7cm]{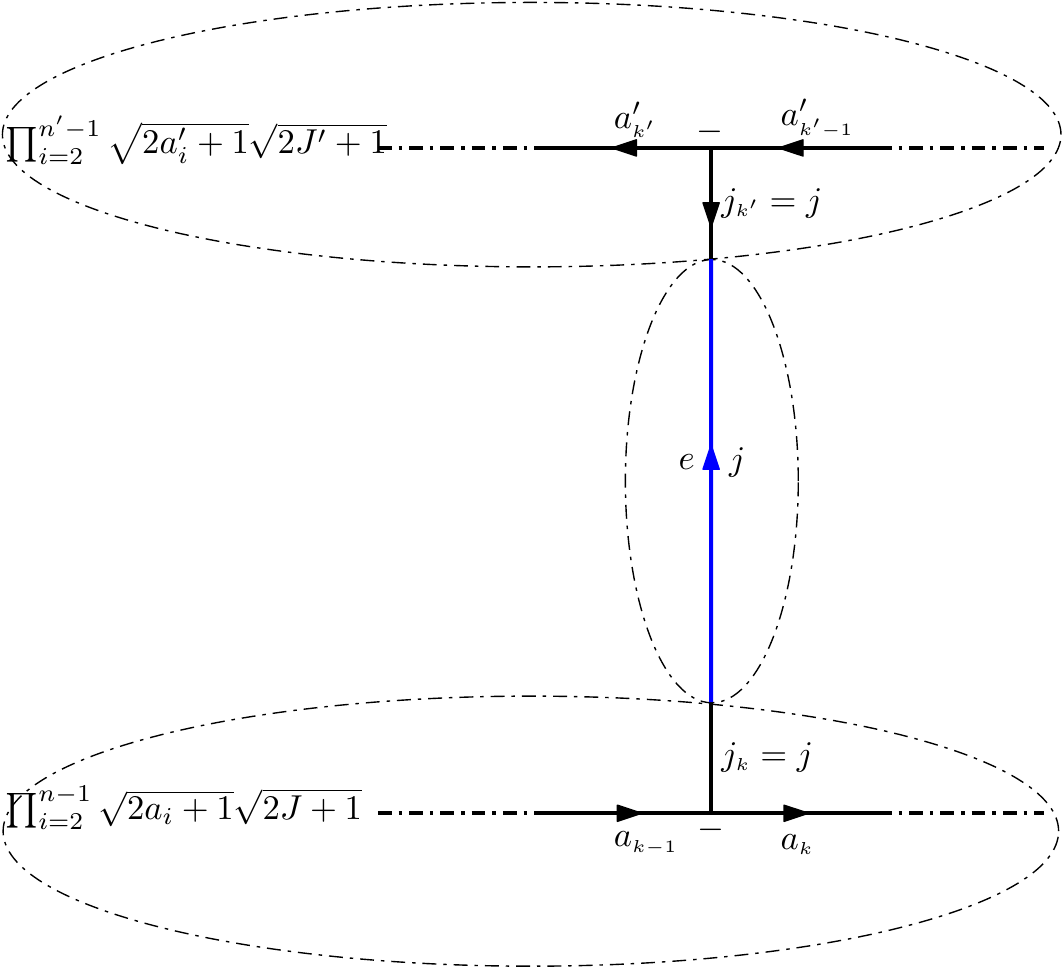}}=i\makeSymbol{\includegraphics[width=7cm]{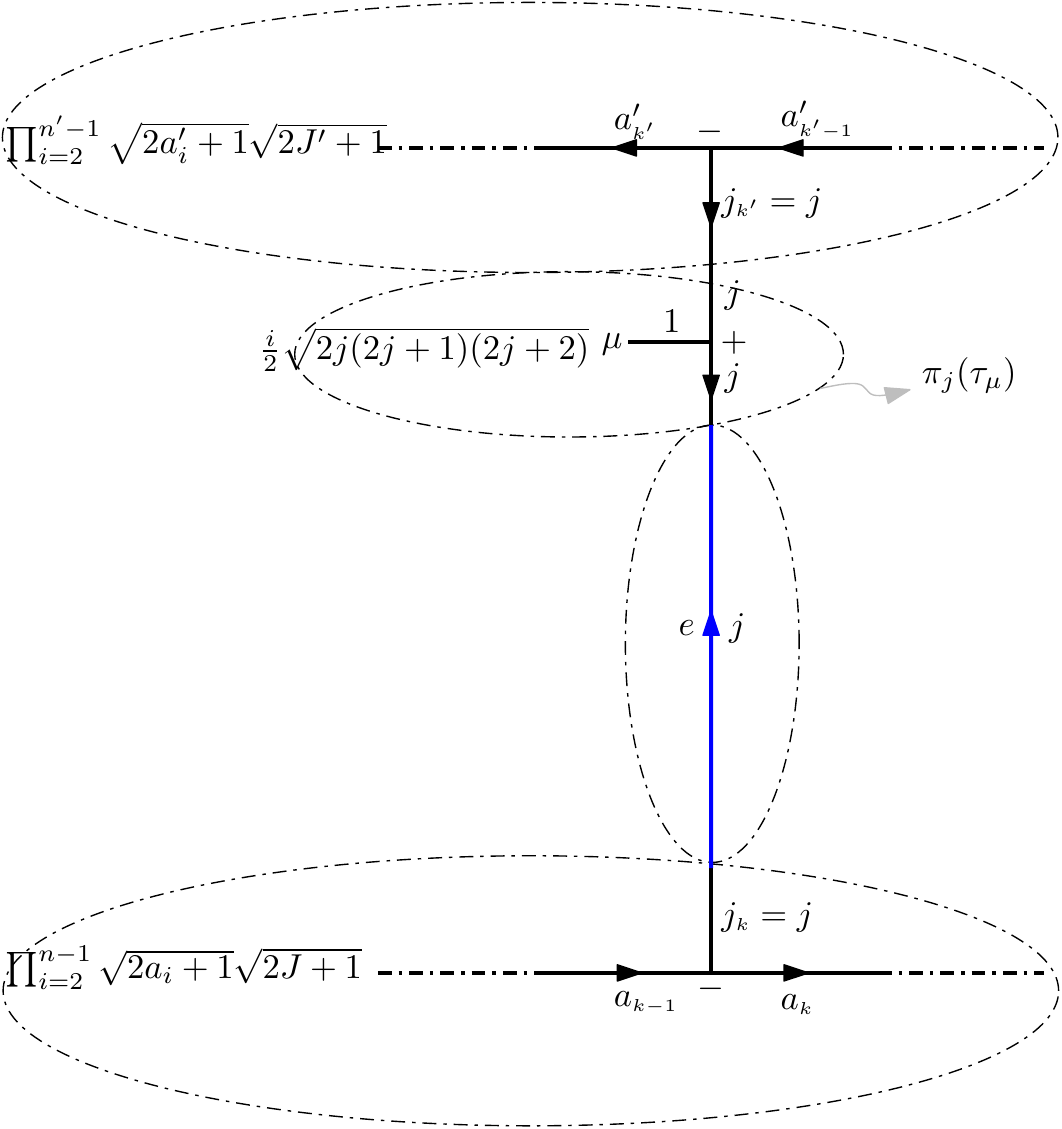}}\notag\\
&=-i\makeSymbol{\includegraphics[width=7cm]{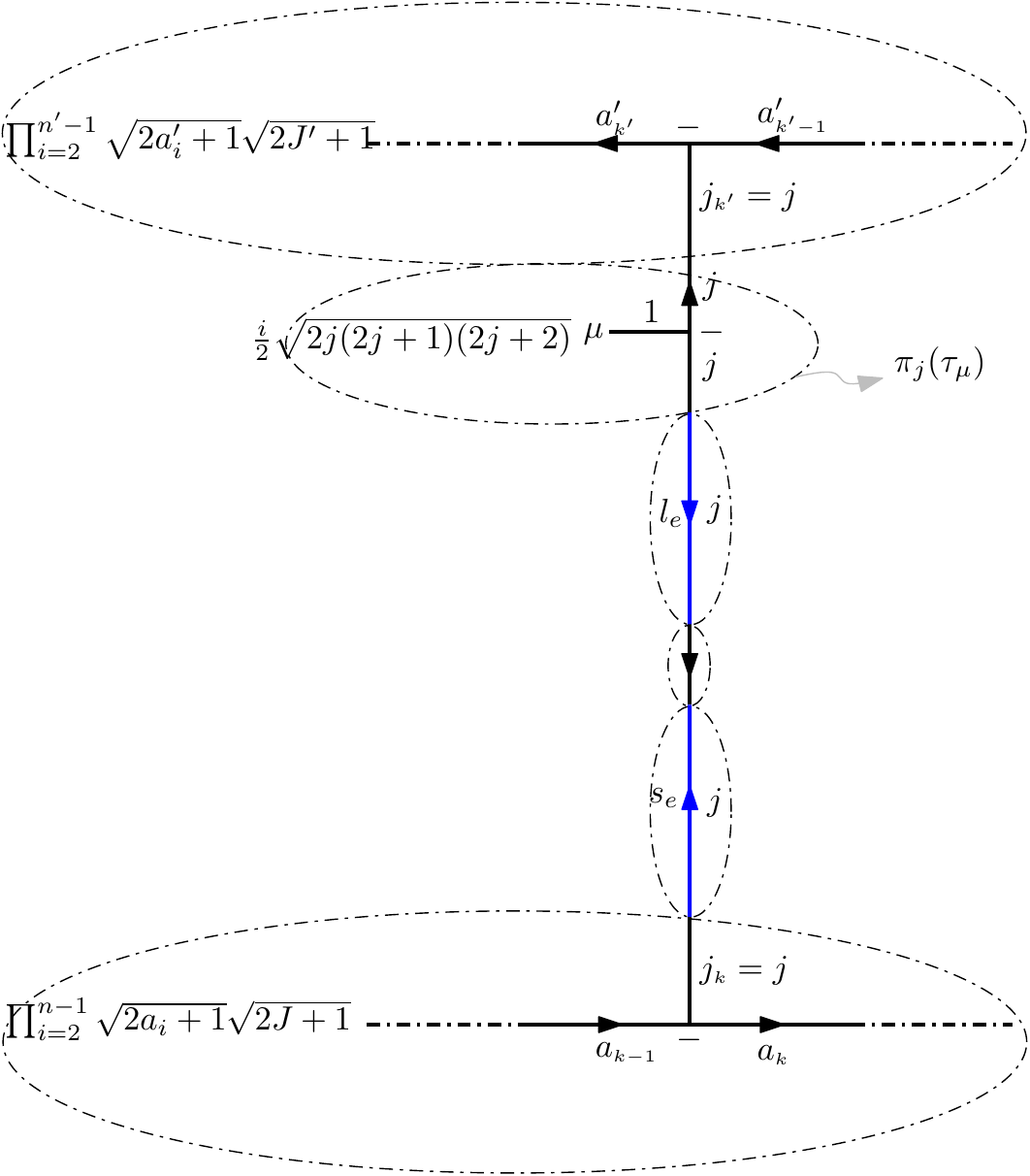}}=J^{\mu}_{v',l_e;\,R}\cdot\makeSymbol{\includegraphics[width=7cm]{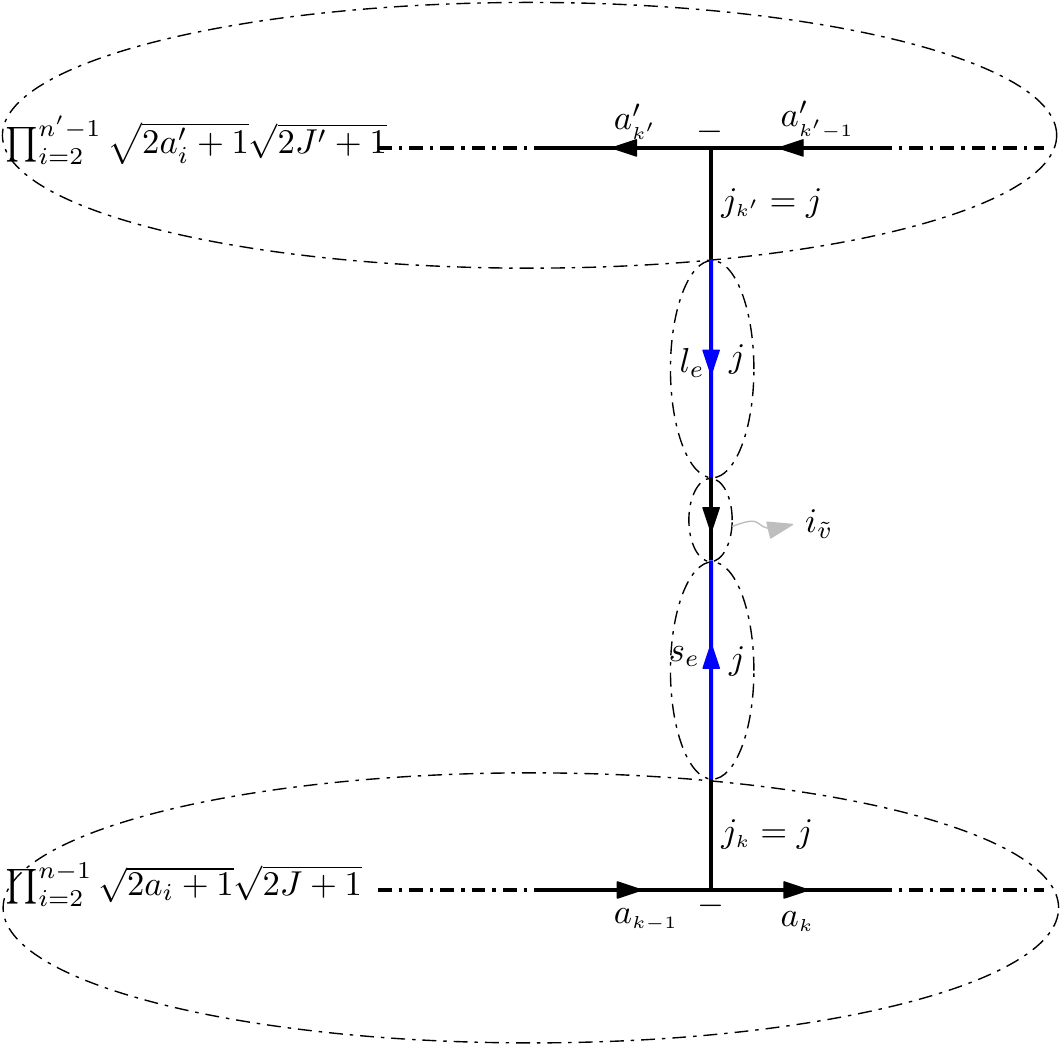}}\,,
\end{align}
where, in the second step, we have used
\begin{align}
\makeSymbol{\includegraphics[width=3.8cm]{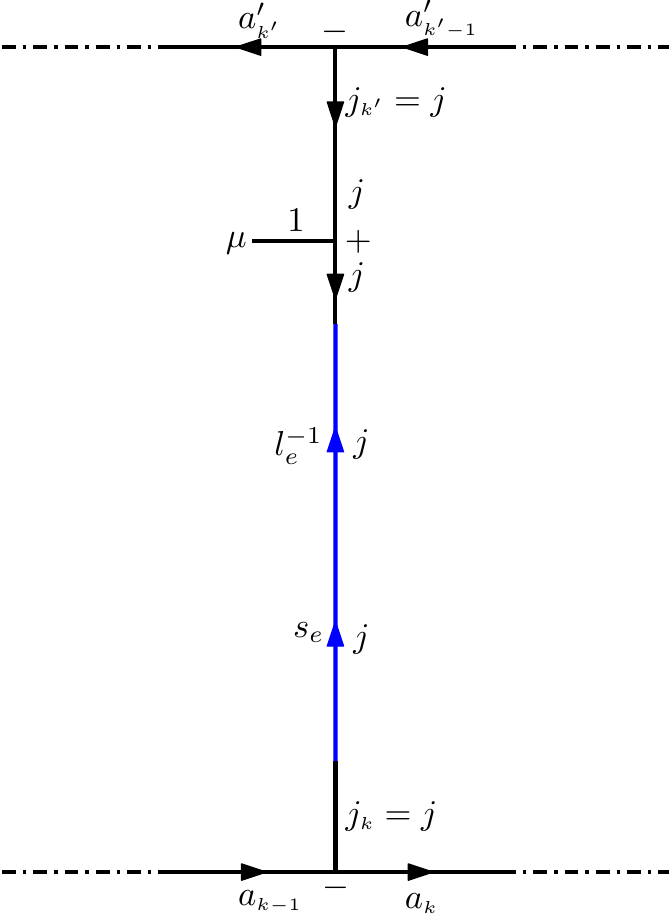}}&=\makeSymbol{\includegraphics[width=3.8cm]{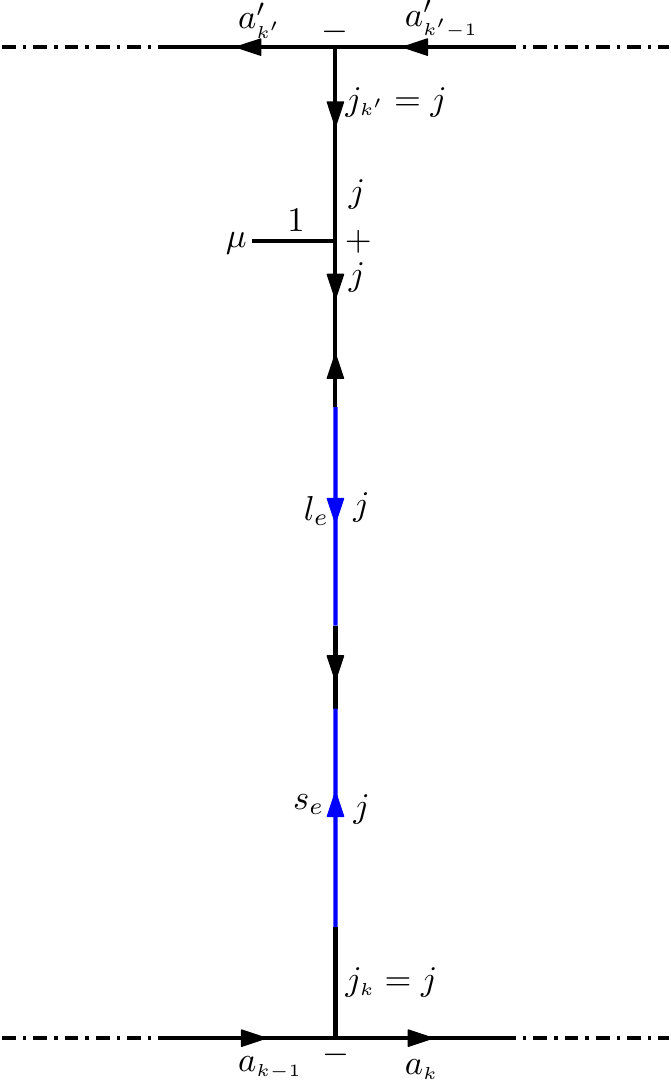}}=(-1)^{2j+1}(-1)^{2j}
\makeSymbol{\includegraphics[width=3.8cm]{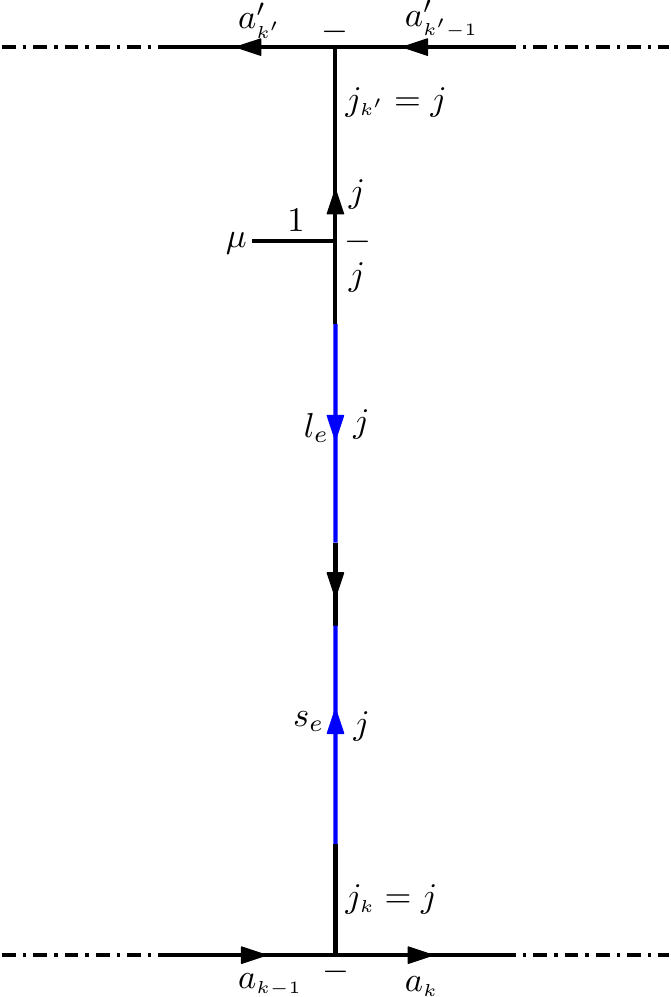}}\,,
\end{align}
here we have used \eqref{holonomy-inverse-arrow} in the first step, and \eqref{3j-orientation-change-graph}, \eqref{two-arrow-cancel} and \eqref{arrow-flip} in the second step. By the above procedure, we can also transform other edges into their corresponding standard edges. Thus we obtain
\begin{align}
\left(J^i_{v',e;\,L}\right)_{\gamma_{\rm org}}=\left(J^i_{v',l_e;\,R}\right)_{\gamma_{\rm std}}\,.
\end{align}
The above calculus and analysis complete our proof.

\section{The matrix elements of volume operator}\label{section-IV}

There are two versions of volume operator in the literature. We only consider the volume operator defined in \cite{Ashtekar:1997fb,Thiemann:1996au}, which passed the consistency check in quantum kinematical framework and was used to define a Hamiltonian constraint operator in LQG \cite{Giesel:2005bk,Giesel:2005bm,Thiemann:1996aw}. The volume operator acts on a spin network state as
\begin{align}\label{volume-operator}
\hat{V}\cdot\,T_{\gamma,\vec{j},\vec{i}}(A)=\ell_{\rm p}^3\,\beta^{\frac32}\sum_{v\in V(\gamma)}\sqrt{\left|\frac{i}{8\times4}\sum_{I<J<K,\,e_I\cap e_J \cap e_K=v}\varsigma(e_I,e_J,e_K)\;\hat{q}_{IJK}\right|}\,\cdot T_{\gamma,\vec{j}\,,\vec{i}}(A)\,,
\end{align}
where $\ell_{\rm p}\equiv\sqrt{\hbar\kappa}$, $\varsigma(e_I,e_J,e_K)\equiv{\rm sgn}(\det(\dot{e}_I(0),\dot{e}_J(0),\dot{e}_K(0)))$, and
\begin{align}\label{q-IJK}
\hat{q}_{IJK}&:=-4i\epsilon_{ijk}J^i_{e_I}J^j_{e_J}J^k_{e_K}=4\left[\delta_{ij}J^i_{e_I}J^j_{e_J},\delta_{lk}J^l_{e_J}J^k_{e_K}\right]=-\frac14\left(16\,\delta_{lk}J^l_{e_J}J^k_{e_K}\delta_{ij}J^i_{e_I}J^j_{e_J}-16\,\delta_{ij}J^i_{e_I}J^j_{e_J}\delta_{lk}J^l_{e_J}J^k_{e_K}\right)=:-\frac14\left(\hat{q}^{<JK;IJ>}_{IJK}-\hat{q}^{<IJ;JK>}_{IJK}\right)\,.
\end{align}
Here $J^i_{e_I}$ defined in \eqref{right-inv-operator-def} is the self-adjoint operator of the right-invariant vector field on the copy of $SU(2)$ corresponding to the $I$-th edge, satisfying
\begin{align}
[J^i_{e_I},J^j_{e_J}]=i\epsilon_{ijk}J^k_{e_I}\delta_{e_I,e_J}\,,
\end{align}
and $\delta_{ij}:=-2{\rm tr}(\tau_i\tau_j)$ is the Cartan-Killing metric on SU(2).
The action of volume operator is local, in the sense that its action is a sum on independent vertices. Therefore, we can focus on its action on a single vertex. The fact that the pseudo-vertices as endpoints of edges are divalent and the self-adjoint operators $J^i_{e_I}$ act only at the beginning points of $e_I$ implies that the summation in Eq. \eqref{volume-operator} is only over the true vertices $v$ of $\gamma$.

Eq. \eqref{J-i-snf-vertex} reveals the fact that the operators $\hat{q}_{IJK}$ and thus $\hat{V}$ only change the intertwiners $\vec{i}$ when evaluated on $T_{\gamma,\vec{j},\vec{i}}(A)$.  The operators $\hat{q}_{IJK}$ acts on an intertwiner by contracting the corresponding matrix elements of $\tau_i$ with the intertwiner. Note that
\begin{align}\label{angular-2RightVectAct}
\delta_{ij}J^i_{e_I}J^j_{e_J}\cdot \;{\left(i^{\,J;\,\vec{a}}_v\right)_{\,m_1\cdots m_I\cdots m_J\cdots m_n}}^M&={\left(i^{\,J;\,\vec{a}}_v\right)_{\,m_1\cdots m'_I\cdots m'_J\cdots m_n}}^M\left(-{[\pi_{j_I}(\tau_i)]^{m'_I}}_{\,m_I}\delta^{ij}\,{[\pi_{j_J}(\tau_j)]^{m'_J}}_{\,m_J}\right)\notag\\
&={\left(i^{\,J;\,\vec{a}}_v\right)_{\,m_1\cdots m'_I\cdots m'_J\cdots m_n}}^M\,{[\pi_{j_I}(\tau_{\mu})]^{m'_I}}_{\,m_I}C^{\nu\mu}_{(1)}\;{[\pi_{j_J}(\tau_{\nu}
)]^{m'_J}}_{\,m_J}\,.\notag\\
&=-C^{(1)}_{\mu\nu}J^\mu_{e_I}J^\nu_{e_J}\cdot \;{\left(i^{\,J;\,\vec{a}}_v\right)_{\,m_1\cdots m_I\cdots m_J\cdots m_n}}^M\,,
\end{align}
where in the second step we have used the following identity (see Appendix \ref{appendix-B-1} for proof)
\begin{align}\label{two-tau}
-{[\pi_{j_I}(\tau_i)]^{m'_I}}_{\,m_I}\delta^{ij}\,{[\pi_{j_J}(\tau_j)]^{m'_J}}_{\,m_J}&={[\pi_{j_I}(\tau_{\mu})]^{m'_I}}_{\,m_I}C^{\nu\mu}_{(1)}\;{[\pi_{j_J}(\tau_{\nu}
)]^{m'_J}}_{\,m_J}\,.
\end{align}
Hence the operator $\hat{q}^{<JK;IJ>}_{IJK}$ and  $\hat{q}^{<IJ;JK>}_{IJK}$ in \eqref{q-IJK} can be represented in terms of $J^\mu$ by
\begin{align}\label{q-JK-IJ-action}
\hat{q}^{<JK;IJ>}_{IJK}\cdot {\left(i^{\,J;\,\vec{a}}_v\right)_{\,m_1\cdots m_I\cdots m_J\cdots m_K\cdots m_n}}^M&=16\,C^{(1)}_{\rho\sigma}J^\rho_{e_J}J^\sigma_{e_K}C^{(1)}_{\mu\nu}J^\mu_{e_I}J^\nu_{e_J}\cdot {\left(i^{\,J;\,\vec{a}}_v\right)_{\,m_1\cdots m_I\cdots m_J\cdots m_K\cdots m_n}}^M\notag\\
&=16{\left(i^{\,J;\,\vec{a}}_v\right)_{\,m_1\cdots m'_I\cdots m'_J\cdots m'_K\cdots m_n}}^M{[\pi_{j_J}(\tau_{\nu})]^{m''_J}}_{\,m_J}C^{\nu'\nu}_{(1)}\;{[\pi_{j_K}(\tau_{\nu'}
)]^{m'_K}}_{\,m_K}\times{[\pi_{j_I}(\tau_{\mu})]^{m'_I}}_{\,m_I}C^{\mu'\mu}_{(1)}\;{[\pi_{j_J}(\tau_{\mu'}
)]^{m'_J}}_{\,m''_J}\,,\\
\hat{q}^{<IJ;JK>}_{IJK}\cdot {\left(i^{\,J;\,\vec{a}}_v\right)_{\,m_1\cdots m_I\cdots m_J\cdots m_K\cdots m_n}}^M&=16\,C^{(1)}_{\mu\nu}J^\mu_{e_I}J^\nu_{e_J}C^{(1)}_{\rho\sigma}J^\rho_{e_J}J^\sigma_{e_K}\cdot {\left(i^{\,J;\,\vec{a}}_v\right)_{\,m_1\cdots m_I\cdots m_J\cdots m_K\cdots m_n}}^M\notag\\
&=16{\left(i^{\,J;\,\vec{a}}_v\right)_{\,m_1\cdots m'_I\cdots m'_J\cdots m'_K\cdots m_n}}^M{[\pi_{j_I}(\tau_{\mu})]^{m'_I}}_{\,m_I}C^{\mu'\mu}_{(1)}\;{[\pi_{j_J}(\tau_{\mu'}
)]^{m''_J}}_{\,m_J}\times{[\pi_{j_J}(\tau_{\nu})]^{m'_J}}_{\,m''_J}C^{\nu'\nu}_{(1)}\;{[\pi_{j_K}(\tau_{\nu'}
)]^{m'_K}}_{\,m_K}\,. \label{q-IJ-JK-action}
\end{align}
With above preparations, we now turn to the action of $\hat{q}_{IJK}$ on the intertwiner ${\left(i^{\,J;\,\vec{a}}_v\right)_{\,m_1\cdots m_I\cdots m_J\cdots m_K\cdots m_n}}^M$ associated to a true vertex $v$ in the graphical method. We first consider the case $I>2$ and $K<n$, where $a_{I-1}$ and $a_K$ will appear in the final result. The other special cases will be dealt with later.
According to Eq. \eqref{q-JK-IJ-action}, the first term in the parenthesis of Eq. \eqref{q-IJK} evaluated on intertwiner \eqref{intertwined-form} can be represented by the following graphical formula (we present only the parts of the graph of the intertwiner which closely connect to the key steps in the following calculations),
\begin{align}\label{action-q}
&\quad\hat{q}^{<JK;IJ>}_{IJK}\cdot {\left(i^{\,J;\,\vec{a}}_v\right)_{\,m_1\cdots m_I\cdots m_J\cdots m_K\cdots m_n}}^M\notag\\
&=\hat{q}^{<JK;IJ>}_{IJK}\cdot\prod_{i=2}^{n-1}\sqrt{2a_i+1}\sqrt{2J+1}\;\makeSymbol{\includegraphics[width=9.5cm]{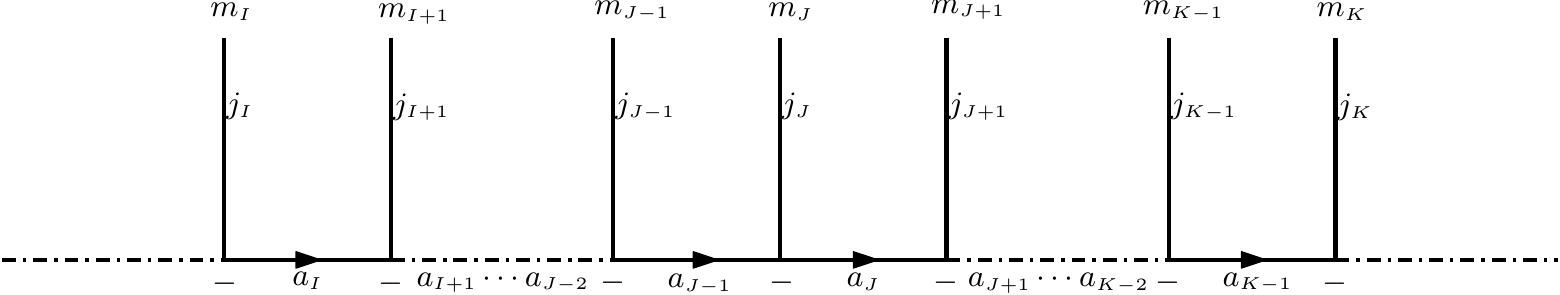}}\notag\\
&=X(j_I,j_J)^{\frac12}X(j_J,j_K)^{\frac12}\prod_{i=2}^{n-1}\sqrt{2a_i+1}\sqrt{2J+1}\makeSymbol{\includegraphics[width=9.5cm]{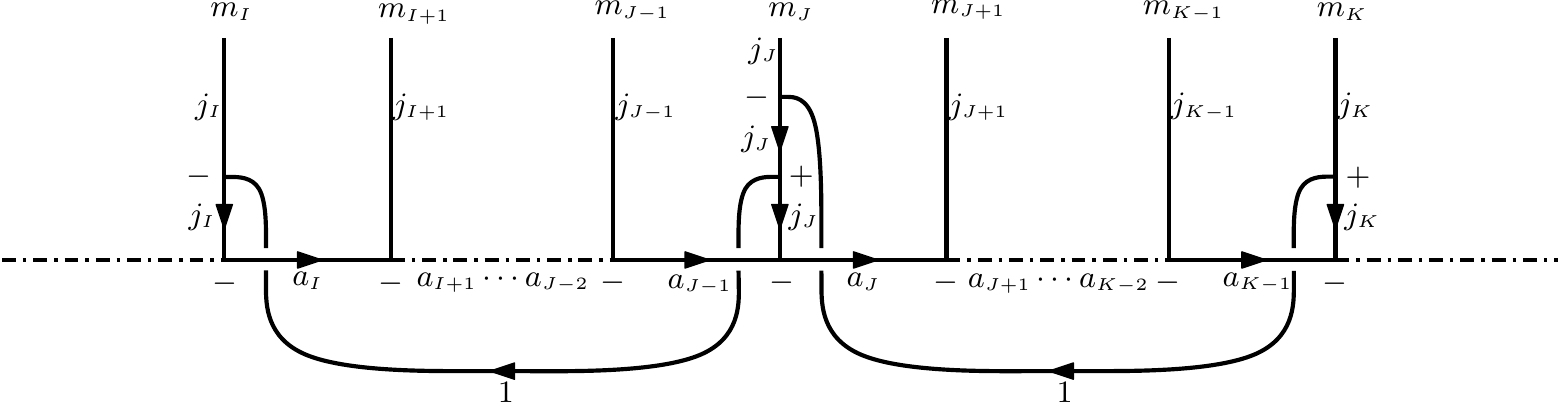}}\,,
\end{align}
where $X(j_1,j_2)\equiv 2j_1(2j_1+1)(2j_1+2)2j_2(2j_2+1)(2j_2+2)$. Similarly, the second term in the parenthesis of Eq. \eqref{q-IJK} acting on the intertwiner can be expressed as
\begin{align}\label{3J-2}
&\quad\hat{q}^{<IJ;JK>}_{IJK}\cdot {\left(i^{\,J;\,\vec{a}}_v\right)_{\,m_1\cdots m_I\cdots m_J\cdots m_K\cdots m_n}}^M\notag\\
&=X(j_I,j_J)^{\frac12}X(j_J,j_K)^{\frac12}\prod_{i=2}^{n-1}\sqrt{2a_i+1}\sqrt{2J+1}\makeSymbol{\includegraphics[width=9.5cm]{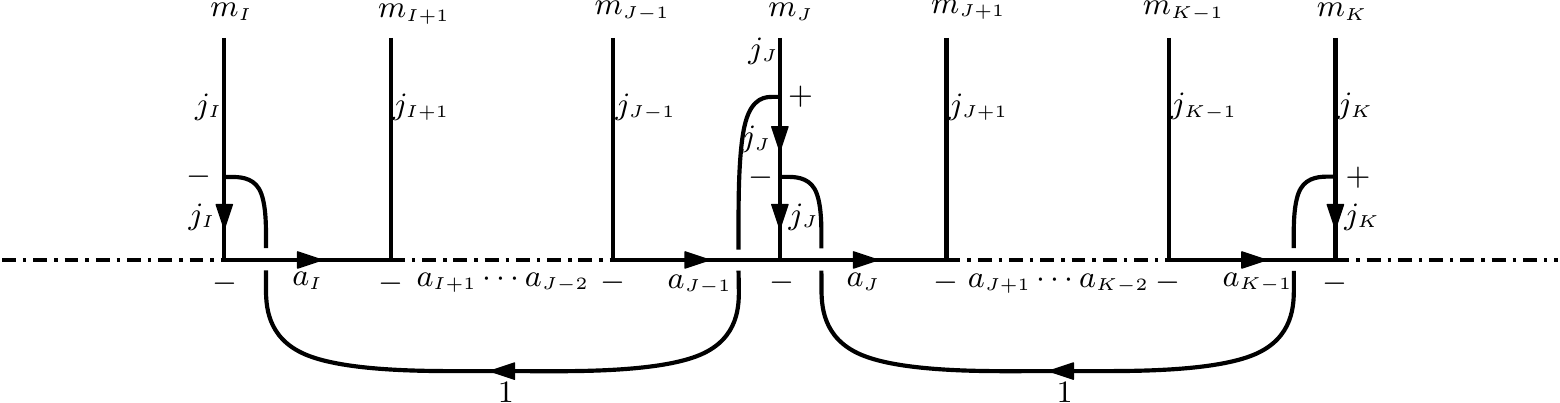}}\,.
\end{align}
It is obvious that the two terms $\hat{q}^{<JK;IJ>}_{IJK}$ and $\hat{q}^{<IJ;JK>}_{IJK}$ in \eqref{q-IJK} are gauge invariant. Hence the operator $\hat{q}_{IJK}$ and the volume operator \eqref{volume-operator} are gauge invariant. Each of them leaves the intertwiner space ${\cal H}^v_{j_1,\cdots,j_n}$ with the intertwiners as its orhtonormal basis, determined by the given $j_1,\cdots,j_n$ and the resulting angular momentum $J$, at the vertex $v$ invariant. Therefore the action of $\hat{q}_{IJK}$ on an intertwiner can be expanded linearly in terms of intertwiners in ${\cal H}^v_{j_1,\cdots,j_n}$ at the vertex $v$. In graphical language, they leave the vertical lines denoted by $j_i$ and the last horizontal line denoted by $J$ invariant, but change the intermediate couplings $a_i$ labelling the intermediate horizontal lines. Hence, in graphical calculation, our task is to drag the endpoints of the two curves with spin $1$ down to attach the horizontal lines, and then yank them away from the horizontal lines following the simple and rigorous rules of transforming graphs presented in  Appendix \ref{appendix-A-2}. Certainly, there are many alternative ways, corresponding to different choices of recoupling, to remove the two curves with spin $1$. The results obtained from different ways are related by unitary transformations. In what follows, we choose a way guided by the simplicity principle that the number of changed intermediate values $a_i$ is as little as possible and the final result is as simple as possible. The calculations of the action of $\hat{q}_{IJK}$ on the given intertwiner in the graphical method consist of the following four steps.

The first step involves dragging the two endpoints of curves with spin $1$ attached to lines with spins $j_I$ and $j_K$ respectively down to join with two horizontal lines denoted by spins $a_{I}$ and $a_{K-1}$. To do this, we use the following recoupling identities (see Appendix \ref{appendix-B-2} for proof),
\begin{align}\label{id-1}
\makeSymbol{\includegraphics[width=1.5cm]{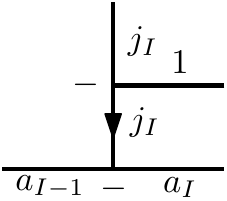}}
&=\sum_{a'_I}(2a'_I+1)(-1)^{a_{I-1}-a'_I+j_I}\begin{Bmatrix} a_{I-1} & j_I &  a_I \\
 1 & a'_I & j_I
\end{Bmatrix}\;\makeSymbol{\includegraphics[width=2cm]{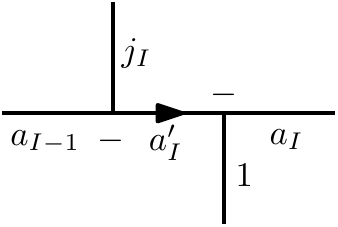}}\,,\\
\makeSymbol{\includegraphics[width=1.5cm]{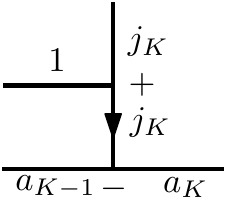}}\label{id-2}
&=\sum_{b'_{K-1}}(2b'_{K-1}+1)(-1)^{a_K-b'_{K-1}+j_K+1}
\begin{Bmatrix}  a_K & j_K & a_{K-1} \\
  1 & b'_{K-1} & j_K
\end{Bmatrix}\makeSymbol{\includegraphics[width=2cm]{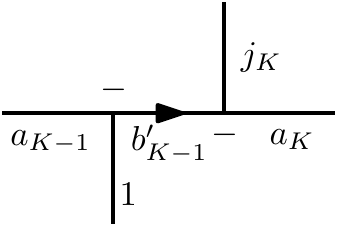}}\,.
\end{align}
Then we have
\begin{align}
&\quad\hat{q}^{<JK;IJ>}_{IJK}\cdot {\left(i^{\,J;\,\vec{a}}_v\right)_{\,m_1\cdots m_I\cdots m_J\cdots m_K\cdots m_n}}^M\notag\\
&=\sum_{a'_I}(2a'_I+1)(-1)^{a_{I-1}-a'_I+j_I}\begin{Bmatrix} a_{I-1} & j_I &  a_I \\
 1 & a'_I & j_I
\end{Bmatrix}\times\sum_{b'_{K-1}}(2b'_{K-1}+1)(-1)^{a_K-b'_{K-1}+j_K+1}
\begin{Bmatrix}  a_K & j_K & a_{K-1} \\
  1 & b'_{K-1} & j_K
\end{Bmatrix}\notag\\
&\quad\times X(j_I,j_J)^{\frac12}X(j_J,j_K)^{\frac12}\prod_{i=2}^{n-1}\sqrt{2a_i+1}\sqrt{2J+1}\makeSymbol{\includegraphics[width=9.5cm]{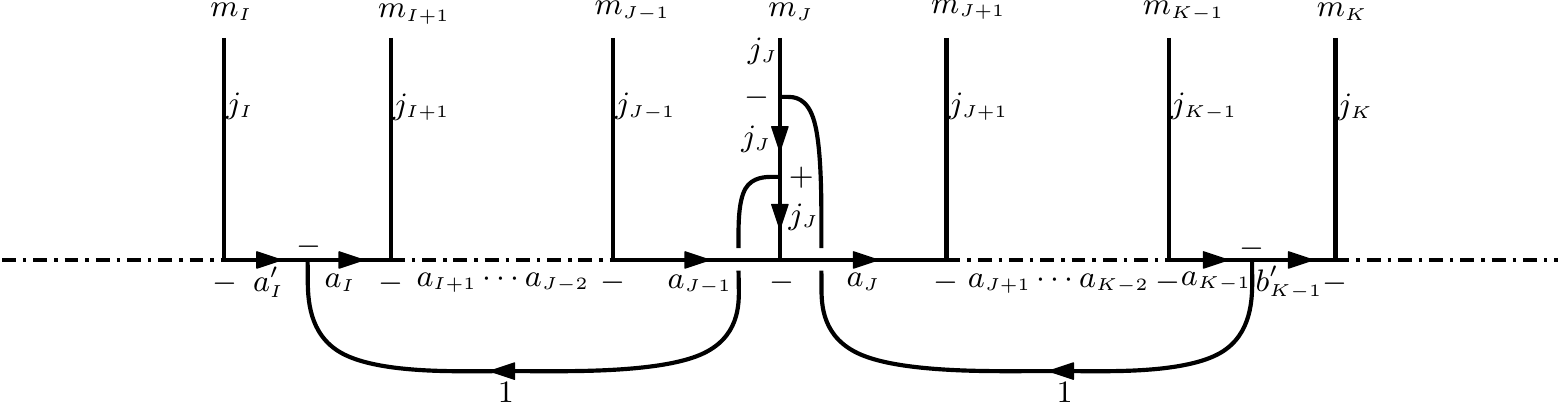}}\,.
\end{align}

Let us turn to the second step. We first consider the case that $J>I+1$ and $K>J+1$. The other cases will be handled later. We move the two points labelled by $(a'_I,a_I,1)$ and $(a_{K-1},b'_{K-1},1)$, step by step, to the right hand side of $(a_{J-2},j_{J-1},a_{J-1})$ and the left hand side of $(a_J, j_{J+1},a_{J+1})$ respectively, by repeatedly applying of the following identies (see Appendix \ref{appendix-B-2} for proof),
\begin{align}\label{id-3}
\makeSymbol{\includegraphics[width=2cm]{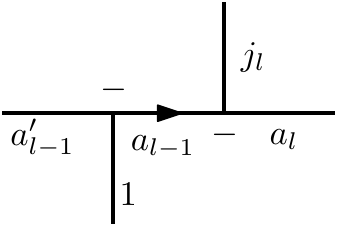}}
&=\sum_{a'_l}(2a'_l+1)(-1)^{a'_{l-1}+a_{l-1}+1}(-1)^{a_{l-1}-a_l+j_l}
\begin{Bmatrix}  j_l & a'_{l-1} & a'_l \\
  1 & a_l & a_{l-1}
\end{Bmatrix}\makeSymbol{\includegraphics[width=2cm]{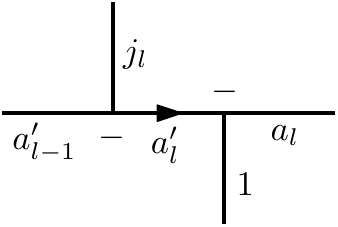}}\,,\\
\makeSymbol{\includegraphics[width=2cm]{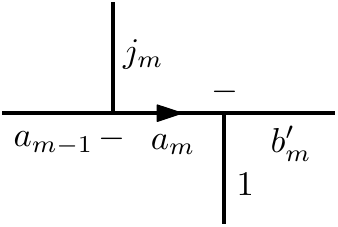}}
&=\sum_{b'_{m-1}}(2b'_{m-1}+1)(-1)^{b'_{m-1}+a_{m-1}+1}(-1)^{b'_m-b'_{m-1}-j_m}
\begin{Bmatrix}  j_m & b'_{m-1} & b'_m\\
  1 & a_m & a_{m-1}
\end{Bmatrix}\makeSymbol{\includegraphics[width=2cm]{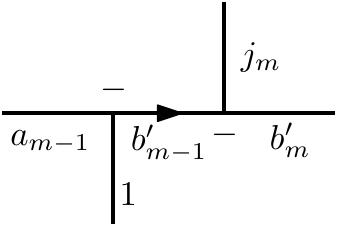}}\label{id-4}\,,
\end{align}
where $l=I+1,\cdots, J-1$ and $m={K-1,\cdots,J+1}$.
Then we have
\begin{align}
&\sum_{(a'_I,a'_{I+1}\cdots,a'_{J-1})}\sum_{(b'_{K-1},\cdots,b'_{J+1},b'_J)}(2a'_I+1)(-1)^{a_{I-1}-a'_I+j_I}\begin{Bmatrix} a_{I-1} & j_I &  a_I \\
 1 & a'_I & j_I
\end{Bmatrix}\times(2b'_{K-1}+1)(-1)^{a_K-b'_{K-1}+j_K+1}
\begin{Bmatrix}  a_K & j_K & a_{K-1} \\
  1 & b'_{K-1} & j_K
\end{Bmatrix}\notag\\
&\hspace{3.1cm}\times\prod_{l=I+1}^{J-1}(2a'_l+1)(-1)^{a'_{l-1}+a_{l-1}+1}(-1)^{a_{l-1}-a_l+j_l}
\begin{Bmatrix}  j_l & a'_{l-1} & a'_l \\
  1 & a_l & a_{l-1}
\end{Bmatrix}\notag\\
&\hspace{3.1cm}\times\prod_{m=J+1}^{K-1}(2b'_{m-1}+1)(-1)^{b'_{m-1}+a_{m-1}+1}(-1)^{b'_m-b'_{m-1}-j_m}
\begin{Bmatrix}  j_m & b'_{m-1} & b'_m\\
  1 & a_m & a_{m-1}
\end{Bmatrix}\notag\\
&\quad\times X(j_I,j_J)^{\frac12}X(j_J,j_K)^{\frac12}\prod_{i=2}^{n-1}\sqrt{2a_i+1}\sqrt{2J+1}\makeSymbol{\includegraphics[width=9.5cm]{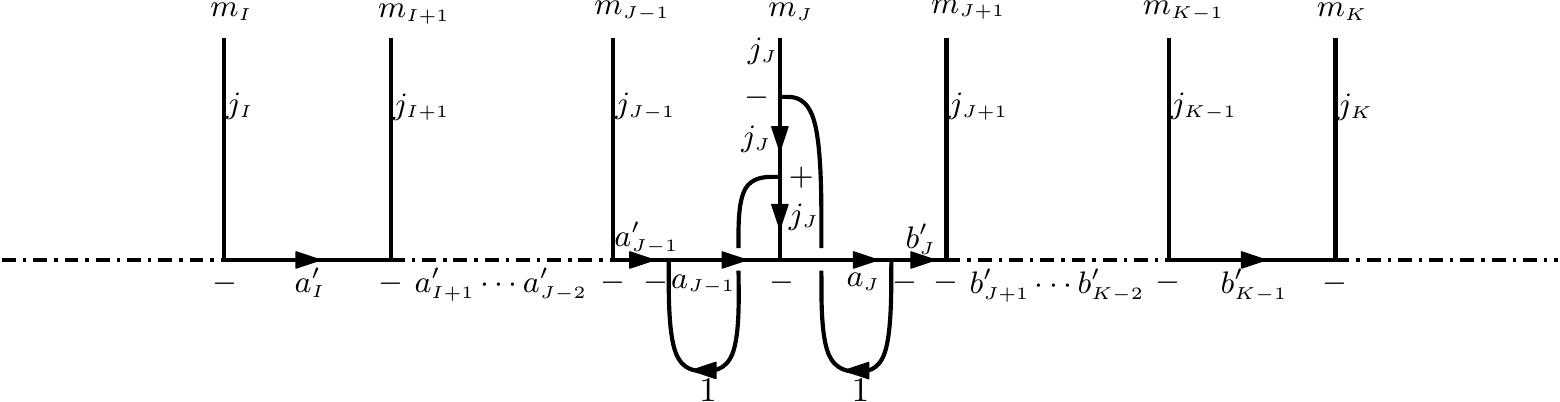}}\,,
\end{align}
where $(a'_I,a'_{I+1}\cdots,a'_{J-1})$ and $(b'_{K-1},\cdots,b'_{J+1},b'_J)$ denote the all allowed tuples, and $a'_l$ and $b'_{m-1}$ are determined by $a'_{l-1}$ and $b'_m$ respectively.

In the thrid step, we dragg the two endpoints of two curves with spin $1$ attached to the line with spin$j_J$ down to join with two horizontal lines denoted by spins $a_{J-1}$ and $a_J$, respectively. To do this,
we use again the identity \eqref{id-2} to pull down the point $(j_J,j_J,1)$ and simplify the two quantities $\prod_{l=I+1}^{J-1}(-1)^{a_{l-1}-a_l+j_l}$ and $\prod_{m=J+1}^{K-1}(-1)^{b'_m-b'_{m-1}-j_m}$. Then we obtain
\begin{align}
&\sum_{(a'_I,a'_{I+1}\cdots,a'_{J-1})}\sum_{(b'_{K-1},\cdots,b'_{J+1},b'_J)}(2a'_I+1)(-1)^{a_{I-1}-a'_I+j_I}\begin{Bmatrix} a_{I-1} & j_I &  a_I \\
 1 & a'_I & j_I
\end{Bmatrix}\times(2b'_{K-1}+1)(-1)^{a_K-b'_{K-1}+j_K+1}
\begin{Bmatrix}  a_K & j_K & a_{K-1} \\
  1 & b'_{K-1} & j_K
\end{Bmatrix}\notag\\
&\hspace{3.1cm}\times(-1)^{a_I-a_{J-1}+\sum_{l=I+1}^{J-1}j_l}\prod_{l=I+1}^{J-1}(2a'_l+1)(-1)^{a'_{l-1}+a_{l-1}+1}
\begin{Bmatrix}  j_l & a'_{l-1} & a'_l \\
  1 & a_l & a_{l-1}
\end{Bmatrix}\notag\\
&\hspace{3.1cm}\times(-1)^{b'_{K-1}-b'_J-\sum_{m=J+1}^{K-1}j_m}\prod_{m=J+1}^{K-1}(2b'_{m-1}+1)(-1)^{b'_{m-1}+a_{m-1}+1}
\begin{Bmatrix}  j_m & b'_{m-1} & b'_m\\
  1 & a_m & a_{m-1}
\end{Bmatrix}\notag\\
&\hspace{3.1cm}\times\sum_{b'_{J-1}}(2b'_{J-1}+1)(-1)^{a_J-b'_{J-1}+j_J+1}
\begin{Bmatrix}  a_J & j_J & a_{J-1} \\
  1 & b'_{J-1} & j_J
\end{Bmatrix}\notag\\
&\quad\times X(j_I,j_J)^{\frac12}X(j_J,j_K)^{\frac12}\prod_{i=2}^{n-1}\sqrt{2a_i+1}\sqrt{2J+1}\makeSymbol{\includegraphics[width=9.5cm]{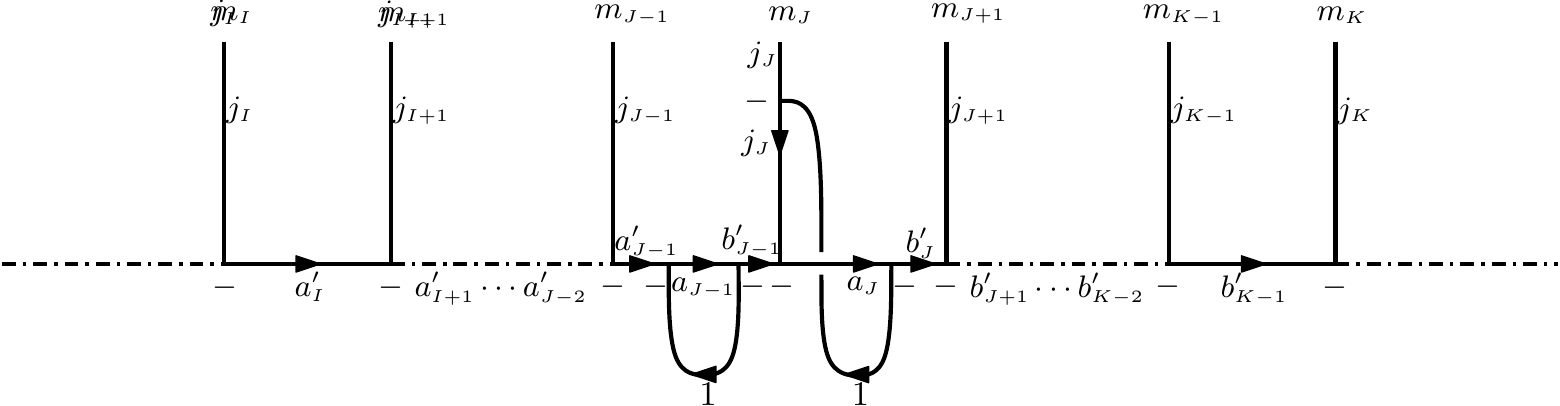}}\,.
\end{align}
The identity in Eq. \eqref{id-1} can be used to drag the remained endpoint, which attaches to the $j_J$ line, down to join the horizontal line denoted by $a_J$ and yield
\begin{align}
&\sum_{(a'_I,a'_{I+1}\cdots,a'_{J-1})}\sum_{(b'_{K-1},\cdots,b'_{J+1},b'_J)}(2a'_I+1)(-1)^{a_{I-1}-a'_I+j_I}\begin{Bmatrix} a_{I-1} & j_I &  a_I \\
 1 & a'_I & j_I
\end{Bmatrix}\times(2b'_{K-1}+1)(-1)^{a_K-b'_{K-1}+j_K+1}
\begin{Bmatrix}  a_K & j_K & a_{K-1} \\
  1 & b'_{K-1} & j_K
\end{Bmatrix}\notag\\
&\hspace{3.1cm}\times(-1)^{a_I-a_{J-1}+\sum_{l=I+1}^{J-1}j_l}\prod_{l=I+1}^{J-1}(2a'_l+1)(-1)^{a'_{l-1}+a_{l-1}+1}
\begin{Bmatrix}  j_l & a'_{l-1} & a'_l \\
  1 & a_l & a_{l-1}
\end{Bmatrix}\notag\\
&\hspace{3.1cm}\times(-1)^{b'_{K-1}-b'_J-\sum_{m=J+1}^{K-1}j_m}\prod_{m=J+1}^{K-1}(2b'_{m-1}+1)(-1)^{b'_{m-1}+a_{m-1}+1}
\begin{Bmatrix}  j_m & b'_{m-1} & b'_m\\
  1 & a_m & a_{m-1}
\end{Bmatrix}\notag\\
&\hspace{3.1cm}\times\sum_{b'_{J-1}}(2b'_{J-1}+1)(-1)^{a_J-b'_{J-1}+j_J+1}
\begin{Bmatrix}  a_J & j_J & a_{J-1} \\
  1 & b'_{J-1} & j_J
\end{Bmatrix}
\times\sum_{a'_J}(2a'_J+1)(-1)^{b'_{J-1}-a'_J+j_J}\begin{Bmatrix} b'_{J-1} & j_J &  a_J \\
 1 & a'_J & j_J
\end{Bmatrix}\notag\\
&\quad\times X(j_I,j_J)^{\frac12}X(j_J,j_K)^{\frac12}\prod_{i=2}^{n-1}\sqrt{2a_i+1}\sqrt{2J+1}\makeSymbol{\includegraphics[width=9.5cm]{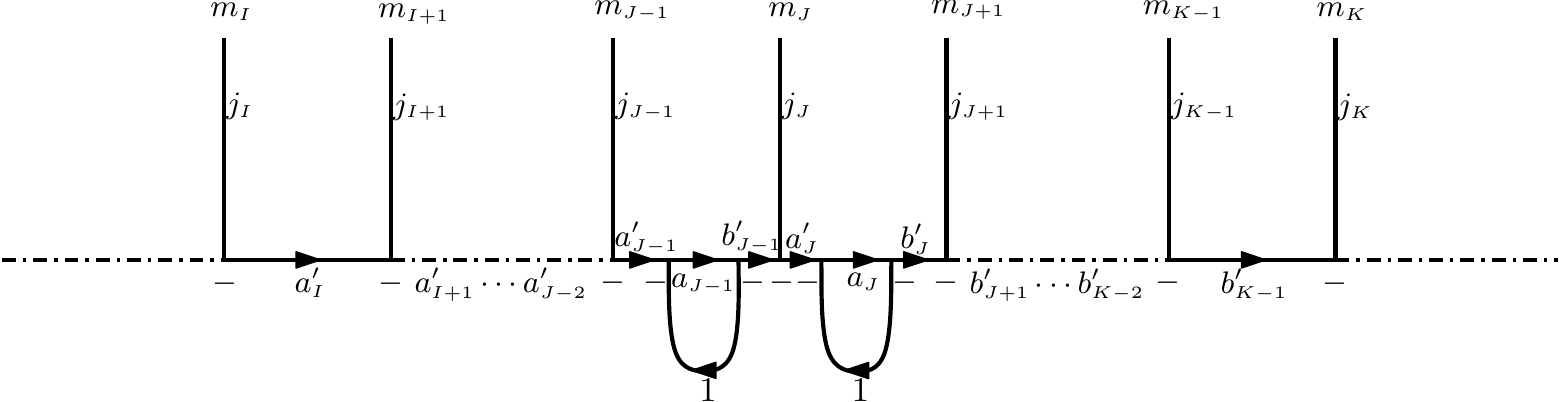}}\,.
\end{align}

Let us go to the forth (last) step. It is the time to remove the two curves with spin $1$. To do this, we use the identity (see Appendix \ref{appendix-B-3} for a proof)
\begin{align}\label{id-circle}
\makeSymbol{\includegraphics[width=2.4cm]{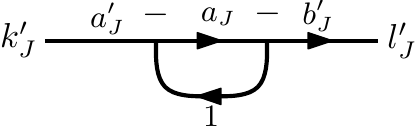}}&
=(-1)^{a'_J-a_J+1}\frac{\delta_{a'_J,b'_J}}{2a'_J+1}\,\makeSymbol{\includegraphics[width=1.6cm]{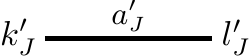}}
\end{align}
to remove the two curves with spin $1$ from intertwiners and obtain
\begin{align}\label{a-b}
&\sum_{(a'_I,a'_{I+1}\cdots,a'_{J-1})}\sum_{(b'_{K-1},\cdots,b'_{J+1},b'_J)}(2a'_I+1)(-1)^{a_{I-1}-a'_I+j_I}\begin{Bmatrix} a_{I-1} & j_I &  a_I \\
 1 & a'_I & j_I
\end{Bmatrix}
\times(2b'_{K-1}+1)(-1)^{a_K-b'_{K-1}+j_K+1}
\begin{Bmatrix}  a_K & j_K & a_{K-1} \\
  1 & b'_{K-1} & j_K
\end{Bmatrix}\notag\\
&\hspace{3.1cm}\times(-1)^{a_I-a_{J-1}+\sum_{l=I+1}^{J-1}j_l}\prod_{l=I+1}^{J-1}(2a'_l+1)(-1)^{a'_{l-1}+a_{l-1}+1}
\begin{Bmatrix}  j_l & a'_{l-1} & a'_l \\
  1 & a_l & a_{l-1}
\end{Bmatrix}\notag\\
&\hspace{3.1cm}\times(-1)^{b'_{K-1}-b'_J-\sum_{m=J+1}^{K-1}j_m}\prod_{m=J+1}^{K-1}(2b'_{m-1}+1)(-1)^{b'_{m-1}+a_{m-1}+1}
\begin{Bmatrix}  j_m & b'_{m-1} & b'_m\\
  1 & a_m & a_{m-1}
\end{Bmatrix}\notag\\
&\hspace{3.1cm}\times\sum_{b'_{J-1}}(2b'_{J-1}+1)(-1)^{a_J-b'_{J-1}+j_J+1}
\begin{Bmatrix}  a_J & j_J & a_{J-1} \\
  1 & b'_{J-1} & j_J
\end{Bmatrix}\times\sum_{a'_J}(2a'_J+1)(-1)^{b'_{J-1}-a'_J+j_J}\begin{Bmatrix} b'_{J-1} & j_J &  a_J \\
 1 & a'_J & j_J
\end{Bmatrix}\notag\\
&\hspace{3.8cm}\times(-1)^{a'_{J-1}-a_{J-1}+1}\frac{1}{2a'_{J-1}+1}\delta_{a'_{J-1},b'_{J-1}}
\times(-1)^{a'_J-a_J+1}\frac{1}{2a'_J+1}\delta_{a'_J,b'_J}\notag\\
&\quad\times X(j_I,j_J)^{\frac12}X(j_J,j_K)^{\frac12}\prod_{i=2}^{n-1}\sqrt{2a_i+1}\sqrt{2J+1}
\makeSymbol{\includegraphics[width=9.5cm]{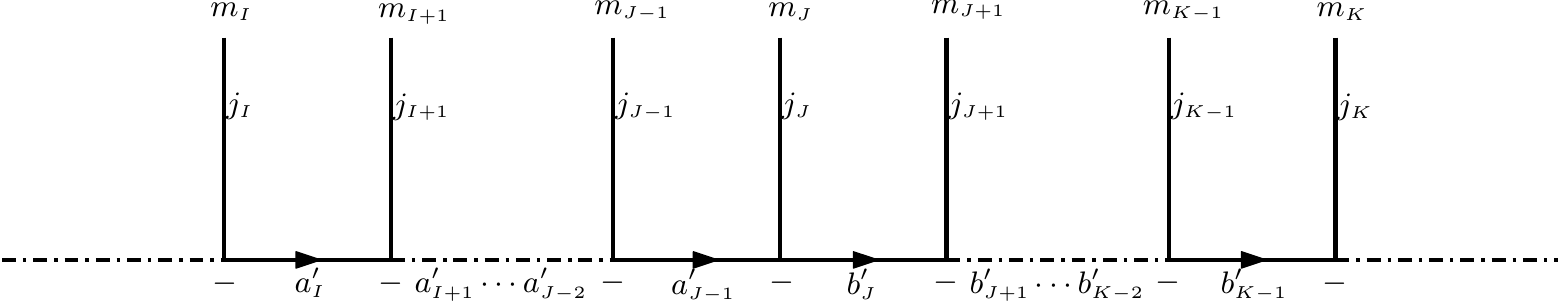}}\,.
\end{align}
Summing over $b'_{J-1}$ and $a'_J$ and relabeling the indices $b'$ by $a'$, we get
\begin{align}\label{general-1}
&\left.
\begin{array}{cc}
\sum\limits_{(a'_I,\cdots,a'_{K-1})}\;(2a'_I+1)(-1)^{a_{I-1}-a'_I+j_I}\begin{Bmatrix} a_{I-1} & j_I &  a_I \\
 1 & a'_I & j_I
\end{Bmatrix}\notag\\
\hspace{2.4cm}\times(2a'_{K-1}+1)(-1)^{a_K-a'_{K-1}+j_K+1}
\begin{Bmatrix}  a_K & j_K & a_{K-1} \\
  1 & a'_{K-1} & j_K
\end{Bmatrix}
\end{array}
\right\}{\text{from the first step}}\notag\\
&\left.
\begin{array}{cc}
\hspace{1.2cm}\times(-1)^{a_I-a_{J-1}+\sum_{l=I+1}^{J-1}j_l}\prod\limits_{l=I+1}^{J-1}(2a'_l+1)(-1)^{a'_{l-1}+a_{l-1}+1}
\begin{Bmatrix}  j_l & a'_{l-1} & a'_l \\
  1 & a_l & a_{l-1}
\end{Bmatrix}\notag\\
\hspace{2.4cm}\times(-1)^{a'_{K-1}-a'_J-\sum_{m=J+1}^{K-1}j_m}\prod\limits_{m=J+1}^{K-1}(2a'_{m-1}+1)(-1)^{a'_{m-1}+a_{m-1}+1}
\begin{Bmatrix}  j_m & a'_{m-1} & a'_m\\
  1 & a_m & a_{m-1}
\end{Bmatrix}
\end{array}
\right\}{\text{from the second step}}\notag\\
&\left.
\begin{array}{cc}
\hspace{2.4cm}\times\begin{Bmatrix}
   a_J & j_J & a_{J-1} \\
  1 & a'_{J-1} & j_J
\end{Bmatrix}(-1)^{a_J-a'_{J-1}+j_J+1}(-1)^{-a'_{J-1}+a_{J-1}-1}
\notag\\
\hspace{1.6cm}\times\begin{Bmatrix}
 a'_{J-1} & j_J &  a_J \\
 1 & a'_J & j_J
\end{Bmatrix}(-1)^{a'_{J-1}-a'_J+j_J}(-1)^{a'_J-a_J+1}
\end{array}
\right\}{\text{from the third and forth steps}}\notag\\
&\hspace{2cm}\times X(j_I,j_J)^{\frac12}X(j_J,j_K)^{\frac12}\prod_{i=2}^{n-1}\sqrt{2a_i+1}\sqrt{2J+1}
\makeSymbol{\includegraphics[width=9.5cm]{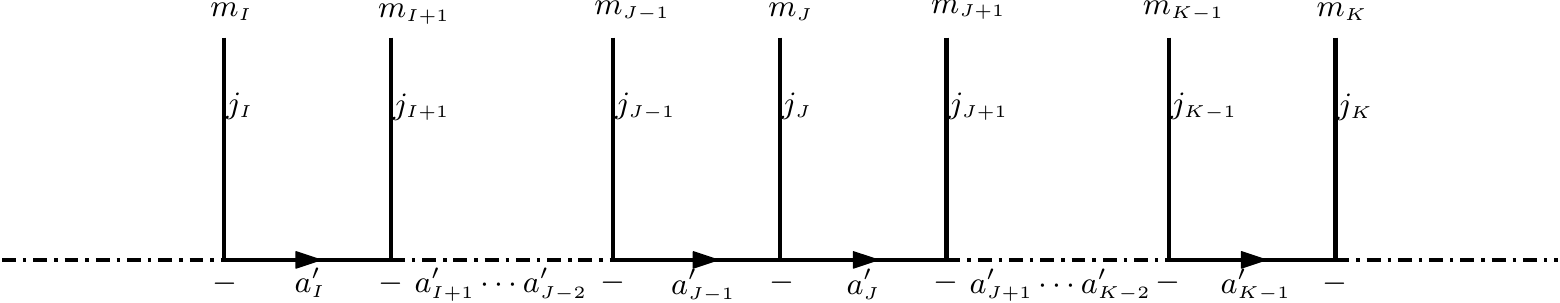}}\,.
\end{align}
Up to now, we have completed the key calculations and obtained the action of the operator $\hat{q}^{<JK;IJ>}_{IJK}$ on the intertwiner ${\left(i^{\,J;\,\vec{a}}_v\right)_{\,m_1\cdots m_n}}^M$ as \eqref{general-1}, which is a linear combination of new intertwiners.

Now we want to simplify the expression \eqref{general-1} under the following two considerations. One is to get more symmetric factors in the two multi-products. The other is to simplify the exponents. Notice that the result \eqref{general-1} was obtained in the case of $J>I+1$ and $K>J+1$. Under this case, the product terms of $(2a'+1)$ in Eq. \eqref{general-1} reduce to
\begin{align}\label{interchange}
(2a'_{K-1}+1)\prod_{m=J+1}^{K-1}(2a'_{m-1}+1)=(2a'_J+1)\prod_{m=J+1}^{K-1}(2a'_m+1)\,,
\end{align}
which enables us to write the multiple product over $m$ as the formula which is more close to the multiple product over $l$ in Eq. \eqref{general-1}. Thus we have
\begin{align}\label{general-1-1}
&\sum_{(a'_I,\cdots,a'_{K-1})}(2a'_I+1)(-1)^{a_{I-1}-a'_I+j_I}\begin{Bmatrix} a_{I-1} & j_I &  a_I \\
 1 & a'_I & j_I
\end{Bmatrix}\
\times(2a'_J+1)(-1)^{a_K-a'_{K-1}+j_K+1}
\begin{Bmatrix}  a_K & j_K & a_{K-1} \\
  1 & a'_{K-1} & j_K
\end{Bmatrix}\notag\\
&\hspace{0.95cm}\times(-1)^{a_I-a_{J-1}+\sum_{l=I+1}^{J-1}j_l}\prod_{l=I+1}^{J-1}(2a'_l+1)(-1)^{a'_{l-1}+a_{l-1}+1}
\begin{Bmatrix}  j_l & a'_{l-1} & a'_l \\
  1 & a_l & a_{l-1}
\end{Bmatrix}\notag\\
&\hspace{0.95cm}\times(-1)^{a'_{K-1}-a'_J-\sum_{m=J+1}^{K-1}j_m}\prod_{m=J+1}^{K-1}(2a'_m+1)(-1)^{a'_{m-1}+a_{m-1}+1}
\begin{Bmatrix}  j_m & a'_{m-1} & a'_m\\
  1 & a_m & a_{m-1}
\end{Bmatrix}\notag\\
&\hspace{0.95cm}\times\begin{Bmatrix}
   a_J & j_J & a_{J-1} \\
  1 & a'_{J-1} & j_J
\end{Bmatrix}(-1)^{a_J-a'_{J-1}+j_J+1}(-1)^{-a'_{J-1}+a_{J-1}-1}
\times\begin{Bmatrix}
 a'_{J-1} & j_J &  a_J \\
 1 & a'_J & j_J
\end{Bmatrix}(-1)^{a'_{J-1}-a'_J+j_J}(-1)^{a'_J-a_J+1}\notag\\
&\quad\times X(j_I,j_J)^{\frac12}X(j_J,j_K)^{\frac12}\prod_{i=2}^{n-1}\sqrt{2a_i+1}\sqrt{2J+1}
\makeSymbol{\includegraphics[width=9.5cm]{graph/volume/volumeoperator-IJK-m-8}}\,.
\end{align}
The exponents in \eqref{general-1-1} can be simplified as
\begin{align}
&(-1)^{a_{I-1}-a'_I+j_I}(-1)^{a_K\bcancel{-a'_{K-1}}+j_K+1}(-1)^{a_I\cancel{-a_{J-1}}+\sum_{l=I+1}^{J-1}j_l}(-1)^{\bcancel{a'_{K-1}}-a'_J-\sum_{m=J+1}^{K-1}j_m}(-1)^{\xcancel{a_J}-a'_{J-1}+j_J+1}(-1)^{-a'_{J-1}\cancel{+a_{J-1}}-1}(-1)^{a'_{J-1}-a'_J+j_J}(-1)^{a'_J\xcancel{-a_J}+1}\notag\\
&=(-1)^{a_{I-1}+j_I+a_K+j_K}(-1)^{a_I-a'_I}(-1)^{\sum_{l=I+1}^{J-1}j_l}(-1)^{-\sum_{m=J+1}^{K-1}j_m}(-1)^{a'_{J-1}+a'_J}(-1)^{-2(a'_J+a'_{J-1}-j_J)}\notag\\
&=(-1)^{a_{I-1}+j_I+a_K+j_K}(-1)^{a_I-a'_I}(-1)^{\sum_{l=I+1}^{J-1}j_l}(-1)^{-\sum_{m=J+1}^{K-1}j_m}(-1)^{a'_{J-1}+a'_J}\,,
\end{align}
where we have used, in the second step, the fact $a'_J+a'_{J-1}-j_J\in {\rm Z}$ due to the allowed triple $(a'_J,a'_{J-1},j_J)$ satisfying triangular condition. Considering the simplified factor and properly adjusting the ordering of multi-products of $\sqrt{2a+1}$, we finally have the compact result
\begin{align}\label{1-general-final}
&\quad\hat{q}^{<JK;IJ>}_{IJK}\cdot {\left(i^{\,J;\,\vec{a}}_v\right)_{\,m_1\cdots m_I\cdots m_J\cdots m_K\cdots m_n}}^M\notag\\
&=\sum_{(a'_I\cdots a'_{K-1})}(-1)^{a_{I-1}+j_I+a_K+j_K}(-1)^{a_I-a'_I}(-1)^{\sum_{l=I+1}^{J-1}j_l}(-1)^{-\sum_{m=J+1}^{K-1}j_m}X(j_I,j_J)^{\frac12}X(j_J,j_K)^{\frac12}\notag\\
&\quad\times\sqrt{(2a'_I+1)(2a_I+1)}\sqrt{(2a'_J+1)(2a_J+1)}
\begin{Bmatrix} a_{I-1} & j_I &  a_I \\
 1 & a'_I & j_I
\end{Bmatrix}\;\begin{Bmatrix}  a_K & j_K & a_{K-1} \\
  1 & a'_{K-1} & j_K
\end{Bmatrix}\notag\\
&\quad\times\prod_{l=I+1}^{J-1}\sqrt{(2a'_l+1)(2a_l+1)}(-1)^{a'_{l-1}+a_{l-1}+1}
\begin{Bmatrix}  j_l & a'_{l-1} & a'_l \\
  1 & a_l & a_{l-1}
\end{Bmatrix}\notag\\
&\quad\times\prod_{m=J+1}^{K-1}\sqrt{(2a'_m+1)(2a_m+1)}(-1)^{a'_{m-1}+a_{m-1}+1}
\begin{Bmatrix}  j_m & a'_{m-1} & a'_m\\
  1 & a_m & a_{m-1}
\end{Bmatrix}\notag\\
&\quad\times(-1)^{a'_{J-1}+a'_J}\begin{Bmatrix}  a_J & j_J & a_{J-1} \\
  1 & a'_{J-1} & j_J
\end{Bmatrix}\begin{Bmatrix} a'_{J-1} & j_J &  a_J \\
 1 & a'_J & j_J
\end{Bmatrix}\notag\\
&\quad\times\prod_{i=2}^{I-1}\sqrt{2a_i+1}\prod_{s=I}^{K-1}\sqrt{2a'_s+1}\prod_{k=K}^{n-1}\sqrt{2a_k+1}\sqrt{2J+1}
\makeSymbol{\includegraphics[width=9.5cm]{graph/volume/volumeoperator-IJK-m-8}}\,.
\end{align}
The above result is in the case of $J>I+1$ and $K>J+1$, which assures the limitations of the summations and multi-product exist, i.e, the upper limitations are always not less than the low limitations. Intuitively the results for the other cases can be seen as special cases of the above result by omitting the corresponding summations and multiplications which do not exit. The following analysis in step by step will confirm this intuition.\\\\
(I). The first case of $J=I+1$ and $K=J+1$\\
In this case, the corresponding processes can be simplified and shown in the following way
\begin{align}
&\makeSymbol{\includegraphics[width=6cm]{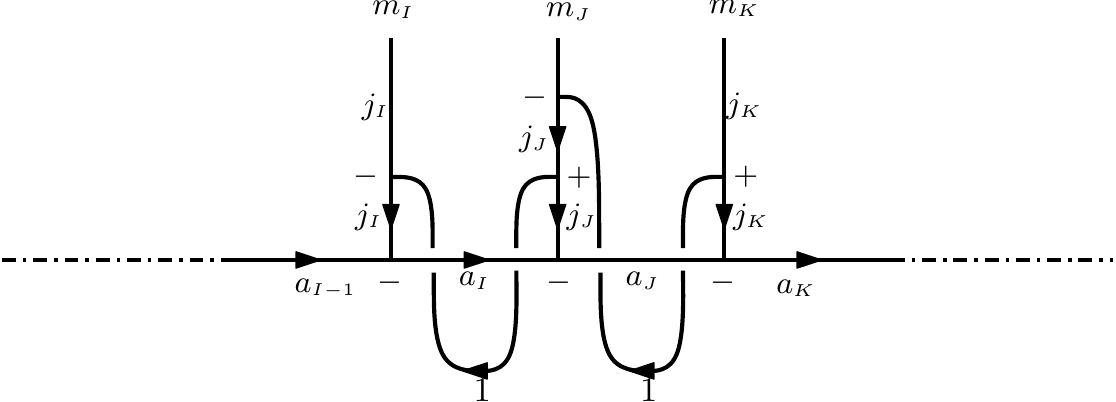}}\quad\mapsto\quad\makeSymbol{\includegraphics[width=6cm]{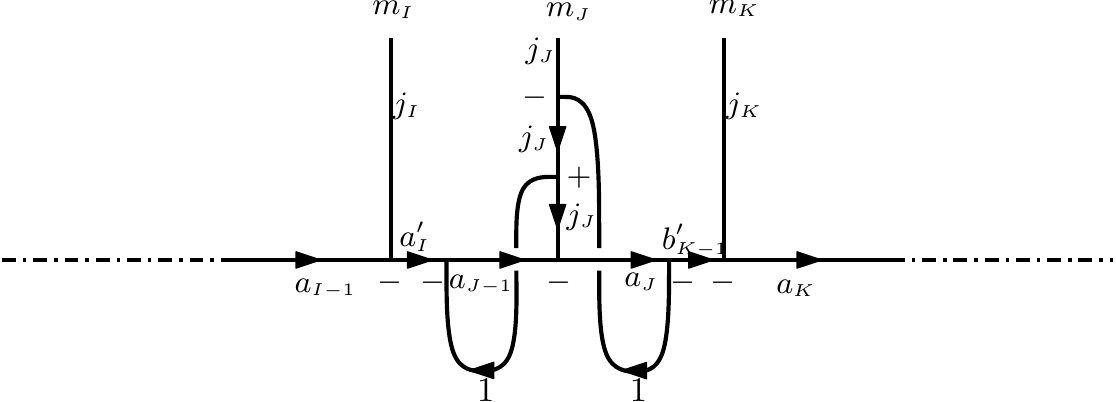}}\notag\\
&\mapsto\quad\makeSymbol{\includegraphics[width=6cm]{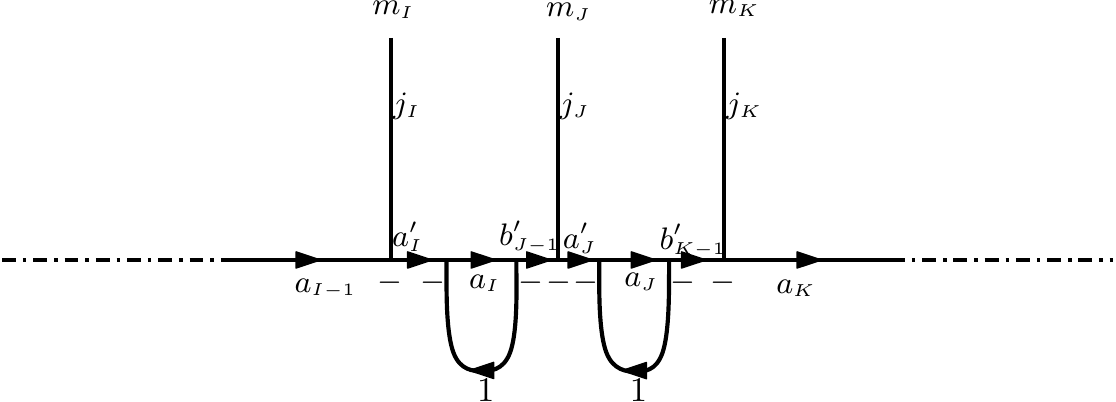}}\quad\mapsto\quad\makeSymbol{\includegraphics[width=6cm]{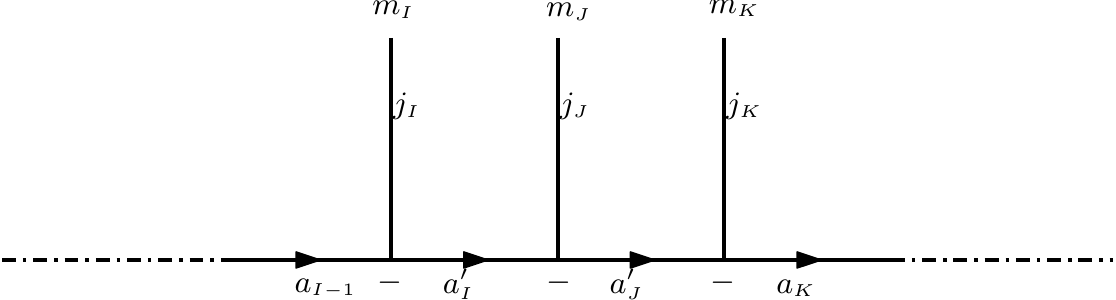}}\,.
\end{align}
The result can be directly obtained from \eqref{general-1} by omitting those terms appeared in the second step as
\begin{align}\label{1-case-1-1}
&\sum_{(a'_I,\cdots,a'_{K-1})}(2a'_I+1)(-1)^{a_{I-1}-a'_I+j_I}\begin{Bmatrix} a_{I-1} & j_I &  a_I \\
 1 & a'_I & j_I
\end{Bmatrix}
\times(2a'_{K-1}+1)(-1)^{a_K-a'_{K-1}+j_K+1}
\begin{Bmatrix}  a_K & j_K & a_{K-1} \\
  1 & a'_{K-1} & j_K
\end{Bmatrix}\notag\\
&\hspace{0.95cm}\times\begin{Bmatrix}
   a_J & j_J & a_{J-1} \\
  1 & a'_{J-1} & j_J
\end{Bmatrix}(-1)^{a_J-a'_{J-1}+j_J+1}(-1)^{-a'_{J-1}+a_{J-1}-1}
\times\begin{Bmatrix}
 a'_{J-1} & j_J &  a_J \\
 1 & a'_J & j_J
\end{Bmatrix}(-1)^{a'_{J-1}-a'_J+j_J}(-1)^{a'_J-a_J+1}\notag\\
&\quad\times X(j_I,j_J)^{\frac12}X(j_J,j_K)^{\frac12}\prod_{i=2}^{n-1}\sqrt{2a_i+1}\sqrt{2J+1}
\makeSymbol{\includegraphics[width=6.5cm]{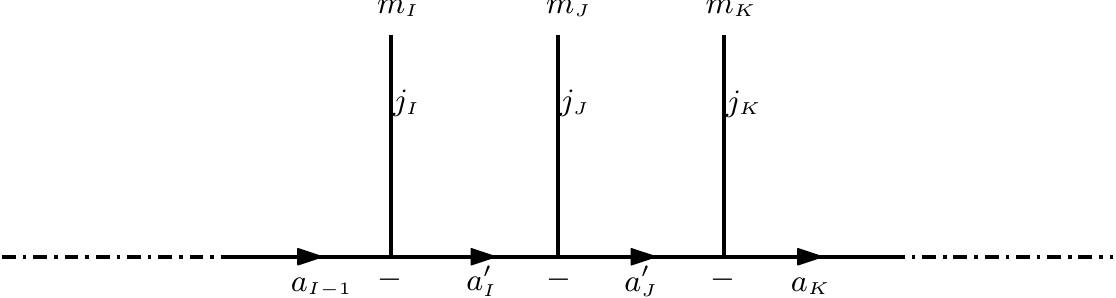}}\,.
\end{align}
The exponents can be simplified as
\begin{align}
&(-1)^{a_{I-1}-a'_I+j_I}(-1)^{a_K-a'_{K-1}+j_K+1}(-1)^{\bcancel{a_J}-a'_{J-1}+j_J+1}(-1)^{-a'_{J-1}+a_{J-1}-1}(-1)^{a'_{J-1}-a'_J+j_J}(-1)^{a'_J\bcancel{-a_J}+1}\notag\\
&=(-1)^{a_{I-1}+j_I+a_K+j_K}(-1)^{a_{J-1}-a'_I}(-1)^{a'_{J-1}+a'_J}(-1)^{-a'_{K-1}-a'_J-2a'_{J-1}+2j_J}\notag\\
&=(-1)^{a_{I-1}+j_I+a_K+j_K}(-1)^{a_I-a'_I}(-1)^{a'_{J-1}+a'_J}(-1)^{-2(a'_J+a'_{J-1}-j_J)}\notag\\
&=(-1)^{a_{I-1}+j_I+a_K+j_K}(-1)^{a_I-a'_I}(-1)^{a'_{J-1}+a'_J}\,,
\end{align}
where, in the second step, we have made the replacements $a_{J-1}=a_I$ and $a'_{K-1}=a'_J$. After replacing $(2a'_{K-1}+1)$ by $(2a'_J+1)$ and properly adjusting the ordering of multi-products of $\sqrt{2a+1}$, Eq. \eqref{1-case-1-1} reduces to
\begin{align}\label{1-case-1-final}
&\sum_{(a'_I\cdots a'_{K-1})}(-1)^{a_{I-1}+j_I+a_K+j_K}(-1)^{a_I-a'_I}X(j_I,j_J)^{\frac12}X(j_J,j_K)^{\frac12}\notag\\
&\quad\times\sqrt{(2a'_I+1)(2a_I+1)}\sqrt{(2a'_J+1)(2a_J+1)}
\begin{Bmatrix} a_{I-1} & j_I &  a_I \\
 1 & a'_I & j_I
\end{Bmatrix}\;\begin{Bmatrix}  a_K & j_K & a_{K-1} \\
  1 & a'_{K-1} & j_K
\end{Bmatrix}\notag\\
&\quad\times(-1)^{a'_{J-1}+a'_J}\begin{Bmatrix}  a_J & j_J & a_{J-1} \\
  1 & a'_{J-1} & j_J
\end{Bmatrix}\begin{Bmatrix} a'_{J-1} & j_J &  a_J \\
 1 & a'_J & j_J
\end{Bmatrix}\notag\\
&\quad\times\prod_{i=2}^{I-1}\sqrt{2a_i+1}\prod_{s=I}^{K-1}\sqrt{2a'_s+1}\prod_{k=K}^{n-1}\sqrt{2a_k+1}\sqrt{2J+1}
\makeSymbol{\includegraphics[width=6.5cm]{graph/volume/volumeoperator-IJK-case-I}}\,.
\end{align}
As a special case, the result \eqref{1-case-1-final} can also be directly written down from Eq. \eqref{1-general-final}. In \eqref{1-general-final}, the multi-products $\prod_{l=I+1}^{J-1}$ and $\prod_{m=J+1}^{K-1}$ and summations $\sum_{l=I+1}^{J-1}$ and $\sum_{m=J+1}^{K-1}$ do not exist for $J=I+1,K=J+1$. Thus the terms involve these multi-products and summations can be omitted, and this yields the result \eqref{1-case-1-final}.\\\\
(II). The second case of $J=I+1$ and $K>J+1$\\
The result can be directly obtained from \eqref{general-1} by omitting some terms appeared in the first line from the second step. Taking account of Eq. \eqref{interchange}, we obtain
\begin{align}
&\sum_{(a'_I,\cdots,a'_{K-1})}(2a'_I+1)(-1)^{a_{I-1}-a'_I+j_I}\begin{Bmatrix} a_{I-1} & j_I &  a_I \\
 1 & a'_I & j_I
\end{Bmatrix}
\times(2a'_J+1)(-1)^{a_K-a'_{K-1}+j_K+1}
\begin{Bmatrix}  a_K & j_K & a_{K-1} \\
  1 & a'_{K-1} & j_K
\end{Bmatrix}\notag\\
&\hspace{0.95cm}\times(-1)^{a'_{K-1}-a'_J-\sum_{m=J+1}^{K-1}j_m}\prod_{m=J+1}^{K-1}(2a'_m+1)(-1)^{a'_{m-1}+a_{m-1}+1}
\begin{Bmatrix}  j_m & a'_{m-1} & a'_m\\
  1 & a_m & a_{m-1}
\end{Bmatrix}\notag\\
&\hspace{0.95cm}\times\begin{Bmatrix}
   a_J & j_J & a_{J-1} \\
  1 & a'_{J-1} & j_J
\end{Bmatrix}(-1)^{a_J-a'_{J-1}+j_J+1}(-1)^{-a'_{J-1}+a_{J-1}-1}
\times\begin{Bmatrix}
 a'_{J-1} & j_J &  a_J \\
 1 & a'_J & j_J
\end{Bmatrix}(-1)^{a'_{J-1}-a'_J+j_J}(-1)^{a'_J-a_J+1}\notag\\
&\quad\times X(j_I,j_J)^{\frac12}X(j_J,j_K)^{\frac12}\prod_{i=2}^{n-1}\sqrt{2a_i+1}\sqrt{2J+1}
\makeSymbol{\includegraphics[width=8.5cm]{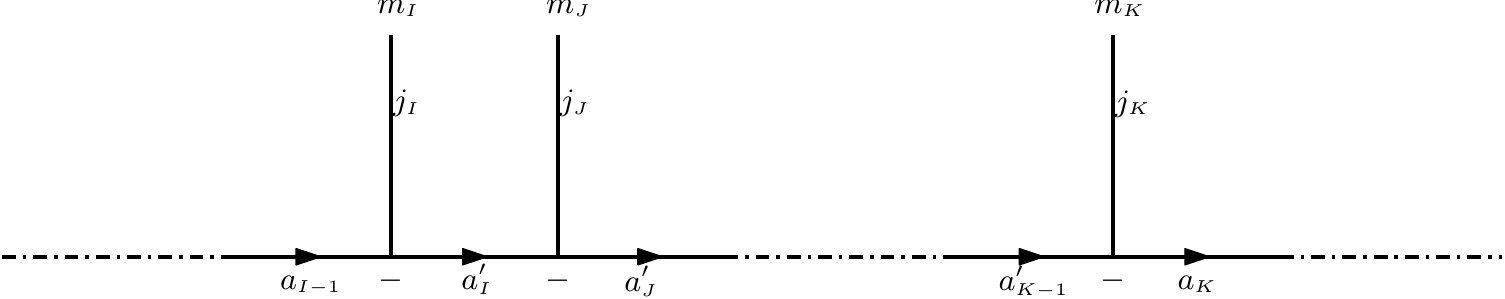}}\,.
\end{align}
The exponents can be simplified as
\begin{align}
&(-1)^{a_{I-1}-a'_I+j_I}(-1)^{a_K\bcancel{-a'_{K-1}}+j_K+1}(-1)^{\bcancel{a'_{K-1}}-a'_J-\sum_{m=J+1}^{K-1}j_m}(-1)^{\cancel{a_J}-a'_{J-1}+j_J+1}(-1)^{-a'_{J-1}+a_{J-1}-1}(-1)^{a'_{J-1}-a'_J+j_J}(-1)^{a'_J\cancel{-a_J}+1}\notag\\
&=(-1)^{a_{I-1}+j_I+a_K+j_K}(-1)^{a_{J-1}-a'_I}(-1)^{-\sum_{m=J+1}^{K-1}j_m}(-1)^{a'_{J-1}+a'_J}(-1)^{-2(a'_J+a'_{J-1}-j_J)}\notag\\
&=(-1)^{a_{I-1}+j_I+a_K+j_K}(-1)^{a_I-a'_I}(-1)^{-\sum_{m=J+1}^{K-1}j_m}(-1)^{a'_{J-1}+a'_J}\,,
\end{align}
where, in the second step, we have replaced $a_{J-1}$ by $a_I$. After properly adjusting the ordering of multi-products of $\sqrt{2a+1}$, we have
\begin{align}\label{1-case-2-final}
&\sum_{(a'_I\cdots a'_{K-1})}(-1)^{a_{I-1}+j_I+a_K+j_K}(-1)^{a_I-a'_I}(-1)^{-\sum_{m=J+1}^{K-1}j_m}X(j_I,j_J)^{\frac12}X(j_J,j_K)^{\frac12}\notag\\
&\quad\times\sqrt{(2a'_I+1)(2a_I+1)}\sqrt{(2a'_J+1)(2a_J+1)}
\begin{Bmatrix} a_{I-1} & j_I &  a_I \\
 1 & a'_I & j_I
\end{Bmatrix}\;\begin{Bmatrix}  a_K & j_K & a_{K-1} \\
  1 & a'_{K-1} & j_K
\end{Bmatrix}\notag\\
&\quad\times\prod_{m=J+1}^{K-1}\sqrt{(2a'_m+1)(2a_m+1)}(-1)^{a'_{m-1}+a_{m-1}+1}
\begin{Bmatrix}  j_m & a'_{m-1} & a'_m\\
  1 & a_m & a_{m-1}
\end{Bmatrix}\notag\\
&\quad\times(-1)^{a'_{J-1}+a'_J}\begin{Bmatrix}  a_J & j_J & a_{J-1} \\
  1 & a'_{J-1} & j_J
\end{Bmatrix}\begin{Bmatrix} a'_{J-1} & j_J &  a_J \\
 1 & a'_J & j_J
\end{Bmatrix}\notag\\
&\quad\times\prod_{i=2}^{I-1}\sqrt{2a_i+1}\prod_{s=I}^{K-1}\sqrt{2a'_s+1}\prod_{k=K}^{n-1}\sqrt{2a_k+1}\sqrt{2J+1}
\makeSymbol{\includegraphics[width=8.5cm]{graph/volume/volumeoperator-IJK-case-II}}\,.
\end{align}
Again, the above result can also be directly written down from Eq. \eqref{1-general-final}. In \eqref{1-general-final}, the multi-products $\prod_{l=I+1}^{J-1}$ and summations $\sum_{l=I+1}^{J-1}$ do not exist for $J=I+1$. Hence the terms involve these multi-products and summations can be omitted, and this yields the result \eqref{1-case-2-final}.\\\\
(III). The third case of $J>I+1$ and $K=J+1$\\
The result can be directly obtained from \eqref{general-1} by omitting some terms appeared in the second line from the second step as
\begin{align}
&\sum_{(a'_I,\cdots,a'_{K-1})}(2a'_I+1)(-1)^{a_{I-1}-a'_I+j_I}\begin{Bmatrix} a_{I-1} & j_I &  a_I \\
 1 & a'_I & j_I
\end{Bmatrix}
\times(2a'_{K-1}+1)(-1)^{a_K-a'_{K-1}+j_K+1}
\begin{Bmatrix}  a_K & j_K & a_{K-1} \\
  1 & a'_{K-1} & j_K
\end{Bmatrix}\notag\\
&\hspace{0.95cm}\times(-1)^{a_I-a_{J-1}+\sum_{l=I+1}^{J-1}j_l}\prod_{l=I+1}^{J-1}(2a'_l+1)(-1)^{a'_{l-1}+a_{l-1}+1}
\begin{Bmatrix}  j_l & a'_{l-1} & a'_l \\
  1 & a_l & a_{l-1}
\end{Bmatrix}\notag\\
&\hspace{0.95cm}\times\begin{Bmatrix}
   a_J & j_J & a_{J-1} \\
  1 & a'_{J-1} & j_J
\end{Bmatrix}(-1)^{a_J-a'_{J-1}+j_J+1}(-1)^{-a'_{J-1}+a_{J-1}-1}
\times\begin{Bmatrix}
 a'_{J-1} & j_J &  a_J \\
 1 & a'_J & j_J
\end{Bmatrix}(-1)^{a'_{J-1}-a'_J+j_J}(-1)^{a'_J-a_J+1}\notag\\
&\quad\times X(j_I,j_J)^{\frac12}X(j_J,j_K)^{\frac12}\prod_{i=2}^{n-1}\sqrt{2a_i+1}\sqrt{2J+1}
\makeSymbol{\includegraphics[width=8.5cm]{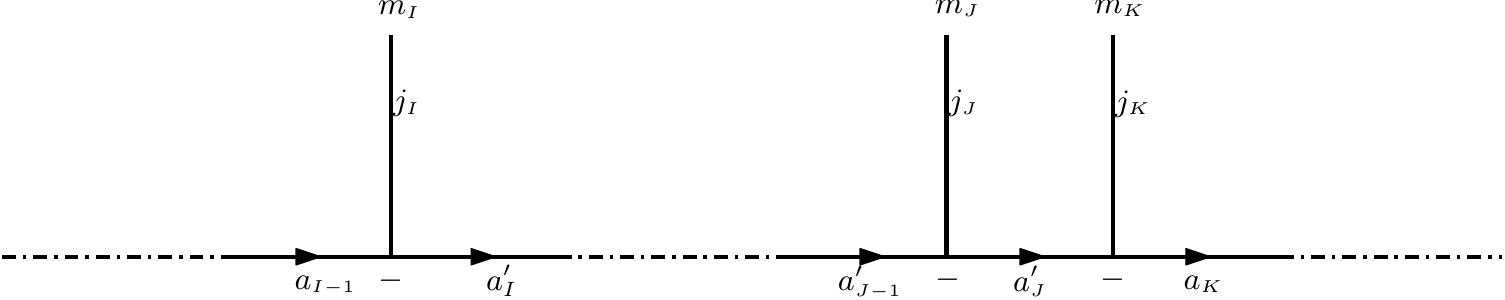}}\,.
\end{align}
The exponents can be simplified as
\begin{align}
&(-1)^{a_{I-1}-a'_I+j_I}(-1)^{a_K-a'_{K-1}+j_K+1}(-1)^{a_I\bcancel{-a_{J-1}}+\sum_{l=I+1}^{J-1}j_l}(-1)^{\cancel{a_J}-a'_{J-1}+j_J+1}(-1)^{-a'_{J-1}\bcancel{+a_{J-1}}-1}(-1)^{a'_{J-1}-a'_J+j_J}(-1)^{a'_J\cancel{-a_J}+1}\notag\\
&=(-1)^{a_{I-1}+j_I+a_K+j_K}(-1)^{a_I-a'_I}(-1)^{\sum_{l=I+1}^{J-1}j_l}(-1)^{a'_{J-1}+a'_J}(-1)^{-a'_{K-1}-a'_J-2a'_{J-1}+2j_J}\notag\\
&=(-1)^{a_{I-1}+j_I+a_K+j_K}(-1)^{a_I-a'_I}(-1)^{\sum_{l=I+1}^{J-1}j_l}(-1)^{a'_{J-1}+a'_J}(-1)^{-2(a'_J+a'_{J-1}-j_J)}\notag\\
&=(-1)^{a_{I-1}+j_I+a_K+j_K}(-1)^{a_I-a'_I}(-1)^{\sum_{l=I+1}^{J-1}j_l}(-1)^{a'_{J-1}+a'_J}\,,
\end{align}
where, in the second step, $a'_{K-1}=a'_J$ was used. After replacing $(2a'_{K-1}+1)$ by $(2a'_J+1)$ and properly adjusting the ordering of multi-products of $\sqrt{2a+1}$, we have
\begin{align}\label{1-case-3-final}
&\sum_{(a'_I\cdots a'_{K-1})}(-1)^{a_{I-1}+j_I+a_K+j_K}(-1)^{a_I-a'_I}(-1)^{\sum_{l=I+1}^{J-1}j_l}X(j_I,j_J)^{\frac12}X(j_J,j_K)^{\frac12}\notag\\
&\quad\times\sqrt{(2a'_I+1)(2a_I+1)}\sqrt{(2a'_J+1)(2a_J+1)}
\begin{Bmatrix} a_{I-1} & j_I &  a_I \\
 1 & a'_I & j_I
\end{Bmatrix}\;\begin{Bmatrix}  a_K & j_K & a_{K-1} \\
  1 & a'_{K-1} & j_K
\end{Bmatrix}\notag\\
&\quad\times\prod_{l=I+1}^{J-1}\sqrt{(2a'_l+1)(2a_l+1)}(-1)^{a'_{l-1}+a_{l-1}+1}
\begin{Bmatrix}  j_l & a'_{l-1} & a'_l \\
  1 & a_l & a_{l-1}
\end{Bmatrix}\notag\\
&\quad\times(-1)^{a'_{J-1}+a'_J}\begin{Bmatrix}  a_J & j_J & a_{J-1} \\
  1 & a'_{J-1} & j_J
\end{Bmatrix}\begin{Bmatrix} a'_{J-1} & j_J &  a_J \\
 1 & a'_J & j_J
\end{Bmatrix}\notag\\
&\quad\times\prod_{i=2}^{I-1}\sqrt{2a_i+1}\prod_{s=I}^{K-1}\sqrt{2a'_s+1}\prod_{k=K}^{n-1}\sqrt{2a_k+1}\sqrt{2J+1}
\makeSymbol{\includegraphics[width=8.5cm]{graph/volume/volumeoperator-IJK-case-III}}\,.
\end{align}
The above result can be also directly written down from Eq. \eqref{1-general-final}. In \eqref{1-general-final}, the multi-products $\prod_{m=J+1}^{K-1}$ and summations $\sum_{m=J+1}^{K-1}$ do not exist for $K=J+1$. Hence the terms involve these multi-products and summations can be omitted, and this yields the result \eqref{1-case-3-final}.

The above discussions in case by case indicate that the general form of the operation can be written as
\begin{align}\label{1-general-final-re}
\hat{q}^{<JK;IJ>}_{IJK}\cdot {\left(i^{\,J;\,\vec{a}}_v\right)_{\,m_1\cdots m_I\cdots m_J\cdots m_K\cdots m_n}}^M
&=\sum_{(a'_I\cdots a'_{K-1})}(-1)^{a_{I-1}+j_I+a_K+j_K}(-1)^{a_I-a'_I}(-1)^{\sum_{l=I+1}^{J-1}j_l}(-1)^{-\sum_{m=J+1}^{K-1}j_m}X(j_I,j_J)^{\frac12}X(j_J,j_K)^{\frac12}\notag\\
&\quad\times\sqrt{(2a'_I+1)(2a_I+1)}\sqrt{(2a'_J+1)(2a_J+1)}
\begin{Bmatrix} a_{I-1} & j_I &  a_I \\
 1 & a'_I & j_I
\end{Bmatrix}\;\begin{Bmatrix}  a_K & j_K & a_{K-1} \\
  1 & a'_{K-1} & j_K
\end{Bmatrix}\notag\\
&\quad\times\prod_{l=I+1}^{J-1}\sqrt{(2a'_l+1)(2a_l+1)}(-1)^{a'_{l-1}+a_{l-1}+1}
\begin{Bmatrix}  j_l & a'_{l-1} & a'_l \\
  1 & a_l & a_{l-1}
\end{Bmatrix}\notag\\
&\quad\times\prod_{m=J+1}^{K-1}\sqrt{(2a'_m+1)(2a_m+1)}(-1)^{a'_{m-1}+a_{m-1}+1}
\begin{Bmatrix}  j_m & a'_{m-1} & a'_m\\
  1 & a_m & a_{m-1}
\end{Bmatrix}\notag\\
&\quad\times(-1)^{a'_{J-1}+a'_J}\begin{Bmatrix}  a_J & j_J & a_{J-1} \\
  1 & a'_{J-1} & j_J
\end{Bmatrix}\begin{Bmatrix} a'_{J-1} & j_J &  a_J \\
 1 & a'_J & j_J
\end{Bmatrix}\notag\\
&\quad\times{\left(i^{\,J;\,\vec{\tilde{a}}}_v\right)_{\,m_1\cdots m_I\cdots m_J\cdots m_K\cdots m_n}}^M\,,
\end{align}
where the tuple $\vec{\tilde{a}}$ in the new intertwiner ${\left(i^{\,J;\,\vec{\tilde{a}}}_v\right)_{\,m_1\cdots m_I\cdots m_J\cdots m_K\cdots m_n}}^M$ is given by
\begin{align}
\vec{\tilde{a}}=(a_2,\cdots,a_{I-1},a'_{I},\cdots,a'_{K-1},a_K,\cdots,a_{n-1})\,.
\end{align}
For special cases of $J-1<I+1$ and $K-1<J+1$, the final results can be obtained from \eqref{1-general-final-re} by omitting the corresponding multi-products $\prod_{l=I+1}^{J-1}$ and $\prod_{m=J+1}^{K-1}$ and summations $\sum_{l=I+1}^{J-1}$ and $\sum_{m=J+1}^{K-1}$.

The second term in parenthesis of Eq. \eqref{q-IJK} can be calculated similarly to  the former one. Here we omit the intermediate steps and directly write down the result as (a complete calculation is also shown in \ref{appendix-C})
\begin{align}\label{2-general-final-re}
\hat{q}^{<IJ;JK>}_{IJK}\cdot {\left(i^{\,J;\,\vec{a}}_v\right)_{\,m_1\cdots m_I\cdots m_J\cdots m_K\cdots m_n}}^M&=\sum_{(a'_I,\cdots,a'_{K-1})}(-1)^{a_{I-1}+j_I+a_K+j_K}(-1)^{a_I-a'_I}(-1)^{\sum_{l=I+1}^{J-1}j_l}(-1)^{-\sum_{m=J+1}^{K-1}}X(j_I,j_J)^{\frac12}X(j_J,j_K)^{\frac12}\notag\\&\quad\times\sqrt{(2a'_I+1)(2a_I+1)}\sqrt{(2a'_J+1)(2a_J+1)}\begin{Bmatrix} a_{I-1} & j_I &  a_I \\
 1 & a'_I & j_I
\end{Bmatrix}
\begin{Bmatrix}  a_K & j_K & a_{K-1} \\
  1 & a'_{K-1} & j_K
\end{Bmatrix}\notag\\
&\quad\times\prod_{l=I+1}^{J-1}\sqrt{(2a'_l+1)(2a_l+1)}(-1)^{a'_{l-1}+a_{l-1}+1}
\begin{Bmatrix}  j_l & a'_{l-1} & a'_l \\
  1 & a_l & a_{l-1}
\end{Bmatrix}\notag\\
&\quad\times\prod_{m=J+1}^{K-1}\sqrt{(2a'_m+1)(2a_m+1)}(-1)^{a'_{m-1}+a_{m-1}+1}
\begin{Bmatrix}  j_m & a'_{m-1} & a'_m\\
  1 & a_m & a_{m-1}
\end{Bmatrix}\notag\\
&\quad\times
(-1)^{a_{J-1}+a_J}\begin{Bmatrix}  a'_J & j_J & a_{J-1} \\
  1 & a'_{J-1} & j_J
\end{Bmatrix}\begin{Bmatrix} a_{J-1} & j_J &  a_J \\
 1 & a'_J & j_J
\end{Bmatrix}
\notag\\
&\quad\times{\left(i^{\,J;\,\vec{\tilde{a}}}_v\right)_{\,m_1\cdots m_I\cdots m_J\cdots m_K\cdots m_n}}^M\,.
\end{align}
Again, the final results for the special cases of $J-1<I+1$ and $K-1<J+1$ can be obtained from \eqref{2-general-final-re} by omitting the corresponding multi-products $\prod_{l=I+1}^{J-1}$ and $\prod_{m=J+1}^{K-1}$ and summations $\sum_{l=I+1}^{J-1}$ and $\sum_{m=J+1}^{K-1}$.

Combining the results \eqref{1-general-final-re} with \eqref{2-general-final-re},  for the case of $I>2$ and $K<n$, the action of $\hat{q}_{IJK}$ in Eq. \eqref{q-IJK} on the intertwiner can be explicitly written down as
\begin{align}\label{q-IJK-result-general}
\hat{q}_{IJK}\cdot {\left(i^{\,J;\,\vec{a}}_v\right)_{\,m_1\cdots m_I\cdots m_J\cdots m_K\cdots m_n}}^M
&=-\frac14\sum_{(a'_I\cdots a'_{K-1})}(-1)^{a_K+j_K+a_{I-1}+j_I}(-1)^{a_I-a'_I}(-1)^{\sum_{l=I+1}^{J-1}j_l}(-1)^{-\sum_{m=J+1}^{K-1}j_m}X(j_I,j_J)^{\frac12}X(j_J,j_K)^{\frac12}\notag\\
&\quad\times\sqrt{(2a'_I+1)(2a_I+1)}\sqrt{(2a'_J+1)(2a_J+1)}
\begin{Bmatrix} a_{I-1} & j_I &  a_I \\
 1 & a'_I & j_I
\end{Bmatrix}\;\begin{Bmatrix}  a_K & j_K & a_{K-1} \\
  1 & a'_{K-1} & j_K
\end{Bmatrix}\notag\\
&\quad\times\prod_{l=I+1}^{J-1}\sqrt{(2a'_l+1)(2a_l+1)}(-1)^{a'_{l-1}+a_{l-1}+1}
\begin{Bmatrix}  j_l & a'_{l-1} & a'_l \\
  1 & a_l & a_{l-1}
\end{Bmatrix}\notag\\
&\quad\times\prod_{m=J+1}^{K-1}\sqrt{(2a'_m+1)(2a_m+1)}(-1)^{a'_{m-1}+a_{m-1}+1}
\begin{Bmatrix}  j_m & a'_{m-1} & a'_m\\
  1 & a_m & a_{m-1}
\end{Bmatrix}\notag\\
&\quad\times\left[(-1)^{a'_{J-1}+a'_J}\begin{Bmatrix}  a_J & j_J & a_{J-1} \\
  1 & a'_{J-1} & j_J
\end{Bmatrix}\begin{Bmatrix} a'_{J-1} & j_J &  a_J \\
 1 & a'_J & j_J
\end{Bmatrix}-(-1)^{a_{J-1}+a_J}\begin{Bmatrix}  a'_J & j_J & a_{J-1} \\
  1 & a'_{J-1} & j_J
\end{Bmatrix}\begin{Bmatrix} a_{J-1} & j_J &  a_J \\
 1 & a'_J & j_J
\end{Bmatrix}
\right]\notag\\
&\quad\times{\left(i^{\,J;\,\vec{\tilde{a}}}_v\right)_{\,m_1\cdots m_I\cdots m_J\cdots m_K\cdots m_n}}^M\,.
\end{align}

Again, we expect that the result \eqref{q-IJK-result-general} is general and also suitable for the remained special cases of $0<I\leqslant2$ and $K=n$ when $a_{I-1}$ and $a_K$ do not exist. While $a_{I-1}$ and $a_K$ do not exist in the intertwiner, we can `create' them via (see Appendix \ref{appendix-B-3} for proof)
\begin{align}\label{line-to-3lines}
\makeSymbol{\includegraphics[width=1.3cm]{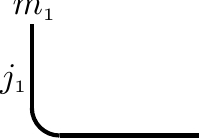}}=(1)^{2j_1}\sqrt{2a_0+1}\sqrt{2a_1+1}\;\;\makeSymbol{\includegraphics[width=2cm]{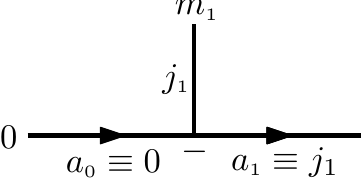}}\,,
\end{align}
where $a_0\equiv 0,a_1\equiv j_1$, and $J$ is relabeled by $a_n$. Then the intertwiner in \eqref{intertwined-form} can be extended as
\begin{align}
{\left(i^{\,J;\,\vec{a}}_v\right)_{\,m_1\cdots m_I\cdots m_J\cdots m_K\cdots m_n}}^M&=(1)^{2j_1}\;\prod_{i=0}^{n}\sqrt{2a_i+1}\;\;
\makeSymbol{\includegraphics[width=8cm]{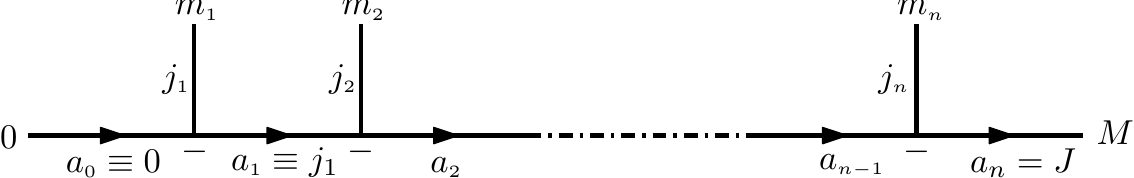}}\,.
\end{align}
We are immediately awake to the fact that \eqref{q-IJK-result-general} is suitable for these special cases just by taking $a_0=0$,  $a_1=a'_1=j_1$, and $a_n=J$.

Note that the operator $\hat{q}_{IJK}$ at $v$ only changes the intermediate angular momenta $a_I,\cdots,a_{K-1}$ between $I$ and $K$ of the intertwiner $i_v$, but leaves the magnetic quantum numbers $m_1,\cdots,m_n,M$ invariant. The matrix elements of $\hat{q}_{IJK}$ with respect to two given normalized spin network states $T^{\rm norm}_{\gamma,\vec{j},\vec{i}}(A):=\prod_{e\in E(\gamma)}\sqrt{2j_e+1}T_{\gamma,\vec{j},\vec{i}}(A)$ read
\begin{align}\label{q-IJK-norm-inner-prod}
\left(T^{\rm norm}_{\gamma',\vec{j}',\vec{i}'}, \hat{q}_{IJK}\cdot T^{\rm norm}_{\gamma,\vec{j},\vec{i}}\right)_{{\cal H}_{\rm kin}}&=\int_{SU(2)^{|E(\tilde{\gamma})|}}\prod_{e\in E(\tilde{\gamma})}{\rm d}\mu_H(h_e)
\overline{T^{\rm norm}_{\gamma',\vec{j}',\vec{i}'}(A)}\,\hat{q}_{IJK}\cdot T^{\rm norm}_{\gamma,\vec{j},\vec{i}}(A)\notag\\
&=\delta_{\gamma,\gamma'}\int_{SU(2)^{|E(\gamma)|}}\prod_{e\in E(\gamma)}{\rm d}\mu_H(h_e)
\overline{T^{\rm norm}_{\gamma,\vec{j}',\vec{i}'}(A)}\,\hat{q}_{IJK}\cdot T^{\rm norm}_{\gamma,\vec{j},\vec{i}}(A)\notag\\
&=\delta_{\gamma,\gamma'}\prod_{e\in E(\gamma)}\delta_{j_e,j'_e}\prod_{v'\in V(\gamma),v'\neq v}\delta_{i_{v'},i'_{v'}}\cdot\sum_{m_1,\cdots,m_n}\overline{{\left(i^{\,J';\,\vec{a}'}_{v}\right)_{m_1\cdots m_n}}^{M'}}\,\hat{q}_{IJK}\cdot {\left(i^{\,J;\,\vec{a}}_{v}\right)_{m_1\cdots m_n}}^M\notag\\
&=\delta_{\gamma,\gamma'}\prod_{e\in E(\gamma)}\delta_{j_e,j'_e}\prod_{v'\in V(\gamma),v'\neq v}\delta_{i_{v'},i'_{v'}}\cdot\sum_{m_1,\cdots,m_n}{\left[\left(i^{\,J';\,\vec{a}'}_{v}\right)^\dag\right]_{M'}}^{m_1\cdots m_n}\,\hat{q}_{IJK}\cdot {\left(i^{\,J;\,\vec{a}}_{v}\right)_{m_1\cdots m_n}}^M\notag\\
&=\delta_{\gamma,\gamma'}\prod_{e\in E(\gamma)}\delta_{j_e,j'_e}\prod_{v'\in V(\gamma),v'\neq v}\delta_{i_{v'},i'_{v'}}\cdot\langle \vec{a}';J',M'|\hat{q}_{IJK}|\vec{a};J,M\rangle_{{\cal H}^{v}_{j_1,\cdots,j_n}}\notag\\
&=\delta_{\gamma,\gamma'}\prod_{e\in E(\gamma)}\delta_{j_e,j'_e}\prod_{v'\in V(\gamma),v'\neq v}\delta_{i_{v'},i'_{v'}}\cdot\langle \vec{a}';J,M|\hat{q}_{IJK}|\vec{a};J,M\rangle_{{\cal H}^{v}_{j_1,\cdots,j_n}}\delta_{J,J'}\delta_{M,M'}\notag\\
&=:\delta_{\gamma,\gamma'}\prod_{e\in E(\gamma)}\delta_{j_e,j'_e}\prod_{v'\in V(\gamma),v'\neq v}\delta_{i_{v'},i'_{v'}}\cdot\langle \vec{a}'|\hat{q}_{IJK}|\vec{a}\rangle\delta_{J,J'}\delta_{M,M'}\,,
\end{align}
where $\tilde{\gamma}$ is any graph bigger than $\gamma$ and $\gamma'$, $|E(\tilde{\gamma})|$ denotes the number of the edges in $\tilde{\gamma}$, and ${\rm d}\mu_H(g)$ is the Haar measure on $SU(2)$. In the second step of Eq. \eqref{q-IJK-norm-inner-prod}, we notice that $\hat{q}_{IJK}$ at $v$ only change the intertwiner $i_v$, and if $\gamma$ differs from $\gamma'$, the integration with respect to the Haar measure gives a zero result. In the third step, the integration for the holonomies along the same edges but with different spins yields delta functions and the contractions of the intertwiner with its conjugate. In the forth and fifth steps, we have used the definition \eqref{intert-space-inner-product} of the inner product of the intertwiner space ${\cal H}^v_{j_1,\cdots,j_n}$ associated to $v$. Thus the general expression \eqref{q-IJK-result-general} allows us to {\em uniformly} write the matrix elements of $\hat{q}_{IJK}$ in the gauge-variant and gauge-invariant intertwiner, corresponding to resulting angular momentum $J\neq0$ and $J=0$ respectively, as
\begin{align}\label{q-matrix-element}
\langle \vec{a}'|\hat{q}_{IJK}|\vec{a}\rangle&\equiv\sum_{m_1,\cdots,m_n}\overline{{\left(i^{\,J;\,\vec{a}'}_{v}\right)_{m_1\cdots m_n}}^{M}}\,\hat{q}_{IJK}\cdot {\left(i^{\,J;\,\vec{a}}_{v}\right)_{m_1\cdots m_n}}^M\notag\\
&=-\frac14\sum_{(a'_I\cdots a'_{K-1})}(-1)^{a_K+j_K+a_{I-1}+j_I}(-1)^{a_I-a'_I}(-1)^{\sum_{l=I+1}^{J-1}j_l}(-1)^{-\sum_{m=J+1}^{K-1}j_m}X(j_I,j_J)^{\frac12}X(j_J,j_K)^{\frac12}\notag\\
&\quad\times\sqrt{(2a'_I+1)(2a_I+1)}\sqrt{(2a'_J+1)(2a_J+1)}
\begin{Bmatrix} a_{I-1} & j_I &  a_I \\
 1 & a'_I & j_I
\end{Bmatrix}\;\begin{Bmatrix}  a_K & j_K & a_{K-1} \\
  1 & a'_{K-1} & j_K
\end{Bmatrix}\notag\\
&\quad\times\prod_{l=I+1}^{J-1}\sqrt{(2a'_l+1)(2a_l+1)}(-1)^{a'_{l-1}+a_{l-1}+1}
\begin{Bmatrix}  j_l & a'_{l-1} & a'_l \\
  1 & a_l & a_{l-1}
\end{Bmatrix}\notag\\
&\quad\times\prod_{m=J+1}^{K-1}\sqrt{(2a'_m+1)(2a_m+1)}(-1)^{a'_{m-1}+a_{m-1}+1}
\begin{Bmatrix}  j_m & a'_{m-1} & a'_m\\
  1 & a_m & a_{m-1}
\end{Bmatrix}\notag\\
&\quad\times\left[(-1)^{a'_{J-1}+a'_J}\begin{Bmatrix}  a_J & j_J & a_{J-1} \\
  1 & a'_{J-1} & j_J
\end{Bmatrix}\begin{Bmatrix} a'_{J-1} & j_J &  a_J \\
 1 & a'_J & j_J
\end{Bmatrix}-(-1)^{a_{J-1}+a_J}\begin{Bmatrix}  a'_J & j_J & a_{J-1} \\
  1 & a'_{J-1} & j_J
\end{Bmatrix}\begin{Bmatrix} a_{J-1} & j_J &  a_J \\
 1 & a'_J & j_J
\end{Bmatrix}
\right]\notag\\
&\quad\times\sum_{m_1,\cdots,m_n}\overline{{\left(i^{\,J;\,\vec{a}'}_{v}\right)_{m_1\cdots m_n}}^{M}}{\left(i^{\,J;\,\vec{\tilde{a}}}_v\right)_{\,m_1\cdots m_I\cdots m_J\cdots m_K\cdots m_n}}^M\notag\\
&=-\frac14(-1)^{a_K+j_K+a_{I-1}+j_I}(-1)^{a_I-a'_I}(-1)^{\sum_{l=I+1}^{J-1}j_l}(-1)^{-\sum_{m=J+1}^{K-1}j_m}X(j_I,j_J)^{\frac12}X(j_J,j_K)^{\frac12}\notag\\
&\quad\times\sqrt{(2a'_I+1)(2a_I+1)}\sqrt{(2a'_J+1)(2a_J+1)}
\begin{Bmatrix} a_{I-1} & j_I &  a_I \\
 1 & a'_I & j_I
\end{Bmatrix}\;\begin{Bmatrix}  a_K & j_K & a_{K-1} \\
  1 & a'_{K-1} & j_K
\end{Bmatrix}\notag\\
&\quad\times\prod_{l=I+1}^{J-1}\sqrt{(2a'_l+1)(2a_l+1)}(-1)^{a'_{l-1}+a_{l-1}+1}
\begin{Bmatrix}  j_l & a'_{l-1} & a'_l \\
  1 & a_l & a_{l-1}
\end{Bmatrix}\notag\\
&\quad\times\prod_{m=J+1}^{K-1}\sqrt{(2a'_m+1)(2a_m+1)}(-1)^{a'_{m-1}+a_{m-1}+1}
\begin{Bmatrix}  j_m & a'_{m-1} & a'_m\\
  1 & a_m & a_{m-1}
\end{Bmatrix}\notag\\
&\quad\times\left[(-1)^{a'_{J-1}+a'_J}\begin{Bmatrix}  a_J & j_J & a_{J-1} \\
  1 & a'_{J-1} & j_J
\end{Bmatrix}\begin{Bmatrix} a'_{J-1} & j_J &  a_J \\
 1 & a'_J & j_J
\end{Bmatrix}-(-1)^{a_{J-1}+a_J}\begin{Bmatrix}  a'_J & j_J & a_{J-1} \\
  1 & a'_{J-1} & j_J
\end{Bmatrix}\begin{Bmatrix} a_{J-1} & j_J &  a_J \\
 1 & a'_J & j_J
\end{Bmatrix}
\right]\notag\\
&\quad\times
\left\{
\begin{array}{cc}
\prod\limits_{s=2}^{I-1}\delta_{a'_s,a_s}\prod\limits_{t=K}^{n-1}\delta_{a'_t,a_t}, &\text{for}\; I>2\;\text{and}\;K<n \\
\prod\limits_{s=2}^{I-1}\delta_{a'_s,a_s}, &\text{for}\; I>2\;\text{and}\;K=n \\
\prod\limits_{t=K}^{n-1}\delta_{a'_t,a_t}, &\text{for}\; I\leqslant2\;\text{and}\;K<n \\
1, &\text{for}\; I\leqslant2\;\text{and}\;K=n \\
\end{array}
\right.\,,
\end{align}
where, in the third step, we have used the result \eqref{orth-intert-graph-proof}. Note that the multi-products $\prod_{l=I+1}^{J-1}$ and $\prod_{m=J+1}^{K-1}$ and summations $\sum_{l=I+1}^{J-1}$ and $\sum_{m=J+1}^{K-1}$ should be omitted for the cases of $J<I+2$ and $K<J+2$, and we need to set $a_0=0$,  $a_1=a'_1=j_1$, and $a_n=J$ when $0<I\leqslant2$ and $K=n$, respectively. Under exchange $a\leftrightarrow a'$, the expression in the square bracket of \eqref{q-matrix-element} is antisymmetric, while the other terms leave invariant because of the symmetric properties of $6j$-symbol, symmetry of the delta function and the fact that $(-1)^{a_I-a'_I}=(-1)^{a'_I-a_I}$. Hence the matrix elements of $\hat{q}_{IJK}$ are antisymmetric, i.e.,
\begin{align}
\langle \vec{a}'|\hat{q}_{IJK}|\vec{a}\rangle=-\langle \vec{a}|\hat{q}_{IJK}|\vec{a}'\rangle\,.
\end{align}
The matrix element formula \eqref{q-matrix-element} derived in graphical method is the same as the formula obtained from algebraic manipulation for the case of $I>1$ and $J>I+1$ in \cite{Brunnemann:2004xi}, although different ways were adopted to deal with the recoupling problem. Moreover, as shown in above discussions, the formula \eqref{q-matrix-element} is also valid for other cases and hence can be regarded as a general expression.

Finally, we consider some special cases which usually appear. With the following values of $6j$-symbols \cite{Edmonds},
\begin{align}
\begin{Bmatrix}
0 & b & c\\
d & e & f
\end{Bmatrix}&=(-1)^{b+e+d}\frac{\delta_{b,c}\delta_{e,f}}{\sqrt{(2b+1)(2e+1)}}\label{6j-0}\,,\\
\begin{Bmatrix}
  a & b & c \\
  1 & c & b
\end{Bmatrix}
&=(-1)^{s+1}\frac{2\left[b(b+1)+c(c+1)-a(a+1)\right]}{X(b,c)^{1/2}}\label{6j-1}\,,\\
\begin{Bmatrix}
a & b & c-1\\
1 & c & b
\end{Bmatrix}&=(-1)^s\left[\frac{2(s+1)(s-2a)(s-2b)(s-2c+1)}{2b(2b+1)(2b+2)(2c-1)2c(2c+1)}\right]^{1/2}\label{6j-2}\,,
\end{align}
where $s\equiv a+b+c$, the general matrix element formula \eqref{q-matrix-element} can be simplified in the following special cases.\\\\
(I) $I=1,J=2,K=3$\\
In this case, the general matrix element formula \eqref{q-matrix-element} reduces to
\begin{align}\label{q-123-matrix-element}
\langle \vec{a}'|\hat{q}_{123}|\vec{a}\rangle
&=-\frac14(-1)^{a_3+j_3+j_1}X(j_1,j_2)^{\frac12}X(j_2,j_3)^{\frac12}\notag\\
&\quad\times\sqrt{(2j_1+1)(2j_1+1)}\sqrt{(2a'_2+1)(2a_2+1)}
\begin{Bmatrix} 0 & j_1 &  j_1 \\
 1 & j_1 & j_1
\end{Bmatrix}\;\begin{Bmatrix}  a_3 & j_3 & a_2 \\
  1 & a'_2 & j_3
\end{Bmatrix}\notag\\
&\quad\times\begin{Bmatrix} j_1 & j_2 &  a_2 \\
 1 & a'_2 & j_2
\end{Bmatrix}\left[(-1)^{j_1+a'_2}\begin{Bmatrix}  a_2 & j_2 & j_1 \\
  1 & j_1 & j_2
\end{Bmatrix}-(-1)^{j_1+a_2}\begin{Bmatrix}  a'_2 & j_2 & j_1 \\
  1 & j_1 & j_2
\end{Bmatrix}
\right]\
\times
\left\{
\begin{array}{cc}
\prod\limits_{t=3}^{n-1}\delta_{a'_t,a_t}, &\text{for}\; n>3 \\
1, &\text{for}\; n=3 \\
\end{array}
\right.\notag\\
&=-\frac12(-1)^{j_1+j_2+j_3+a_2+a'_2+a_3}X(j_2,j_3)^{\frac12}\sqrt{(2a'_2+1)(2a_2+1)}
\begin{Bmatrix} j_1 & j_2 &  a_2 \\
 1 & a'_2 & j_2
\end{Bmatrix}
\begin{Bmatrix}  a_3 & j_3 & a_2 \\
  1 & a'_2 & j_3
\end{Bmatrix}\notag\\
&\quad\times[a'_2(a'_2+1)-a_2(a_2+1)]\times
\left\{
\begin{array}{cc}
\prod\limits_{t=3}^{n-1}\delta_{a'_t,a_t}, &\text{for}\; n>3 \\
1, &\text{for}\; n=3 \\
\end{array}
\right.\,,
\end{align}
where we have used $a_0=0,a_1=a'_1=j_1$ for $I=1$ in the first step, and \eqref{6j-0} and \eqref{6j-1} in the second step.  Moreover, we can further simplify the result \eqref{q-123-matrix-element}, since the triangular conditions on the $6j$-symbols will constrain the values of $a'$ in \eqref{q-123-matrix-element} as
\begin{align}
a'_2=
\begin{cases}
a_2-1\\
a_2\\
a_2+1
\end{cases}\,.
\end{align}
Denoting $|a_2\rangle\equiv|a_2,a_3,\cdots\rangle$ and $|a_2-1\rangle\equiv|a_2-1,a_3,\cdots\rangle$, we get
\begin{align}
\langle a_2-1|\hat{q}_{123}|a_2\rangle&=-\frac12(-1)^{j_1+j_2+j_3+a_2+a_2-1+a_3}X(j_2,j_3)^{\frac12}\sqrt{[2(a_2-1)+1](2a_2+1)}
\begin{Bmatrix} j_1 & j_2 &  a_2 \\
 1 & a_2-1 & j_2
\end{Bmatrix}
\begin{Bmatrix}  a_3 & j_3 & a_2 \\
  1 & a_2-1 & j_3
\end{Bmatrix}\notag\\
&\quad\times[(a_2-1)(a_2-1+1)-a_2(a_2+1)]\notag\\
&=-\frac{(-1)^{2j_1+2j_2+2j_3+4a_2+2a_3}}{\sqrt{(2a_2-1)(2a_2+1)}}\left[(j_1+j_2+a_2+1)(-j_1+j_2+a_2)(j_1-j_2+a_2)(j_1+j_2-a_2+1)\right.\notag\\
&\hspace{4cm}\left.(a_3+j_3+a_2+1)(-a_3+j_3+a_2)(a_3-j_3+a_2)(a_3+j_3-a_2+1)\right]^{1/2}\notag\\
&=-\frac{1}{\sqrt{(2a_2-1)(2a_2+1)}}\left[(j_1+j_2+a_2+1)(-j_1+j_2+a_2)(j_1-j_2+a_2)(j_1+j_2-a_2+1)\right.\notag\\
&\hspace{4cm}\left.(a_3+j_3+a_2+1)(-a_3+j_3+a_2)(a_3-j_3+a_2)(a_3+j_3-a_2+1)\right]^{1/2}\,,
\end{align}
where we have used the fact that $(-1)^{2j_1+2j_2+2a_2}(-1)^{2j_3+2a_3+2a_2}=1$ due to the triangle condition for $(j_1,j_2,a_2)$ and $(j_3,a_3,a_2)$.
\\\\
(II) $I=1,J=2,K=4$\\
In this case, the general matrix element formula \eqref{q-matrix-element} reduces to
\begin{align}
\langle \vec{a}'|\hat{q}_{124}|\vec{a}\rangle&=-\frac14(-1)^{a_4+j_4+j_1}(-1)^{-j_3}X(j_1,j_2)^{\frac12}X(j_2,j_4)^{\frac12}\notag\\
&\quad\times(2j_1+1)\sqrt{(2a'_2+1)(2a_2+1)}
\begin{Bmatrix} 0 & j_1 &  j_1 \\
 1 & j_1 & j_1
\end{Bmatrix}\;\begin{Bmatrix}  a_4 & j_4 & a_3 \\
  1 & a'_3 & j_4
\end{Bmatrix}
\sqrt{(2a'_3+1)(2a_3+1)}(-1)^{a'_2+a_2+1}
\begin{Bmatrix}  j_3 & a'_2 & a'_3\\
  1 & a_3 & a_2
\end{Bmatrix}\notag\\
&\quad\times\begin{Bmatrix} j_1 & j_2 &  a_2 \\
 1 & a'_2 & j_2
\end{Bmatrix}\left[(-1)^{j_1+a'_2}\begin{Bmatrix}  a_2 & j_2 & j_1 \\
  1 & j_1 & j_2
\end{Bmatrix}-(-1)^{j_1+a_2}\begin{Bmatrix}  a'_2 & j_2 & j_1 \\
  1 & j_1 & j_2
\end{Bmatrix}
\right]\times
\left\{
\begin{array}{cc}
\prod\limits_{t=4}^{n-1}\delta_{a'_t,a_t}, &\text{for}\; n>4 \\
1, &\text{for}\; n=4 \\
\end{array}
\right.\notag\\
&=-\frac12(-1)^{j_1-j_3+j_4+a_4}(-1)^{2j_1+1}(-1)^{a'_2+a_2+1}(-1)^{2j_1+j_2+a_2+a'_2+1}
X(j_2,j_4)^{\frac12}\sqrt{(2a'_2+1)(2a_2+1)}\sqrt{(2a'_3+1)(2a_3+1)}\notag\\
&\quad\times\begin{Bmatrix} j_1 & j_2 &  a_2 \\
 1 & a'_2 & j_2
\end{Bmatrix}\begin{Bmatrix}  j_3 & a'_2 & a'_3\\
  1 & a_3 & a_2
\end{Bmatrix}\begin{Bmatrix}  a_4 & j_4 & a_3 \\
  1 & a'_3 & j_4
\end{Bmatrix}
[a'_2(a'_2+1)-a_2(a_2+1)]\times
\left\{
\begin{array}{cc}
\prod\limits_{t=4}^{n-1}\delta_{a'_t,a_t}, &\text{for}\; n>4 \\
1, &\text{for}\; n=4 \\
\end{array}
\right.\notag\\
&=\frac12(-1)^{j_1+j_2-j_3+j_4+a_4} X(j_2,j_4)^{\frac12}\sqrt{(2a'_2+1)(2a_2+1)}\sqrt{(2a'_3+1)(2a_3+1)}
\begin{Bmatrix} j_1 & j_2 &  a_2 \\
 1 & a'_2 & j_2
\end{Bmatrix}\begin{Bmatrix}  j_3 & a'_2 & a'_3\\
  1 & a_3 & a_2
\end{Bmatrix}\begin{Bmatrix}  a_4 & j_4 & a_3 \\
  1 & a'_3 & j_4
\end{Bmatrix}\notag\\
&\quad\times[a'_2(a'_2+1)-a_2(a_2+1)]\times
\left\{
\begin{array}{cc}
\prod\limits_{t=4}^{n-1}\delta_{a'_t,a_t}, &\text{for}\; n>4 \\
1, &\text{for}\; n=4 \\
\end{array}
\right.\,,
\end{align}
where we have used $a_0=0,a_1=a'_1=j_1$ for $I=1$ in the first step, \eqref{6j-0} and \eqref{6j-1} in the second step, and the fact that $a_2+a'_2\in{\mathbb N}\Rightarrow(-1)^{2(a'_2+a_2+1)}=1$ in the last step.
\\\\
(III) $I=1,J=3,K=4$\\
In this case, the general matrix element formula \eqref{q-matrix-element} reduces to
\begin{align}\label{q-134-result-general}
\langle \vec{a}'|\hat{q}_{134}|\vec{a}\rangle&=-\frac14(-1)^{j_1+j_2+j_4+a_4}(-1)^{2j_1+1}X(j_1,j_3)^{\frac12}X(j_3,j_4)^{\frac12}\notag\\
&\quad\times(2j_1+1)\sqrt{(2a'_3+1)(2a_3+1)}
\begin{Bmatrix} 0 & j_1 &  j_1 \\
 1 & j_1 & j_1
\end{Bmatrix}\;\begin{Bmatrix}  a_4 & j_4 & a_3 \\
  1 & a'_3 & j_4
\end{Bmatrix}
\sqrt{(2a'_2+1)(2a_2+1)}
\begin{Bmatrix}  j_2 & j_1 & a'_2 \\
  1 & a_2 & j_1
\end{Bmatrix}\notag\\
&\quad\times\left[(-1)^{a'_2+a'_3}\begin{Bmatrix}  a_3 & j_3 & a_2 \\
  1 & a'_2 & j_3
\end{Bmatrix}\begin{Bmatrix} a'_2 & j_3 &  a_3 \\
 1 & a'_3 & j_3
\end{Bmatrix}-(-1)^{a_2+a_3}\begin{Bmatrix}  a'_3 & j_3 & a_2 \\
  1 & a'_2 & j_3
\end{Bmatrix}\begin{Bmatrix} a_2 & j_3 &  a_3 \\
 1 & a'_3 & j_3
\end{Bmatrix}
\right]\times
\left\{
\begin{array}{cc}
\prod\limits_{t=4}^{n-1}\delta_{a'_t,a_t}, &\text{for}\; n>4 \\
1, &\text{for}\; n=4 \\
\end{array}
\right.
\notag\\
&=-\frac14(-1)^{j_1+j_2+j_4+a_4}X(j_1,j_3)^{\frac12}X(j_3,j_4)^{\frac12}\sqrt{(2a'_2+1)(2a_2+1)}\sqrt{(2a'_3+1)(2a_3+1)}
\begin{Bmatrix}  j_2 & j_1 & a'_2 \\
  1 & a_2 & j_1
\end{Bmatrix}
\begin{Bmatrix}  a_4 & j_4 & a_3 \\
  1 & a'_3 & j_4
\end{Bmatrix}
\notag\\
&\quad\times\left[(-1)^{a'_2+a'_3}\begin{Bmatrix}  a_3 & j_3 & a_2 \\
  1 & a'_2 & j_3
\end{Bmatrix}\begin{Bmatrix} a'_2 & j_3 &  a_3 \\
 1 & a'_3 & j_3
\end{Bmatrix}-(-1)^{a_2+a_3}\begin{Bmatrix}  a'_3 & j_3 & a_2 \\
  1 & a'_2 & j_3
\end{Bmatrix}\begin{Bmatrix} a_2 & j_3 &  a_3 \\
 1 & a'_3 & j_3
\end{Bmatrix}
\right]\times
\left\{
\begin{array}{cc}
\prod\limits_{t=4}^{n-1}\delta_{a'_t,a_t}, &\text{for}\; n>4 \\
1, &\text{for}\; n=4 \\
\end{array}
\right.\,,
\end{align}
where $a_0=0$ and $a_1=a'_1=j_1$ have been used in the first step, and the $6j$-symbol in Eq. \eqref{6j-0} has been used in the second step.\\\\
(IV) $I=2,J=3,K=4$\\
In this case, we have $a_{I-1}=a_1=j_1$. Then the general matrix element formula \eqref{q-matrix-element} reduces to
\begin{align}\label{q-234-result-general}
\langle \vec{a}'|\hat{q}_{234}|\vec{a}\rangle
&=-\frac14(-1)^{j_1+j_2+j_4+a_4}(-1)^{a_2-a'_2}X(j_2,j_3)^{\frac12}X(j_3,j_4)^{\frac12}\notag\\
&\quad\times\sqrt{(2a'_2+1)(2a_2+1)}\sqrt{(2a'_3+1)(2a_3+1)}
\begin{Bmatrix} j_1 & j_2 &  a_2 \\
 1 & a'_2 & j_2
\end{Bmatrix}\;\begin{Bmatrix}  a_4 & j_4 & a_3 \\
  1 & a'_3 & j_4
\end{Bmatrix}\notag\\
&\quad\times\left[(-1)^{a'_2+a'_3}\begin{Bmatrix}  a_3 & j_3 & a_2 \\
  1 & a'_2 & j_3
\end{Bmatrix}\begin{Bmatrix} a'_2 & j_3 &  a_3 \\
 1 & a'_3 & j_3
\end{Bmatrix}-(-1)^{a_2+a_3}\begin{Bmatrix}  a'_3 & j_3 & a_2 \\
  1 & a'_2 & j_3
\end{Bmatrix}\begin{Bmatrix} a_2 & j_3 &  a_3 \\
 1 & a'_3 & j_3
\end{Bmatrix}
\right]\times
\left\{
\begin{array}{cc}
\prod\limits_{t=4}^{n-1}\delta_{a'_t,a_t}, &\text{for}\; n>4 \\
1, &\text{for}\; n=4 \\
\end{array}
\right.\\
&=(-1)^{a_2-a'_2}\frac{X(j_2,j_3)^{\frac12}}{X(j_1,j_3)^{\frac12}}\frac{\begin{Bmatrix} j_1 & j_2 &  a_2 \\
 1 & a'_2 & j_2
\end{Bmatrix}}{\begin{Bmatrix} j_2 & j_1 &  a'_2 \\
 1 & a_2 & j_1
\end{Bmatrix}}\;\langle \vec{a}'|\hat{q}_{134}|\vec{a}\rangle\,.
\end{align}

\section{Summary and discussion}\label{sec-summary}
In the previous sections, the graphical method developed by Yutsis and Brink and their extensions, which suit the requirement of representing the holonomies and the intertwiners, are introduced to LQG. We firstly represent the algebraic formula by its corresponding graphical formula in an unique and unambiguous way. Then the matrix elements of the operator $\hat{q}_{IJK}$, which is the basic building block of the volume operator, are calculated via the simple rules of transforming graphs. Note that the calculations that we did by the graphical method can also be performed by conventional algebraic techniques. Also, to every graphical reduction, there is a corresponding algebraic reduction because of the correspondence between the graphical and algebraic formulae. However, it is obvious that our graphical method is more concise, intuitive and visual.

Note that in our graphical representation, a gauge-invariant intertwiner associated to a vertex $v$ of a standard graph at which $n$ edges with spin $j_1,\cdots, j_n$ incident is represented by
\begin{align}\label{gauge-invariant-intertwiner-graph-1}
\prod_{i=2}^{n-1}\sqrt{2a_i+1}\;
\makeSymbol{\includegraphics[width=6cm]{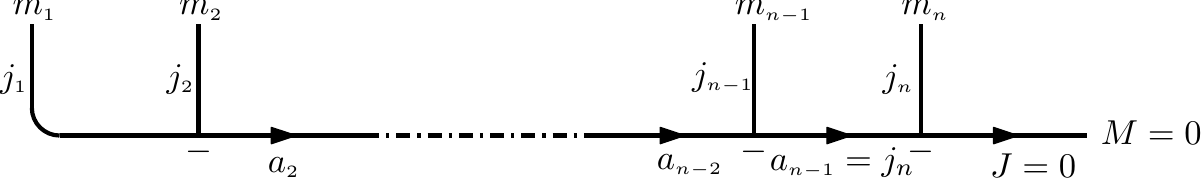}}\,.
\end{align}
Taking account of Eqs. \eqref{arrow-3j} and \eqref{two-arrow-cancel}, the formula \eqref{gauge-invariant-intertwiner-graph-1} is equal to $(-1)^{2j_n}$ times of the formula
\begin{align}\label{gauge-invariant-intertwiner-graph-2}
\prod_{i=2}^{n-2}\sqrt{2a_i+1}\;
\makeSymbol{\includegraphics[width=5cm]{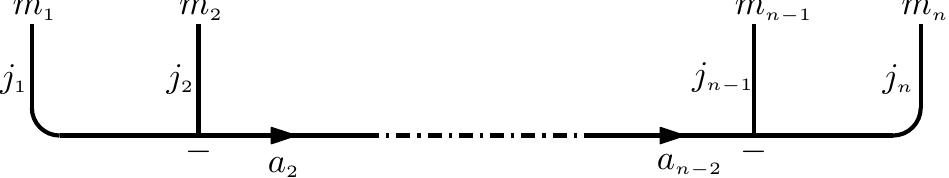}}\,.
\end{align}
Since the only difference between \eqref{gauge-invariant-intertwiner-graph-2} and \eqref{gauge-invariant-intertwiner-graph-1} is a factor $(-1)^{2j_n}$, the formula \eqref{gauge-invariant-intertwiner-graph-2} is also used to represent the gauge-invariant intertwiner in literatures.

Let us compare our calculation with those existing in literature. The operator $\hat{q}_{IJK}$ can be represented by the following three forms \cite{Thiemann:1996au}
\begin{align}\label{3forms}
\hat{q}_{IJK}&:=-4i\epsilon_{ijk}J^i_{e_I}J^j_{e_J}J^k_{e_K}=4\left[J^i_{e_I}J^i_{e_J},J^j_{e_J}J^j_{e_K}\right]=\left[(J^i_{e_I}+J^i_{e_J})^2,(J^j_{e_J}+J^j_{e_K})^2\right]\,.
\end{align}
The first and the third forms (equalities) of the expression \eqref{3forms} were adopted as the starting points respectively in \cite{DePietri:1996pja} and in \cite{Thiemann:1996au,Brunnemann:2004xi}, and their matrix elements are calculated by graphical and algebraic methods respectively. In this paper, we considered the second expression (equality) of $\hat{q}_{IJK}$ and derived its matrix elements by the graphical method introduced in section \ref{section-III}. In \cite{DePietri:1996pja}, to compute the closed formula, Pietri and Rovelli adopted the graphical Penrose binor calulus and tangle-theoretic recoupling theory to deal with recoupling problems in the gauge-invariant context. However the rules of transforming graphs were not shown in \cite{DePietri:1996pja}. So it is not obvious whether the rules uniquely correspond to the algebraic manipulation of formula. Note that the idea in \cite{DePietri:1996pja} to employ the first equality of \eqref{3forms} to calculate the volume operator can also be carried out by our unique and unambiguous rule of graphical calculation. From \eqref{action-inva}, we have
\begin{align}\label{q-epsilon-form}
\hat{q}_{IJK}\cdot \;{\left(i^{\,J;\,\vec{a}}_v\right)_{\,m_1\cdots m_I\cdots m_J\cdots m_K\cdots m_n}}^M&=-4i\epsilon_{ijk}J^i_{e_I}J^j_{e_J}J^k_{e_K}\cdot \;{\left(i^{\,J;\,\vec{a}}_v\right)_{\,m_1\cdots m_I\cdots m_J\cdots m_K\cdots m_n}}^M\notag\\
&={\left(i^{\,J;\,\vec{a}}_v\right)_{\,m_1\cdots m'_I\cdots m'_J\cdots m'_K\cdots m_n}}^M4\epsilon_{ijk}{[\pi_{j_I}(\tau_i)]^{m'_I}}_{\,m_I}{[\pi_{j_J}(\tau_j)]^{m'_J}}_{\,m_J}{[\pi_{j_K}(\tau_k)]^{m'_K}}_{\,m_K}\notag\\
&={\left(i^{\,J;\,\vec{a}}_v\right)_{\,m_1\cdots m'_I\cdots m'_J\cdots m'_K\cdots m_n}}^M(-4i)\epsilon_{\mu\nu\rho}{[\pi_{j_I}(\tau_\mu)]^{m'_I}}_{\,m_I}{[\pi_{j_J}(\tau_\nu)]^{m'_J}}_{\,m_J}{[\pi_{j_K}(\tau_\rho)]^{m'_K}}_{\,m_K}\,,
\end{align}
where, in the last step, we have used the following identity (see Appendix \ref{appendix-B-4} for proof)
\begin{align}\label{q-IJK-munurho}
\epsilon_{ijk}{[\pi_{j_I}(\tau_i)]^{m'_I}}_{\,m_I}{[\pi_{j_J}(\tau_j)]^{m'_J}}_{\,m_J}{[\pi_{j_K}(\tau_k)]^{m'_K}}_{\,m_K}=-i\epsilon_{\mu\nu\rho}{[\pi_{j_I}(\tau_\mu)]^{m'_I}}_{\,m_I}{[\pi_{j_J}(\tau_\nu)]^{m'_J}}_{\,m_J}{[\pi_{j_K}(\tau_\rho)]^{m'_K}}_{\,m_K}\,,
\end{align}
with
\begin{align}\label{two-epsilon}
\epsilon_{\mu\nu\rho}&=\sqrt{6}\begin{pmatrix}
 1  & 1 & 1 \\
 \mu  & \nu & \rho \\
\end{pmatrix}\,.
\end{align}
Note that here one has $\epsilon_{-1\,0\,+1}=1$ (see also Appendix \ref{appendix-B-4} for proof). Notice that both ${[\pi_j(\tau_\mu)]^{m'}}_{\,m}$ and $\epsilon_{\mu\nu\rho}$ (given by a special $3j$-symbol)  in Eq. \eqref{q-epsilon-form} have corresponding graphical representations. The action of $\hat{q}_{IJK}$, corresponding to \eqref{q-epsilon-form}, on an intertwiner is given by
\begin{align}\label{q-epsilon-form-action}
&\hat{q}_{IJK}\cdot \prod_{i=2}^{n-1}\sqrt{2a_i+1}\sqrt{2J+1}\makeSymbol{\includegraphics[width=9.5cm]{graph/volume/volumeoperator-IJK-m-0}}\notag\\
=&-\frac{\sqrt{6}}{2}X(j_I,j_J,j_K)^{\frac12}\prod_{i=2}^{n-1}\sqrt{2a_i+1}\sqrt{2J+1}\makeSymbol{\includegraphics[width=9.5cm]{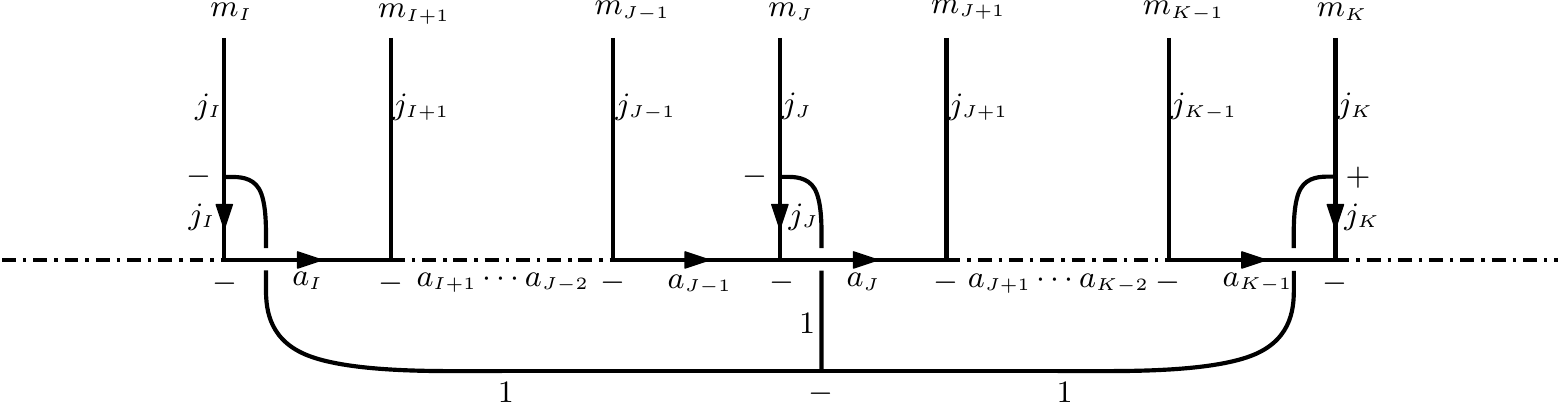}}\,,
\end{align}
where $X(j_I,j_J,j_K)\equiv 2j_I(2j_I+1)(2j_I+2)2j_J(2j_J+1)(2j_J+2)2j_K(2j_K+1)(2j_K+2)$. The derivation of the action of $\hat{q}_{IJK}$ on the intertwiner in the graphical method is to remove the three curves with spin $1$ in \eqref{q-epsilon-form-action} by using the previous rules of transforming graphs. The identities in Eqs. \eqref{id-1}, \eqref{id-2}, \eqref{id-3} and \eqref{id-4} enable us to reduce the graphical formula \eqref{q-epsilon-form-action} as
\begin{align}
\sum_{a',a'',a'''}F(a,a',a'',a''')\makeSymbol{\includegraphics[width=9.5cm]{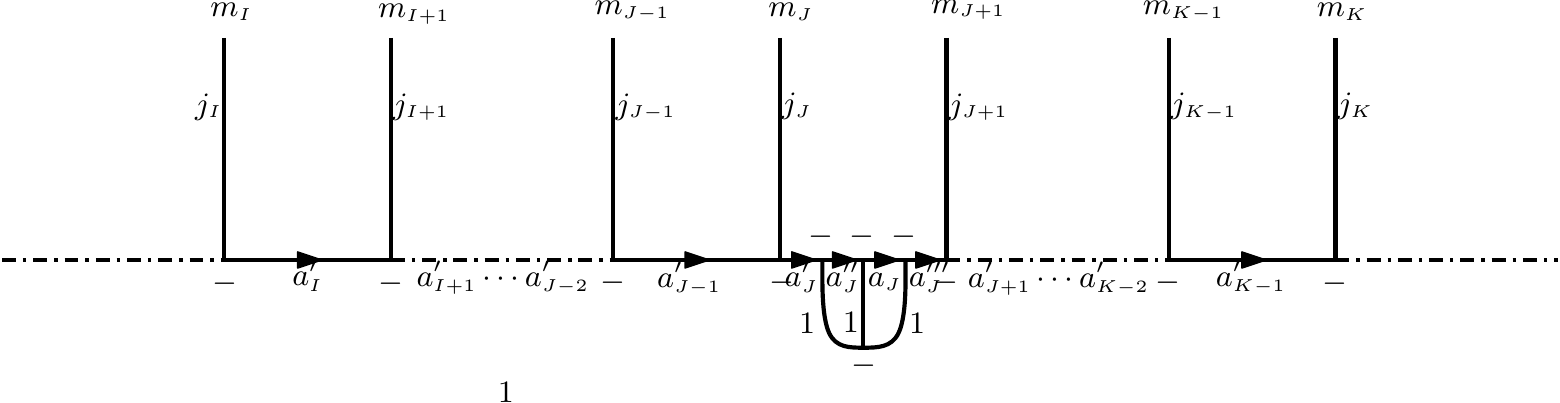}}\,,
\end{align}
where the factor $F(a,a',a'',a''')$ involves the intermediate momenta $a,a',a''$ and $a'''$ in the intertwiner. By the following graphical identity
\begin{align}
\makeSymbol{\includegraphics[width=1.5cm]{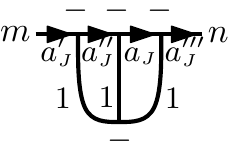}}&=\frac{\delta_{a'_J,a'''_J}}{2a'_J+1}\makeSymbol{\includegraphics[width=1.5cm]{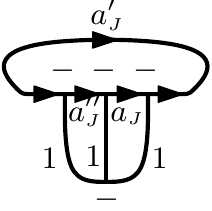}}\makeSymbol{\includegraphics[width=1.5cm]{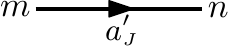}}=\frac{\delta_{a'_J,a'''_J}}{2a'_J+1}\makeSymbol{\includegraphics[width=1.5cm]{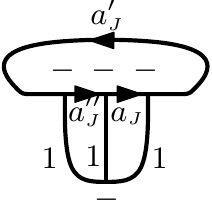}}\makeSymbol{\includegraphics[width=1.5cm]{graph/volume/volumeoperator-IJK-ori-red-identity-3}}=\frac{\delta_{a'_J,a'''_J}}{2a'_J+1}\makeSymbol{\includegraphics[width=1.5cm]{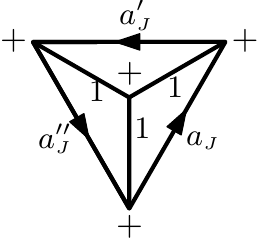}}\makeSymbol{\includegraphics[width=1.5cm]{graph/volume/volumeoperator-IJK-ori-red-identity-3}}\notag\\
&=\frac{\delta_{a'_J,a'''_J}}{2a'_J+1}\begin{Bmatrix}
1 & 1 & 1\\
a_J & a'_J & a''_J
\end{Bmatrix}\makeSymbol{\includegraphics[width=1.5cm]{graph/volume/volumeoperator-IJK-ori-red-identity-3}}\,,
\end{align}
we can remove the three curves with spin $1$ and obtain the final result, which coincides with \eqref{q-IJK-result-general}. The closed formula of the volume operator was also derived by Brunnemann and Thiemann in \cite{Thiemann:1996au,Brunnemann:2004xi} using the algebraic techniques. The derivation process in \cite{Thiemann:1996au,Brunnemann:2004xi} is rigorous but rather abstract and awkward. Our graphical method is convenient and visual, and our result \eqref{q-matrix-element} coincide with the formula derived by the algebraic calculation for the case of $I>1$ and $J>I+1$ in \cite{Brunnemann:2004xi}. Moreover, our analysis shows that the formula \eqref{q-matrix-element} is also valid for other cases and hence can be regarded as a general expression.

In principle, the matrix elements of $\hat{q}_{IJK}$ in Eq. \eqref{q-matrix-element} enable us finally write down the action of the volume operator on the spin network states. We denote
\begin{align}
\hat{q}_v\equiv\frac{i\ell_{\rm p}^6\,\beta^3}{8\times4}\sum_{I<J<K,\,e_I\cap e_J \cap e_K=v}\varsigma(e_I,e_J,e_K)\;\hat{q}_{IJK}\,.
\end{align}
With the matrix elements of $\hat{q}_{IJK}$, we can get the eigenvalues and corresponding eigenstates of $\hat{q}_v$ as
\begin{align}
\hat{q}_v|\lambda_{\hat{q}_v}\rangle=\lambda_{\hat{q}_v}|\lambda_{\hat{q}_v}\rangle\,.
\end{align}
Then we can write down the action of $\hat{V}_v$ on the intertwiner $|i_v\rangle$ associated to $v$ as
\begin{align}
\hat{V}_v\,|i_v\rangle=\sqrt{|\hat{q}_v|}\;|i_v\rangle=\sum_{\lambda_{\hat{q}_v}}\sqrt{|\hat{q}_v|}\;|\lambda_{\hat{q}_v}\rangle\langle\lambda_{\hat{q}_v}|i_v\rangle=\sum_{\lambda_{\hat{q}_v}}\left[\sqrt{|\lambda_{\hat{q}_v}|}\;\langle\lambda_{\hat{q}_v}|i_v\rangle\right]|\lambda_{\hat{q}_v}\rangle\,.
\end{align}
However, when the dimension of the intertwiner space associated to $v$ is bigger than nine, we can not diagonalize $\hat{q}_v$ analytically. This prevents us from explicitly writing down the whole formula for the action of $\hat{V}_v$.

\section*{Acknowledgments}
The authors would like to thank Antonia Zipfel for helpful discussions. J. Y. would also like to thank Chopin Soo and Hoi-Lai Yu for useful discussions. J. Y. is supported in part by NSFC No. 11347006, by the Institute of Physics, Academia Sinica, Taiwan, and by the Natural Science Foundation of Guizhou University (No. 47 in 2013). Y. M. is supported in part by the NSFC (Grant Nos. 11235003 and 11475023) and the Research Fund for the Doctoral Program of Higher Education of China.

\appendix
\renewcommand\thesection{\appendixname~\Alph{section}}
\renewcommand\thesubsection{\Alph{section}.\arabic{subsection}}
\renewcommand\theequation{\Alph{section}.\arabic{equation}}

\section{Elements of graphical representation and calculation}\label{appendix-A}
\subsection{Representations of SU(2), Clebsch-Gordan decomposition, and the intertwiner}\label{appendix-A-1}
To every non-negative integer or half-integer $j$ (i.e., for $j=0,\frac12,1,\frac32,\cdots$), there exists an irreducible representation $\pi_j$ of $SU(2)$, specified by $j$, on a Hilbert space ${\cal H}_j$ with dimension $2j+1$. The orthonormal basis of ${\cal H}_j$ may be denoted by $\{e^{(j)}_m\}$, or $\{|jm\rangle\}$ in Dirac's notation, where $m=-j,-j+1,\cdots,j$. Given two irreducible representations $\pi_{j_1}$ and $\pi_{j_2}$ of $SU(2)$ on ${\cal H}_{j_1}$ and ${\cal H}_{j_2}$, the tensor product representation $\pi_{j_1}\otimes\pi_{j_2}$ on ${\cal H}_{j_1}\otimes{\cal H}_{j_2}$ is $(2j_1+1)(2j_2+1)$-dimensional reducible representation of $SU(2)$.\ Clebsch-Gordan theorem tells us that the representation $\pi_{j_1}\otimes\pi_{j_2}$ can be decomposed into a direct sum of irreducible representations $\pi_J$ on ${\cal H}_J$, where $J\in\{|j_1-j_2|,\cdots,j_1+j_2\}$, formally,
\begin{align}\label{decomp-two}
I_{j_1j_2}\left(\pi_{j_1}(g)\otimes\pi_{j_2}(g)\right)(I_{j_1j_2})^{-1}=\bigoplus_{J=|j_1-j_2|}^{j_1+j_2}\pi_J(g)\,, \qquad \forall g\in SU(2)\,,
\end{align}
where ${I_{j_1j_2}}$ is called the {\em intertwining operator} in representation theory of groups \cite{Fecko}. Given bases $e^{(j_1)}_{m_1}\otimes e^{(j_2)}_{m_2}\in {\cal H}_{j_1}\otimes{\cal H}_{j_2}$ (or $|j_1m_1j_2m_2\rangle\equiv|j_1m_1\rangle\otimes|j_2m_2\rangle\in {\cal H}_{j_1}\otimes{\cal H}_{j_2}$) and $e^{(J)}_M\in {\cal H}_J$ (or $|JM\rangle\in {\cal H}_J$), the components of $I_{j_1j_2}$, as matrix elements, are given by
\begin{align}\label{I-J-j-j}
{\left(I_{j_1j_2}\right)_{m_1m_2}}^{JM}=(e^{(J)}_M,e^{(j_1)}_{m_1}\otimes e^{(j_2)}_{m_2})_{{\cal H}_{j_1}\otimes{\cal H}_{j_2}}=e^M_{(J)}\left(e^{(j_1)}_{m_1}\otimes e^{(j_2)}_{m_2}\right)\,,\quad \text{or}\quad {\left(I_{j_1j_2}\right)_{m_1m_2}}^{JM}=\langle JM|j_1m_1j_2m_2\rangle\,,
\end{align}
where $e^{M}_{(J)}$ is the dual basis or the basis of ${\cal H}_J^*$.
${\left(I_{j_1j_2}\right)_{m_1m_2}}^{JM}$ are called the complex conjugates of the  {\em Clebsch-Gordan coefficients}  (CGCs) or the {\em intertwiners}. In matrix form, the row and column indices of $I_{j_1j_2}$ are denoted by the latter (upper) and former (down) indices $JM$ and $m_1m_2$ respectively. Our convention enable us to regard ${\left(I_{j_1j_2}\right)_{m_1m_2}}^{JM}$ as components of the intertwiner tensor whose indices can be lowered and raised  by a ``metric" which will be introduced by \eqref{metric-tensor}.

The representation $\pi_j$ of $SU(2)$ on ${\cal H}_j$ induces a {\em conjugate} representation $\pi^*_j$ of $SU(2)$ on ${\cal H}_j^*$ via
\begin{align}\label{rep-conj-def}
\left[\pi(g)^*e^m_{(j)}\right]\left(e^{(j)}_n\right):=e^m_{(j)}\left(\pi(g^{-1})e^{(j)}_n\right)\,,
\end{align}
where $\{e^m_{(j)}\}$ is the orthogonal basis of ${\cal H}_j^*$. Furthermore, the irreducible and unitary properties of $\pi_j$ are preserved to its conjugate representation $\pi_j^*$. If the representation $\pi_j$ is unitary, there exists an unitary operator $C^{(j)}: {\cal H}_j\rightarrow {\cal H}_j^*$, such that \cite{Wigner-bk}
\begin{align}\label{representation-conjugate}
C^{(j)}\,\pi_j(g)=\pi_j(g)^*\,C^{(j)}\qquad\Leftrightarrow\qquad C^{(j)}\,\pi(g)\,{C^{(j)}}^{-1}=\pi_j(g)^*\,.
\end{align}
The operator $C^{(j)}$ in fact defines an isomorphism between ${\cal H}_j$ and ${\cal H}^*_j$ by
\begin{align}
e^{(j)}_m&=C^{(j)}_{mm'}\,e^{m'}_{(j)}\,,\qquad \text{or}\quad |jm\rangle=C^{(j)}_{mm'}\,|jm'\rangle^*\label{metric-C}\,,\\
e^m_{(j)}&=C^{mm'}_{(j)}\,e^{(j)}_{m'}\,,\qquad \text{or}\quad |jm\rangle^*=C^{mm'}_{(j)}\,|jm'\rangle\,,\label{metric-inverse-C}
\end{align}
where $C^{mn}_{(j)}\equiv{(C^{(j)}}^{-1})^{mn}$. The operator $C^{(j)}$ and its inverse ${C^{(j)}}^{-1}$ play an important role also in quantum field theories, whose components in the bases of ${\cal H}_j$ and ${\cal H}^*_j$  are given by \cite{Wigner-bk}
\begin{align}\label{metric-tensor}
C^{(j)}_{mn}&:=(-1)^{j-n}\delta_{n,-m}=(-1)^{j+m}\delta_{m,-n}\,,\\
C^{mn}_{(j)}&\equiv{(C^{(j)}}^{-1})^{mn}:=(-1)^{j-m}\delta_{m,-n}=(-1)^{j+n}\delta_{n,-m}\,,\label{metric-tensor-inverse}
\end{align}
satisfying
\begin{align}
C^{(j)}_{mn}C^{nm'}_{(j)}&=C^{m'n}_{(j)}C^{(j)}_{nm}=\delta^{m'}_m\label{metric-tensor-delta}\,,\\
C^{(j)}_{nm}C^{nm'}_{(j)}&=C^{m'n}_{(j)}C^{(j)}_{mn}=(-1)^{2j}\delta^{m'}_m\label{metric-tensor-delta-ex}\,.
\end{align}
Obviously, $C^{(j)}_{mn}$ and $C^{mn}_{(j)}$ satisfy
\begin{align}\label{metric-tensor-exchange}
C^{(j)}_{m'm}=(-1)^{2j}C^{(j)}_{mm'}\,,\quad C^{mm'}_{(j)}=(-1)^{2j}C^{m'm}_{(j)}\,,
\end{align}
which implies that $C^{(j)}_{mn}$ and $C^{mn}_{(j)}$ are symmetric for integer $j$, and anti-symmetric for half-odd integer $j$. The operator $C^{(j)}$ and its inverse ${C^{(j)}}^{-1}$ can be used to lower and raise indices of the tensors on ${\cal H}_j$.
Hence $C^{(j)}_{mn}$ behaves like a {\em metric} tensor. Eq.\eqref{representation-conjugate} can be written in the form of its components as
\begin{align}\label{rep-to-conjugate}
C^{(j)}_{mm'}\,{[\pi_j(g)]^{m'}}_{n'}C^{n'n}_{(j)}={[\pi_j(g)^*]_m}^n\,.
\end{align}
The fact that the representation $\pi_j$ is unitary implies
\begin{align}
\pi_j(g)^\dag=\pi_j(g)^{-1}=\pi_j(g^{-1})\,.
\end{align}
Hence
\begin{align}\label{rep-inverse}
{[\pi_j(g^{-1})]^n}_{\,m}={[\pi_j(g)^{-1}]^n}_{\,m}={[\pi_j(g)^\dag]^n}_{\,m}=\overline{[{\pi_j(g)]^m}_{\,n}}={[\pi_j(g)^*]_m}^{\,n}=C^{(j)}_{mm'}\,{[\pi_j(g)]^{m'}}_{n'}C^{n'n}_{(j)}\,,
\end{align}
where the overline denotes complex conjugation, and we have used \eqref{rep-conj-def} and \eqref{rep-to-conjugate} in the last two steps. By the map $C^{(j)}$ we can define a natural inner product on ${\cal H}^*_j$ as
\begin{align}
(C^{(j)}f,C^{(j)}g)_{{\cal H}^*_j}:=(g,f)_{{\cal H}_j}\,, \qquad \forall f,g\in {\cal H}_j\,.
\end{align}
Then the base transformation $I_{j_1,j_2}^J$ in ${\cal H}_{j_1}\otimes{\cal H}_{j_2}$ induces the base transformation $I^*_{j_1,j_2}$ in ${\cal H}^*_{j_1}\otimes{\cal H}^*_{j_2}$ by
\begin{align}\label{two-spins-conjugate-intertw}
{\left(I^*_{j_1,j_2}\right)_{JM}}^{m_1m_2}=\overline{{\left(I_{j_1j_2}\right)_{m_1m_2}}^{JM}}=\sum_{m'_1,m'_2}C^{m_1m'_1}_{(j_1)}C^{m_2m'_2}_{(j_2)}{\left(I_{j_1j_2}\right)_{m'_1m'_2}}^{JM'}C^{(J)}_{M'M}\,,
\end{align}
For given $j_1,j_2$ and $J$, we denote ${\left(I^{\,J}_{j_1j_2}\right)_{m_1m_2}}^M\equiv{\left(I_{j_1j_2}\right)_{m_1m_2}}^{JM}$
which projects the bases $e^{(j_1)}_{m_1}\otimes e^{(j_2)}_{m_2}$ of ${\cal H}_{j_1}\otimes{\cal H}_{j_2}$ onto $e^{(J)}_{M}$ of ${\cal H}_J\subset{\cal H}_{j_1}\otimes{\cal H}_{j_2}$.
 The corresponding matrix elements of representations $\pi_{j_1}\otimes\pi_{j_2}$ and $\pi_J$ in the two bases, respectively, are related to each other by the so-called Clebsch-Gordan series
\begin{align}
\sum_{m_1,m_2,n_1,n_2}{\left(I_{j_1j_2}^J\right)_{m_1m_2}}^M{[\pi_{j_1}(g)]^{m_1}}_{\,n_1}{[\pi_{j_2}(g)]^{m_2}}_{\,n_2}{\left((I_{j_1j_2}^J)^{-1}\right)_N}^{n_1n_2}={[\pi_J(g)]^M}_{\,N}\,.
\end{align}

Now let us consider the decomposition of the tensor product  $\pi_{j_1}\otimes\cdots\otimes\pi_{j_n}$ of irreducible representations of $SU(2)$ for $n>2$.  The composition involves $n-1$ decompositions of the tensor products of two representations as \eqref{decomp-two} and the choice of the decomposition schemes. Denote $a_i$ ($i=2,\cdots,n$) the irreducible representations that appeared in the $i$-th decomposition for a given scheme. In what follows, we consider the standard scheme where we firstly decompose $\pi_{j_1}\otimes \pi_{j_2}$ into $\bigoplus \pi_{a_2}$, and then decompose $\pi_{a_2}\otimes \pi_{j_3}$ into $\bigoplus\pi_{a_3}$, and so on. We also denote $\vec{a}\equiv(a_2,\cdots a_{n-1})$. For given $j_1\cdots,j_n$, allowable $J$, and compatible vector $\vec{a}\equiv(a_2,\cdots a_{n-1})$, the corresponding Clebsch-Gordan series read
\begin{align}\label{I-C-G-seriers}
\sum_{m_1,\cdots, m_n,n_1,\cdots, n_n}{\left(I^{\,J;\,\vec{a}}_{j_1\cdots j_n}\right)_{m_1\cdots m_n}}^M{[\pi_{j_1}(g)]^{m_1}}_{\,n_1}\cdots{[\pi_{j_n}(g)]^{m_n}}_{n_n}{\left((I^{\,J;\,\vec{a}}_{j_1\cdots j_n})^{-1}\right)_N}^{n_1\cdots n_n}&={[\pi_J(g)]^M}_{\,N}\,,
\end{align}
where
\begin{align}
{\left(I^{\,J;\,\vec{a}}_{j_1\cdots j_n}\right)_{m_1\cdots m_n}}^M\equiv{\left(I^{\,\vec{a}}_{j_1\cdots j_n}\right)_{m_1\cdots m_n}}^{JM}\equiv\sum_{k_2,\cdots, k_{n-1}}{\left(I_{j_1j_2}\right)_{m_1m_2}}^{a_2k_2}\cdots{\left(I_{a_ij_{i+1}}\right)_{k_im_{i+1}}}^{a_{i+1}k_{i+1}}\cdots{\left(I_{a_{n-1}j_n}\right)_{k_{n-1}m_{n}}}^{JM}\,
\end{align}
are the general CGCs, the intertwiners, and often rewritten in quantum mechanics in the form
\begin{align}
\langle JM;\vec{a}|j_1m_1j_2m_2\cdots j_nm_n\rangle&=\sum_{k_2,\cdots,k_{n-1}}\langle a_2k_2|j_1m_1j_2m_2\rangle\langle a_3k_3|a_2k_2j_3m_3\rangle\cdots\langle JM|a_{n-1}k_{n-1}j_nm_n\rangle\,.
\end{align}
It is easy to generalize the above results to the decomposition of the tensor product $\pi_{j_1}\otimes\cdots\otimes\pi_{j_k}(g)\otimes\pi_{j_{k+1}}(g)^{-1}\otimes\cdots\otimes\pi_{j_n}(g)^{-1}$ of $k$ repsentations and $n-k$ inverse of representations into a direct sum of irreducible representations $\pi_J$ on ${\cal H}_J$. The corresponding Clebsch-Gordan series read
\begin{align}\label{I-general-C-G-seriers}
\sum_{m_1,\cdots, m_n,n_1,\cdots, n_n}{\left(I^{\,J;\,\vec{a}}_{j_1\cdots j_n}\right)_{m_1\cdots m_k}}^{m_{k+1}\cdots m_{n}M}{[\pi_{j_1}(g)]^{m_1}}_{\,n_1}\cdots{[\pi_{j_k}(g)]^{m_k}}_{n_k}{[\pi_{j_{k+1}}(g)^{-1}]^{n_{k+1}}}_{\,m_{k+1}}\cdots{[\pi_{j_n}(g)^{-1}]^{n_n}}_{m_n}{\left((I^{\,J;\,\vec{a}}_{j_1\cdots j_n})^{-1}\right)_{Nn_{k+1}\cdots n_n}}^{n_1\cdots n_k}&={[\pi_J(g)]^M}_{\,N}\,,
\end{align}
where
\begin{align}
{\left(I^{\,J;\,\vec{a}}_{j_1\cdots j_n}\right)_{m_1\cdots m_k}}^{m_{k+1}\cdots m_{n}M}\equiv{\left(I^{\,J;\,\vec{a}}_{j_1\cdots j_n}\right)_{m_1\cdots m_kn_{k+1}\cdots n_n}}^M\,C_{(j_{k+1})}^{n_{k+1}m_{k+1}}\cdots C_{(j_n)}^{n_nm_n}\,.
\end{align}
Eq. \eqref{I-general-C-G-seriers} can be written as
\begin{align}
&\sum_{m_1,\cdots, m_n}{[\pi_{j_{k+1}}(g)^{-1}]^{n_{k+1}}}_{\,m_{k+1}}\cdots{[\pi_{j_n}(g)^{-1}]^{n_n}}_{m_n}{\left(I^{\,J;\,\vec{a}}_{j_1\cdots j_n}\right)_{m_1\cdots m_k}}^{m_{k+1}\cdots m_{n}M}{[\pi_{j_1}(g)]^{m_1}}_{\,n_1}\cdots{[\pi_{j_k}(g)]^{m_k}}_{n_k}
&=\sum_{N}{[\pi_J(g)]^M}_{\,N}{\left(I^{\,J;\,\vec{a}}_{j_1\cdots j_n}\right)_{n_1\cdots n_k}}^{n_{k+1}\cdots n_{n}N}\,,
\end{align}
which, in the case of $J=0$, reduces to
\begin{align}
&\sum_{m_1,\cdots, m_n}{[\pi_{j_{k+1}}(g)^{-1}]^{n_{k+1}}}_{\,m_{k+1}}\cdots{[\pi_{j_n}(g)^{-1}]^{n_n}}_{m_n}{\left(I^{\,0;\,\vec{a}}_{j_1\cdots j_n}\right)_{m_1\cdots m_k}}^{m_{k+1}\cdots m_{n}0}{[\pi_{j_1}(g)]^{m_1}}_{\,n_1}\cdots{[\pi_{j_k}(g)]^{m_k}}_{n_k}
&={\left(I^{\,0;\,\vec{a}}_{j_1\cdots j_n}\right)_{n_1\cdots n_k}}^{n_{k+1}\cdots n_{n}0}\,.
\end{align}
Hence the tensor ${\left(I^{\,0;\,\vec{a}}_{j_1\cdots j_n}\right)_{m_1\cdots m_k}}^{m_{k+1}\cdots m_{n}0}$ is also called the {\em invariant tensor}. For the convenience of graphical representation, we introduce
\begin{align}\label{intertwiner-def}
{\left(i^{\,J;\,\vec{a}}_{j_1\cdots j_n}\right)_{m_1\cdots m_n}}^M:=(-1)^{j_1-\sum_{i=2}^nj_i-J}{\left(I^{\,J;\,\vec{a}}_{j_1\cdots j_n}\right)_{m_1\cdots m_n}}^M=(-1)^{j_1-\sum_{i=2}^nj_i-J}\langle JM;\vec{a}|j_1m_1j_2m_2\cdots j_nm_n\rangle\,.
\end{align}
Notice that the factor $(-1)^{j_1-\sum_{i=2}^nj_i-J}$ in Eq. \eqref{intertwiner-def} involves only the spins $j_1,\cdots,j_n$ and $J$, not the intermediate momenta $a_1,\cdots,a_{n-1}$. From now on, the inertwiners refer in particular to $i^{\,J;\,\vec{a}}_{j_1\cdots j_n}$. The Clebsch-Gordan series \eqref{I-C-G-seriers} can be written in terms of $i^{\,J;\,\vec{a}}_{j_1\cdots j_n}$ as
\begin{align}
\sum_{m_1,\cdots, m_n,n_1,\cdots, n_n}{\left(i^{\,J;\,\vec{a}}_{j_1\cdots j_n}\right)_{m_1\cdots m_n}}^M{[\pi_{j_1}(g)]^{m_1}}_{\,n_1}\cdots{[\pi_{j_n}(g)]^{m_n}}_{n_n}{\left((i^{\,J;\,\vec{a}}_{j_1\cdots j_n})^{-1}\right)_N}^{\;n_1\cdots n_n}&={[\pi_J(g)]^M}_{\,N}\,,
\end{align}
and its inverse reads
\begin{align}
{[\pi_{j_1}(g)]^{m_1}}_{\,n_1}\cdots{[\pi_{j_n}(g)]^{m_n}}_{n_n}&=\sum_{J,M,N}{\left((i^{\,J;\,\vec{a}}_{j_1\cdots j_n})^{-1}\right)_M}^{\;m_1\cdots m_n}{[\pi_J(g)]^M}_{\,N}{\left(i^{\,J;\,\vec{a}}_{j_1\cdots j_n}\right)_{n_1\cdots n_n}}^N\,.
\end{align}
In the special case of $n=2$, the Clebsch-Gordan series read
\begin{align}\label{reps-couple}
{[\pi_{j_1}(g)]^{m_1}}_{\,n_1}{[\pi_{j_2}(g)]^{m_2}}_{n_2}&=\sum_{J,M,N}{\left((i^{\,J}_{j_1j_2})^{-1}\right)_M}^{\;m_1m_2}{[\pi_J(g)]^M}_{\,N}{\left(i^{\,J}_{j_1j_2}\right)_{n_1n_2}}^{N}\,.
\end{align}
The fact that the operator $i^{\,J;\,\vec{a}}_{j_1,\cdots,j_n}$ is unitary and its matrix elements take real numbers results in
\begin{align}
{\left((i^{\,J;\,\vec{a}}_{j_1\cdots j_n})^{-1}\right)_M}^{m_1\cdots m_n}={\left((i^{\,J;\,\vec{a}}_{j_1\cdots j_n})^\dag\right)_M}^{m_1\cdots m_n}={\left(i^{\,J;\,\vec{a}}_{j_1\cdots j_n}\right)_{m_1\cdots m_n}}^M\,.
\end{align}
Similarly, the relation \eqref{two-spins-conjugate-intertw} can be generalized as
\begin{align}\label{n-spins-conjugate-intertw}
{\left[\left(i^{\,J;\,\vec{a}}_{j_1\cdots j_n}\right)^*\right]_M}^{m_1\cdots m_n}=\overline{{\left(i^{\,J;\,\vec{a}}_{j_1\cdots j_n}\right)_{m_1\cdots m_n}}^M}&=\sum_{m'_1,\cdots,m'_n,M'}C^{m_1m'_1}_{(j_1)}\cdots C^{m_nm'_n}_{(j_n)}{\left(i^{\,J;\,\vec{a}}_{j_1\cdots j_n}\right)_{m'_1\cdots m'_n}}^{M'}C^{(J)}_{M'M}={\left(i^{\,J;\,\vec{a}}_{j_1\cdots j_n}\right)_{m_1\cdots m_n}}^M\,,
\end{align}
which shows that the GCG is real (see Eq. \eqref{relation-GCG-complex-GCG} for proof with graphical method).
Moreover, the Clebsch-Gordan series \eqref{I-general-C-G-seriers} for the general case can also be written in terms of $i^{\,J;\,\vec{a}}_{j_1\cdots j_n}$ as
\begin{align}\label{i-general-C-G-seriers}
\sum_{m_1,\cdots, m_n,n_1,\cdots, n_n}{\left(i^{\,J;\,\vec{a}}_{j_1\cdots j_n}\right)_{m_1\cdots m_k}}^{m_{k+1}\cdots m_{n}M}{[\pi_{j_1}(g)]^{m_1}}_{\,n_1}\cdots{[\pi_{j_k}(g)]^{m_k}}_{n_k}{[\pi_{j_{k+1}}(g)^{-1}]^{n_{k+1}}}_{\,m_{k+1}}\cdots{[\pi_{j_n}(g)^{-1}]^{n_n}}_{m_n}{\left((i^{\,J;\,\vec{a}}_{j_1\cdots j_n})^{-1}\right)_{Nn_{k+1}\cdots n_n}}^{n_1\cdots n_k}&={[\pi_J(g)]^M}_{\,N}\,,
\end{align}
where
\begin{align}\label{i-general-expression}
{\left(i^{\,J;\,\vec{a}}_{j_1\cdots j_n}\right)_{m_1\cdots m_k}}^{m_{k+1}\cdots m_{n}M}\equiv{\left(i^{\,J;\,\vec{a}}_{j_1\cdots j_n}\right)_{m_1\cdots m_kn_{k+1}\cdots n_n}}^M\,C_{(j_{k+1})}^{n_{k+1}m_{k+1}}\cdots C_{(j_n)}^{n_nm_n}\,.
\end{align}

Given $n$ angular momenta $j_1,\cdots,j_n$, the intertwiner space ${\cal H}_{j_1,\cdots,j_n}$ consists of the intertwiners ${\left(i^{\,J;\,\vec{a}}_{j_1\cdots j_n}\right)_{m_1\cdots m_n}}^M$ with the following inner product
\begin{align}\label{intert-space-inner-product}
\langle \vec{a}';J',M'|\vec{a};J,M\rangle_{{\cal H}_{j_1,\cdots,j_n}}&\equiv \sum_{m_1,\cdots,m_n}\left({\left(i^{\,J';\,\vec{a}'}_{j_1\cdots j_n}\right)_{m_1\cdots m_n}}^{M'},{\left(i^{\,J;\,\vec{a}}_{j_1\cdots j_n}\right)_{m_1\cdots m_n}}^M\right)_{{\cal H}_{j_1,\cdots,j_n}}:=\sum_{m_1,\cdots,m_n}{\left((i^{\,J';\,\vec{a}'}_{j_1\cdots j_n})^\dag\right)_{M'}}^{m_1\cdots m_n}{\left(i^{\,J;\,\vec{a}}_{j_1\cdots j_n}\right)_{m_1\cdots m_n}}^M\notag\\
&=\sum_{m_1,\cdots,m_n}\overline{{(i^{\,J';\,\vec{a}'}_{j_1\cdots j_n})_{m_1\cdots m_n}}^{M'}}{\left(i^{\,J;\,\vec{a}}_{j_1\cdots j_n}\right)_{m_1\cdots m_n}}^M\notag\\
&=\delta_{J,J'}\delta_{M,M'}\delta_{\vec{a},\vec{a}'}\,,
\end{align}
where the last step can be arrived by two ways. The first one is to use the fact that the matrix ${\left(i^{\,J;\,\vec{a}}_{j_1\cdots j_n}\right)_{m_1\cdots m_n}}^M$ is unitary. The second one will be shown in Eq. \eqref{orth-intert-graph-proof}.

\subsection{The basic components of the graphical representation and simple rules of transforming graphs}\label{appendix-A-2}
In this subsection, we introduce the basic components of the graphical representation and simple rules of transforming graphs \cite{brink1968angular}. A graphical representation for the matrix elements of irreducible representations of $SU(2)$ is also proposed. A graphical representation is a correspondence between graphical and algebraic formulae. Each term in an algebraic formula is represented by a component of an appropriate graph in a unique and unambiguous way.

The Wigner $3j$-symbol is associated with the coupling of three angular momenta to give zero resultant. The $3j$-symbol has simple symmetric properties and hence is easier to be handled than the CGC. The $3j$-symbol is defined in terms of the CGC by \cite{Wigner,Wigner-bk,Edmonds}
\begin{align}\label{3j-CG}
\begin{pmatrix}
j_1 & j_2 & j_3\\
m_1 & m_2 & m_3
\end{pmatrix}
&=(-1)^{j_1-j_2+j_3}(-1)^{j_3+m_3}(2j_3+1)^{-\frac12}\;\langle
j_1m_1j_2m_2 | j_3-m_3\rangle\,.
\end{align}
The  $3j$-symbol takes non-vanishing value when the parameters of the upper row $(j_1,j_2,j_3)$ satisfy the triangular condition (i.e., $|j_1-j_2|\leqslant j_3\leqslant j_1+j_2$) and when the sum of the parameters of the lower row $(m_1,m_2,m_3)$ is zero. The parameters $j_i$ and $m_i$ are simultaneously integers or half-integers, such that each of the following numbers
\begin{align}
j_i+m_i,\qquad j_i-m_i,\qquad j_1+j_2+j_3,\qquad -j_1+j_2+j_3,\qquad j_1-j_2+j_3,\qquad j_1+j_2-j_3
\end{align}
takes some integer.

The graphical representation of the $3j$-symbol was collected by Yutsis in \cite{Yutsis:1962bk} and slightly modified by Brink in \cite{brink1968angular} for convenience.
The $3j$-symbol is represented by an oriented node with three lines, which stand for three coupling angular momenta $j_1,j_2,j_3$ incident at the node \cite{brink1968angular}.  The orientation of the node is meant the cyclic order of the lines. A clockwise orientation is denoted by a ``$-$" sign and an anti-clockwise orientation by a ``$+$" sign. Rotation of the diagram does not change the cyclic order of lines, and the angles between two lines as well as their lengths at a node have no significance. The $3j$-symbol can be written in the graphical form
\begin{align}\label{3j-def-graph}
\begin{pmatrix}
j_1 & j_2 & j_3\\
m_1 & m_2 & m_3
\end{pmatrix}
&=\makeSymbol{
\includegraphics[width=2cm]{graph/wigner-symbol/wigner-3j-symbol-1}}=\makeSymbol{
\includegraphics[width=2cm]{graph/wigner-symbol/wigner-3j-symbol-2}}\,.
\end{align}
The $3j$-symbol has the following properties. An even permutaion of the columns leaves the numerical value unchanged, while an odd permutation is equivalent to a multiplication by $(-1)^{j_1+j_2+j_3}$, i.e.,
\begin{align}\label{3j-orientation-change-graph}
\begin{pmatrix}
j_1 & j_2 & j_3\\
m_1 & m_2 & m_3
\end{pmatrix}=(-1)^{j_1+j_2+j_3}\begin{pmatrix}
j_1 & j_3 & j_2 \\
m_1 & m_3 & m_2
\end{pmatrix}
\quad \Leftrightarrow \quad
\makeSymbol{
\includegraphics[width=2cm]{graph/wigner-symbol/wigner-3j-symbol-1}}=(-1)^{j_1+j_2+j_3}\makeSymbol{
\includegraphics[width=2cm]{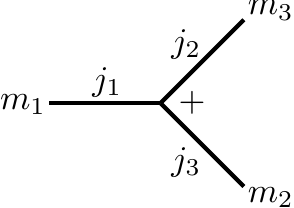}}=(-1)^{j_1+j_2+j_3}\makeSymbol{
\includegraphics[width=2cm]{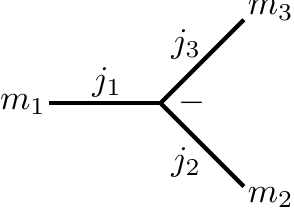}}\,.
\end{align}
Moreover, the $3j$-symbol has the symmetric property
\begin{align}\label{3j-prop}
\begin{pmatrix}
j_1 & j_2 & j_3\\
m_1 & m_2 & m_3
\end{pmatrix}
&=(-1)^{j_1+j_2+j_3}\begin{pmatrix}
j_1 & j_2 & j_3\\
-m_1 & -m_2 & -m_3
\end{pmatrix}\,.
\end{align}
The orthogonality relation for $3j$-symbol is expressed as
\begin{align}\label{3j-orth}
\sum_{m_1,m_2}\begin{pmatrix}
j_1 & j_2 & j_3\\
m_1 & m_2 & m_3
\end{pmatrix}\begin{pmatrix}
j_1 & j_2 & j_3'\\
m_1 & m_2 & m_3'
\end{pmatrix}=\frac{\delta_{j_3,j_3'}}{2j_3+1}\delta_{m_3,m_3'}\,.
\end{align}
Furthermore, the $3j$-symbol is normalized as
\begin{align}\label{3j-norm}
\sum_{m_1,m_2,m_3}\begin{pmatrix}
j_1 & j_2 & j_3\\
m_1 & m_2 & m_3
\end{pmatrix}\begin{pmatrix}
j_1 & j_2 & j_3\\
m_1 & m_2 & m_3
\end{pmatrix}=1\,.
\end{align}
The ``metric'' tensor $C^{(j)}_{m'm}$ in Eq. \eqref{metric-tensor}, which occurs in the contraction of two $3j$-symbols with the same $j$ values, is denoted by a line with an arrow on it as
\begin{align}
C^{(j)}_{m'm}=(-1)^{j-m}\delta_{m,-m'}=(-1)^{j+m'}\delta_{m,-m'}=\makeSymbol{
\includegraphics[width=2cm]{graph/wigner-symbol/wigner-3j-symbol-4}}\,,
\end{align}
and its inverse in Eq. \eqref{metric-tensor-inverse} can be expressed as
\begin{align}
C^{mm'}_{(j)}=(-1)^{j-m}\delta_{m,-m'}=(-1)^{j+m'}\delta_{m,-m'}=\makeSymbol{
\includegraphics[width=2cm]{graph/wigner-symbol/wigner-3j-symbol-4}}\,.
\end{align}
The special $3j$-symbol with one zero-valued angular momentum is related to the ``metric'' tensor by
\begin{align}
\begin{pmatrix}
j & j' & 0\\
m & m' & 0
\end{pmatrix}
&=\frac{\delta_{j,j'}}{\sqrt{2j+1}}C^{(j)}_{m'm}
\qquad \Leftrightarrow \qquad\makeSymbol{
\includegraphics[width=2cm]{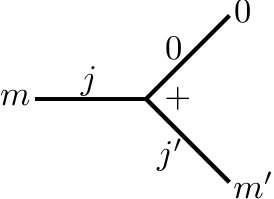}}=\frac{\delta_{j,j'}}{\sqrt{2j+1}}\;\makeSymbol{
\includegraphics[width=2cm]{graph/wigner-symbol/wigner-3j-symbol-4}}\label{arrow-3j}\,.
\end{align}
A line with no arrow represents the expression
\begin{align}
\delta_{m,m'}=\makeSymbol{
\includegraphics[width=2cm]{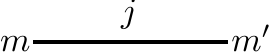}}\,.
\end{align}
The property in Eq. \eqref{3j-prop} means
\begin{align}\label{3j-prop-1}
&\begin{pmatrix}
j_1 & j_2 & j_3\\
m_1 & m_2 & m_3
\end{pmatrix}
=(-1)^{j_1+j_2+j_3}\begin{pmatrix}
j_1 & j_2 & j_3\\
-m_1 & -m_2 & -m_3
\end{pmatrix}=(-1)^{j_1+j_2+j_3}(-1)^{m_1+m_2+m_3}\begin{pmatrix}
j_1 & j_2 & j_3\\
-m_1 & -m_2 & -m_3
\end{pmatrix}\notag\\
&=\sum_{m'_1,m'_2,m'_3}(-1)^{j_1+m_1}\delta_{m_1,-m'_1}(-1)^{j_2+m_2}\delta_{m_2,-m'_2}(-1)^{j_3+m_3}\delta_{m_3,-m'_3}\begin{pmatrix}
j_1 & j_2 & j_3\\
m'_1 & m'_2 & m'_3
\end{pmatrix}=\sum_{m'_1,m'_2,m'_3}\begin{pmatrix}
j_1 & j_2 & j_3\\
m'_1 & m'_2 & m'_3
\end{pmatrix}C^{m'_1m_1}_{(j_1)}C^{m'_2m_2}_{(j_2)}C^{m'_3m_3}_{(j_3)}\notag\\
&=\sum_{m'_1,m'_2,m'_3}(-1)^{j_1-m_1}\delta_{m_1,-m'_1}(-1)^{j_2-m_2}\delta_{m_2,-m'_2}(-1)^{j_3-m_3}\delta_{m_3,-m'_3}\begin{pmatrix}
j_1 & j_2 & j_3\\
m'_1 & m'_2 & m'_3
\end{pmatrix}=\sum_{m'_1,m'_2,m'_3}C^{m_1m'_1}_{(j_1)}C^{m_2m'_2}_{(j_2)}C^{m_3m'_3}_{(j_3)}
\begin{pmatrix}
j_1 & j_2 & j_3\\
m'_1 & m'_2 & m'_3
\end{pmatrix}\,.
\end{align}
In graphical representation, two lines representing the same angular momentum can be joined. Summation over a magnetic quantum number $m$ is graphically represented by joining the free ends of the corresponding lines. Eq. \eqref{3j-CG} implies that the CGC can be represented graphically by
\begin{align}
\langle j_1m_1j_2m_2|j_3m_3\rangle
&=(-1)^{j_1-j_2-j_3}\sqrt{2j_3+1}\makeSymbol{
\includegraphics[width=2.4cm]{graph/wigner-symbol/wigner-3j-symbol-17}}=(-1)^{j_1-j_2-j_3}(2j_3+1)^{1/2}\makeSymbol{
\includegraphics[width=2.6cm]{graph/wigner-symbol/GCG-intertwiner-0}}\,.
\end{align}
Therefore, the graphical representation of the intertwiner defined in Eq. \eqref{intertwiner-def} is
\begin{align}\label{graph-intertwiner}
{\left(i^{\,J;\,\vec{a}}_{j_1\cdots j_n}\right)_{m_1\cdots m_n}}^M:=\prod_{i=2}^{n-1}\sqrt{2a_i+1}\sqrt{2J+1}\;
\makeSymbol{\includegraphics[width=6.2cm]{graph/wigner-symbol/GCG-intertwiner-1}}\,,
\end{align}
and the generalized intertwiner in Eq. \eqref{i-general-expression} can be repressed by
\begin{align}\label{i-general-expression-graph}
{\left(i^{\,J;\,\vec{a}}_{j_1\cdots j_n}\right)_{m_1\cdots m_k}}^{m_{k+1}\cdots m_{n}M}:=\prod_{i=2}^{n-1}\sqrt{2a_i+1}\sqrt{2J+1}\;
\makeSymbol{\includegraphics[width=6.2cm]{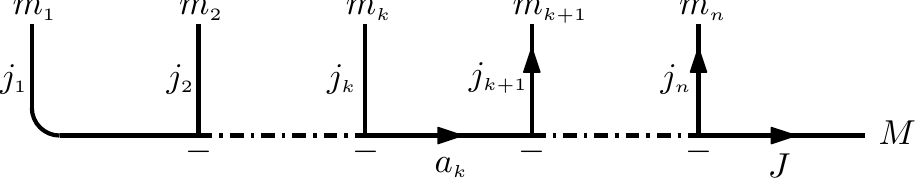}}\,.
\end{align}

As an ingredient of the flux operator, the spherical tensor ${[\pi_j(\tau_\mu)]^{m'}}_m$ in Eq. \eqref{spher-rep} can be represented graphically by
\begin{align}\label{spher-rep-graph-app}
{[\pi_j(\tau_\mu)]^{m'}}_m
&=\frac{i}{2}\sqrt{2j(2j+1)(2j+2)}\makeSymbol{
\includegraphics[width=1.5cm]{graph/wigner-symbol/wigner-3j-symbol-sigma-plus}}=\frac{i}{2}\sqrt{2j(2j+1)(2j+2)}\makeSymbol{
\includegraphics[width=1.5cm]{graph/wigner-symbol/wigner-3j-symbol-sigma-minus}}\,.
\end{align}

Now we outline some useful rules of transforming graphs, which can simplify our calculation of the action of operators on the quantum states in LQG. The frequently used rules for adding or removing arrows in a graph read
\begin{align}
\makeSymbol{
\includegraphics[width=2cm]{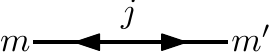}}&=
\makeSymbol{
\includegraphics[width=2cm]{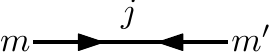}}=\makeSymbol{
\includegraphics[width=2cm]{graph/wigner-symbol/wigner-3j-symbol-9}}\,,\label{two-arrow-cancel}\\
\makeSymbol{
\includegraphics[width=2cm]{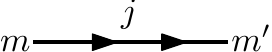}}&=\makeSymbol{
\includegraphics[width=2cm]{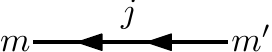}}=(-1)^{2j}\makeSymbol{
\includegraphics[width=2cm]{graph/wigner-symbol/wigner-3j-symbol-9}}\,,\label{two-arrow-result}\\
\makeSymbol{
\includegraphics[width=2cm]{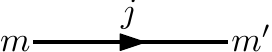}}&=(-1)^{2j}\makeSymbol{
\includegraphics[width=2cm]{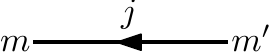}}\,,\label{arrow-flip}\\
\makeSymbol{
\includegraphics[width=2cm]{graph/wigner-symbol/wigner-3j-symbol-1}}&=\makeSymbol{
\includegraphics[width=2cm]{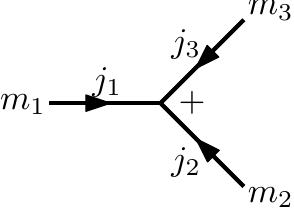}}=\makeSymbol{
\includegraphics[width=2cm]{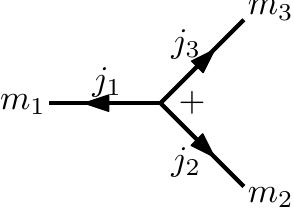}}\,,\label{three-arrow-adding}
\end{align}
which correspond to the algebraic formulae in Eqs. \eqref{metric-tensor-delta}, \eqref{metric-tensor-delta-ex}, \eqref{metric-tensor-exchange} and \eqref{3j-prop-1}. The rule to remove a closed loop in a graph reads
\begin{align}\label{loop-id}
\makeSymbol{
\includegraphics[width=3cm]{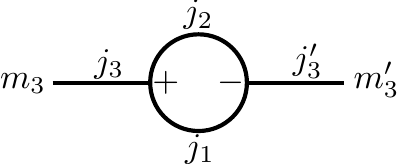}}&=\frac{\delta_{j_3,j_3'}}{2j_3+1}\,
\makeSymbol{
\includegraphics[width=2cm]{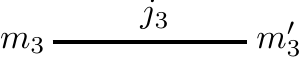}}\,,
\end{align}
which is related to the orthogonality relation for $3j$-symbol in Eq. \eqref{3j-orth}.

Coupling four angular momenta to a zero resultant will involve the $jm$-coefficients. The $jm$-coefficients corresponding to different coupling schemes are related by the $6j$-symbol. The $6j$-symbol is defined by (\cite{Edmonds} p. 94)
\begin{align}\label{6j-symbol-def}
\begin{Bmatrix}
j_1 & j_2 & j_3\\
j_4 & j_5 & j_6
\end{Bmatrix}:=\sum_{\text{all}\, m,m'}\begin{pmatrix}
j_1 & j_2 & j_3\\
m_1 & m_2 & m_3
\end{pmatrix}\begin{pmatrix}
j_1 & j_5 & j_6\\
m'_1 & m_5 & m'_6
\end{pmatrix}\begin{pmatrix}
j_4 & j_2 & j_6\\
m'_4 & m'_2 & m_6
\end{pmatrix}\begin{pmatrix}
j_4 & j_5 & j_3\\
m_4 & m'_5 & m'_3
\end{pmatrix}C^{m'_1m_1}_{(j_1)}C^{m'_2m_2}_{(j_2)}C^{m'_3m_3}_{(j_3)}C^{m'_4m_4}_{(j_4)}C^{m'_5m_5}_{(j_5)}C^{m'_6m_6}_{(j_6)}\,.
\end{align}
The four triangular conditions satisfied by the six angular momenta in the $6j$-symbol may be illustrated in the following way
\begin{align}
\begin{Bmatrix}
\bullet &\bullet   & \bullet\\
 & &
\end{Bmatrix}\qquad
\begin{Bmatrix}\bullet &   & \\
 &\bullet &\bullet
\end{Bmatrix}\qquad
\begin{Bmatrix}&\bullet   & \\
\bullet & & \bullet
\end{Bmatrix}\qquad
\begin{Bmatrix}&   &\bullet\\
\bullet & \bullet &
\end{Bmatrix}\,.
\end{align}
The $6j$-symbol has the following symmetric properties. It is left invariant by any permutation of columns and also by an interchange of the upper and lower arguments in each of any two columns, e.g.,
\begin{align}\label{6j-symmetry-properties}
\begin{Bmatrix}
j_1 & j_2 & j_3\\
j_4 & j_5 & j_6
\end{Bmatrix}&=\begin{Bmatrix}
j_2 & j_1 & j_3\\
j_5 & j_4 & j_6
\end{Bmatrix}=\begin{Bmatrix}
j_3 & j_2 & j_1\\
j_6 & j_5 & j_4
\end{Bmatrix}=\cdots\notag\\
&=\begin{Bmatrix}
j_4 & j_5 & j_3\\
j_1 & j_2 & j_6
\end{Bmatrix}=\begin{Bmatrix}
j_4 & j_2 & j_6\\
j_1 & j_5 & j_3
\end{Bmatrix}=\cdots\,.
\end{align}
Graphically, we can express the $6j$-symbol in Eq. \eqref{6j-symbol-def} as
\begin{align}
\begin{Bmatrix}
j_1 & j_2 & j_3\\
j_4 & j_5 & j_6
\end{Bmatrix}=\begin{Bmatrix}
j_6 & j_2 & j_4\\
j_3 & j_5 & j_1
\end{Bmatrix}=\makeSymbol{
\includegraphics[width=2cm]{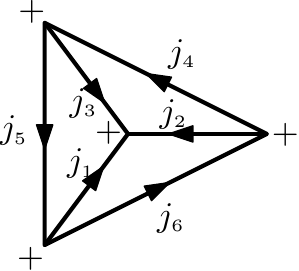}}=\makeSymbol{
\includegraphics[width=2cm]{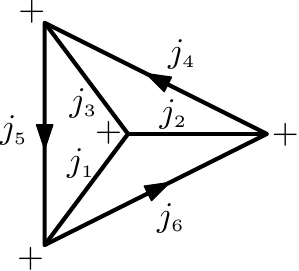}}\,,
\end{align}
where in the last step we have used Eq. \eqref{three-arrow-adding} to remove three arrows.
Taking account of the definition of $6j$-symbol in Eq. \eqref{6j-symbol-def} and the fact that the $3j$-symbol is normalized in Eq. \eqref{3j-norm}, we can easily show the following identity (see \cite{Edmonds} p.95)
\begin{align}
\sum_{m_4,m_5,m_6,m'_4,m'_5,m'_6}\begin{pmatrix}
j_1 & j_5 & j_6\\
m_1 & m_5 & m'_6
\end{pmatrix}\begin{pmatrix}
j_4 & j_2 & j_6\\
m'_4 & m_2 & m_6
\end{pmatrix}\begin{pmatrix}
j_4 & j_5 & j_3\\
m_4 & m'_5 & m_3
\end{pmatrix}C^{m'_4m_4}_{(j_4)}C^{m'_5m_5}_{(j_5)}C^{m'_6m_6}_{(j_6)}=\begin{pmatrix}
j_1 & j_2 & j_3\\
m_1 & m_2 & m_3
\end{pmatrix}\begin{Bmatrix}
j_1 & j_2 & j_3\\
j_4 & j_5 & j_6
\end{Bmatrix}\,.
\end{align}
Hence we have the following graphical identity
\begin{align}
\makeSymbol{
\includegraphics[width=3.5cm]{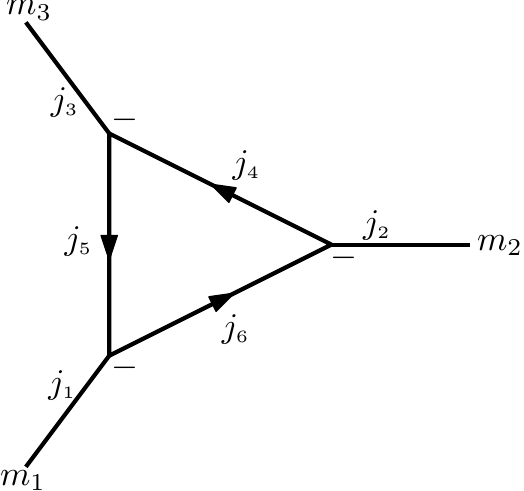}}=\makeSymbol{
\includegraphics[width=2cm]{graph/wigner-symbol/6j-symbol-def-2}}\times\makeSymbol{
\includegraphics[width=2cm]{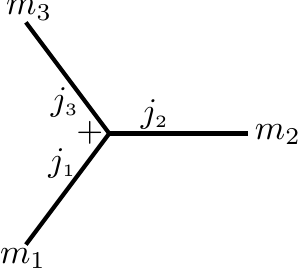}}\,.
\end{align}
The following algebraic relation between two different coupling schemes
\begin{align}
\begin{pmatrix}
j_1 & j_3 & j_5\\
m_1 & m_3 & m_5
\end{pmatrix}C^{m_5m'_5}_{(j_5)}\begin{pmatrix}
j_5 & j_2 & j_4\\
m'_5 & m_2 & m_4
\end{pmatrix}=\sum_{j_6}(2j_6+1)(-1)^{j_2+j_3+j_5+j_6}\begin{Bmatrix}
j_1 & j_2 & j_6\\
j_4 & j_3 & j_5
\end{Bmatrix}\begin{pmatrix}
j_1 & j_2 & j_6\\
m_1 & m_3 & m_6
\end{pmatrix}C^{m_6m'_6}_{(j_6)}\begin{pmatrix}
j_6 & j_3 & j_4\\
m'_6 & m_3 & m_4
\end{pmatrix}
\end{align}
corresponds to the following rule of transforming graphs
\begin{align}\label{6j-interchange}
\makeSymbol{
\includegraphics[width=2cm]{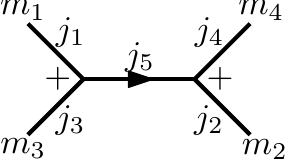}}&=\sum_{j_6}(2j_6+1)(-1)^{j_2+j_3+j_5+j_6}
\begin{Bmatrix}
j_1 & j_2 & j_6\\
j_4 & j_3 & j_5
\end{Bmatrix}
\makeSymbol{
\includegraphics[width=2cm]{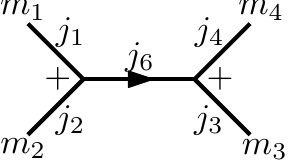}}\,.
\end{align}
Using Eq. \eqref{3j-orientation-change-graph} and the symmetric properties of $6j$-symbol in Eq. \eqref{6j-symmetry-properties}, we can get that from Eq. \eqref{6j-interchange}
\begin{align}\label{6j-interchange-2}
\makeSymbol{
\includegraphics[width=2cm]{graph/wigner-symbol/wigner-3j-symbol-6j-1}}&=\sum_{j_6}(2j_6+1)(-1)^{j_2+j_3-j_5-j_6}
\begin{Bmatrix}
j_1 & j_2 & j_6\\
j_4 & j_3 & j_5
\end{Bmatrix}
\makeSymbol{
\includegraphics[width=2cm]{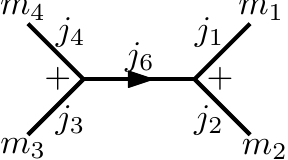}}\,.
\end{align}
Similarly, using Eqs. \eqref{3j-orientation-change-graph} and \eqref{6j-interchange-2}, we have
\begin{align}\label{6j-interchange-3}
\makeSymbol{
\includegraphics[width=2cm]{graph/wigner-symbol/wigner-3j-symbol-6j-1}}&=\sum_{j_6}(2j_6+1)(-1)^{j_1-j_2+j_3+j_4}
\begin{Bmatrix}
j_1 & j_4 & j_6\\
j_2 & j_3 & j_5
\end{Bmatrix}
\makeSymbol{
\includegraphics[width=2cm]{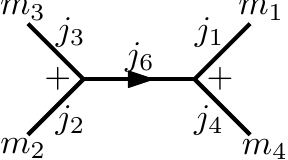}}\,.
\end{align}

Note that Brink's graphical representation in \cite{brink1968angular} does not involve how to represent graphically the matrix element of the representation of $SU(2)$. Here, we will extend the Brink's representation and propose a graphical representation for the unitary irreducible representation $\pi_j$ of $SU(2)$. The matrix element ${[\pi_j(g)]^m}_n$ is denoted by a blue line with a hollow arrow (triangle) in it as
\begin{align}\label{rep-group-graph}
{[\pi_j(g)]^m}_n=:\;\makeSymbol{
\includegraphics[width=2.8cm]{graph/wigner-symbol/wigner-symbol-rep}}\,.
\end{align}
The orientation of the arrow is from its row index $m$ to its column index $n$. The matrix element ${[\pi_j(g^{-1})]^n}_{\,m}$ in Eq. \eqref{rep-inverse} can be presented by
\begin{align}\label{rep-inverse-graph}
{[\pi_j(g^{-1})]^n}_{\,m}=\makeSymbol{
\includegraphics[width=2.8cm]{graph/wigner-symbol/wigner-symbol-rep-inverse-1}}=\makeSymbol{
\includegraphics[width=4cm]{graph/wigner-symbol/wigner-symbol-rep-inverse-2}}\,.
\end{align}
Up to now, we have expressed the quantum states (the spin network states in \eqref{spin-network-state}), and the two elementary operators (the holonomy and flux operators in \eqref{holonomy-operator-def} and \eqref{right-inv-operator-def}
) of LQG in the graphical form. The Clebsch-Gordan series in \eqref{reps-couple} can be represented by
\begin{align}\label{reps-couple-graph}
\makeSymbol{
\includegraphics[width=3cm]{graph/wigner-symbol/wigner-symbol-reps-couple-1}}&=\sum_{j_3}(2j_3+1)\makeSymbol{
\includegraphics[width=5cm]{graph/wigner-symbol/wigner-symbol-reps-couple-2}}
=\sum_{j_3}(2j_3+1)\makeSymbol{
\includegraphics[width=5cm]{graph/wigner-symbol/wigner-symbol-reps-couple-3}}\,,
\end{align}
where, in the second step, we used \eqref{arrow-flip} to flip the orientations of two arrows. Using Eqs. \eqref{rep-inverse-graph} and \eqref{reps-couple-graph}, we have
\begin{align}\label{reps-inverse-couple-graph}
\makeSymbol{
\includegraphics[width=3cm]{graph/wigner-symbol/wigner-symbol-reps-inverse-couple-1}}&=\makeSymbol{
\includegraphics[width=4.1cm]{graph/wigner-symbol/wigner-symbol-reps-inverse-couple-2}}=\sum_{j_3}(2j_3+1)\makeSymbol{
\includegraphics[width=4.8cm]{graph/wigner-symbol/wigner-symbol-reps-inverse-couple-3}}\notag\\
&=\sum_{j_3}(2j_3+1)\makeSymbol{
\includegraphics[width=4.8cm]{graph/wigner-symbol/wigner-symbol-reps-inverse-couple-4}}\,.
\end{align}

\subsection{The spin network states and their orthogonality in graphical method}
In this subsection, we will give a graphical representation for the spin network states and show that these states are orthonormal to each other by the graphical method developed in the previous two subsections. The spin network state consists of a standard graph $\gamma$, the non-trival spins $\vec{j}$ associated to all edges of the graph, and intertwiners $\vec{i}$ to all vertices of the graph \footnote{The relation between the spin network state on the standard graph and the corresponding one on the original graph is discussed in section \ref{subsect-III-2}}.  The normalized spin network state takes the following explicit formula
\begin{align}\label{normlized-spin-network-state}
T^{\rm norm}_{\gamma,\vec{j},\vec{i}}(A):=\bigotimes_{v\in V(\gamma)} i_v\cdot \bigotimes_{e\in E(\gamma)}\;\sqrt{2j_e+1}\,\pi_{j_e}(h_{e}(A))\cdot\bigotimes_{\tilde{v}\in V(\gamma)}i_{\tilde{v}}^*\,,
\end{align}
where $\cdot$ stands for contracting the upper (or former) indices of representation matrices $\pi_{j_e}(h_{e}(A))$ with the indices of intertwines $i_v$ at true vertices $v$, the lower (or later) indices of $\pi_{j_e}(h_{e}(A))$ with indices of the conjugate intertwiners $i_{\tilde{v}}^*$ at pseudo vertices $\tilde{v}$. Comparing to the form \eqref{spin-network-state}, the state \eqref{normlized-spin-network-state} is normalized. The scalar product of the spin network states is defined by the formula
\begin{align}\label{inner-product-def}
\left(T^{\rm norm}_{\gamma,\vec{j},\vec{i}}, T^{\rm norm}_{\gamma',\vec{j}',\vec{i}'}\right)_{{\cal H}_{\rm kin}}:=\int_{SU(2)^{|E(\tilde{\gamma})|}}\prod_{e\in E(\tilde{\gamma})}{\rm d}\mu_H(h_e)\,
\overline{T^{\rm norm}_{\gamma,\vec{j},\vec{i}}(A)}\,T^{\rm norm}_{\gamma',\vec{j}',\vec{i}'}(A)\,,
\end{align}
where $\tilde{\gamma}$ is any graph bigger than $\gamma$ and $\gamma'$, $|E(\tilde{\gamma})|$ denotes the number of the edges in $\tilde{\gamma}$, and ${\rm d}\mu_H(g)$ is the Haar measure on $SU(2)$. If $\gamma$ differs from $\gamma'$, e.g., the edge $e'$ with spin $j'_{e'}$ belongs to $\gamma'$ but does not exist in $\gamma$, then
the orthogonal relation
\begin{align}\label{}
\int_{SU(2)}{\rm d}\mu_H(g)\,\overline{{[\pi_j(g)]^m}_n}{[\pi_{j'}(g)]^{m'}}_{n'}=\frac{\delta_{j,j'}}{2j+1}\delta^{m,m'}\delta_{n,n'}\,,
\end{align}
tells us that the corresponding integration in \eqref{inner-product-def} becomes
\begin{align}
\int_{SU(2)}{\rm d}\mu_H(h_{e'})\,{[\pi_{j'_{e'}}(h_{e'})]^{m'}}_{\,n'}=0\,.
\end{align}
Hence the nontrivial result corresponds to the case $\gamma=\gamma'$. Thus \eqref{inner-product-def} reduces to
\begin{align}
\left(T^{\rm norm}_{\gamma,\vec{j},\vec{i}}, T^{\rm norm}_{\gamma',\vec{j}',\vec{i}'}\right)_{{\cal H}_{\rm kin}}&=\delta_{\gamma,\gamma'}\int_{SU(2)^{|E(\tilde{\gamma})|}}\prod_{e\in E(\tilde{\gamma})}{\rm d}\mu_H(h_e)\,
\overline{T^{\rm norm}_{\gamma,\vec{j},\vec{i}}(A)}\,T^{\rm norm}_{\gamma,\vec{j}',\vec{i}'}(A)\notag\\
&=\delta_{\gamma,\gamma'}\int_{SU(2)^{|E(\gamma)|}}\prod_{e\in E(\gamma)}{\rm d}\mu_H(h_e)\,
\overline{T^{\rm norm}_{\gamma,\vec{j},\vec{i}}(A)}\,T^{\rm norm}_{\gamma,\vec{j}',\vec{i}'}(A)\,,
\end{align}
where, in the second step, we have used the fact that the Haar measure is normalized. By Integrating over all representation functions on the edges, one can obtain the contract of the complex conjugation of intertwiners with the corresponding intertwiners at vertices. Thus we have
\begin{align}\label{inner-product-snf}
\left(T^{\rm norm}_{\gamma,\vec{j},\vec{i}}, T^{\rm norm}_{\gamma',\vec{j}',\vec{i}'}\right)_{{\cal H}_{\rm kin}}
&=\delta_{\gamma,\gamma'}\prod_{e\in E(\gamma)}\delta_{j_e,j'_e}\prod_{v\in V(\gamma)}\sum_{m_1,\cdots,m_n}\overline{{\left(i^{\,J;\,\vec{a}}_{j_1\cdots j_n}\right)_{m_1\cdots m_n}}^M}{\left(i^{\,J';\,\vec{a}'}_{j_1\cdots j_n}\right)_{m_1\cdots m_n}}^{M'}\,,
\end{align}
where $n$ is the number of the edges incident at $v$, and the complex conjugation of intertwiner related to the corresponding intertwiner \eqref{n-spins-conjugate-intertw} satisfies
\begin{align}\label{relation-GCG-complex-GCG}
\overline{{\left(i^{\,J;\,\vec{a}}_{j_1\cdots j_n}\right)_{m_1\cdots m_n}}^M}&=\sum_{m'_1,\cdots,m'_n,M'}C^{m_1m'_1}_{(j_1)}\cdots C^{m_nm'_n}_{(j_n)}{\left(i^{\,J;\,\vec{a}}_{j_1\cdots j_n}\right)_{m'_1\cdots m'_n}}^{M'}C^{(J)}_{M'M}\notag\\
&=\prod_{i=2}^{n-1}\sqrt{2a_i+1}\sqrt{2J+1}\;
\makeSymbol{\includegraphics[width=6.2cm]{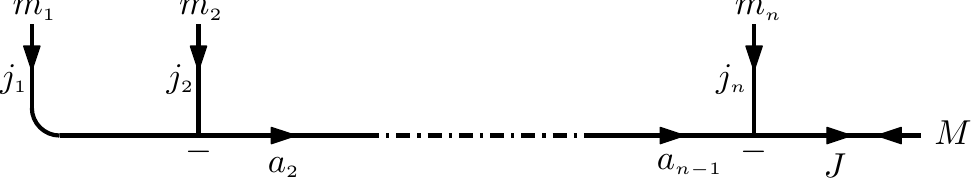}}\notag\\
&=\prod_{i=2}^{n-1}\sqrt{2a_i+1}\sqrt{2J+1}\;
\makeSymbol{\includegraphics[width=6.2cm]{graph/wigner-symbol/GCG-intertwiner-1}}\notag\\
&={\left(i^{\,J;\,\vec{a}}_{j_1\cdots j_n}\right)_{m_1\cdots m_n}}^M\,,
\end{align}
where, in the third step, we have used Eq. \eqref{three-arrow-adding} to remove arrows. Equation \eqref{relation-GCG-complex-GCG} reflects the fact that the CGs (and thus GCGs) are real. The sum in Eq. \eqref{inner-product-snf} can be calculated by graphical method as
\begin{align}\label{orth-intert-graph-proof}
&\sum_{m_1,\cdots,m_n}\overline{{\left(i^{\,J;\,\vec{a}}_{j_1\cdots j_n}\right)_{m_1\cdots m_n}}^M}{\left(i^{\,J';\,\vec{a}'}_{j_1\cdots j_n}\right)_{m_1\cdots m_n}}^{M'}
=\sum_{m_1,\cdots,m_n}{\left(i^{\,J;\,\vec{a}}_{j_1\cdots j_n}\right)_{m_1\cdots m_n}}^M{\left(i^{\,J';\,\vec{a}'}_{j_1\cdots j_n}\right)_{m_1\cdots m_n}}^{M'}\notag\\
=&\prod_{i=2}^{n-1}\sqrt{(2a_i+1)(2a'_i+1)}\sqrt{(2J+1)(2J'+1)}\;\makeSymbol{\includegraphics[width=6.2cm]{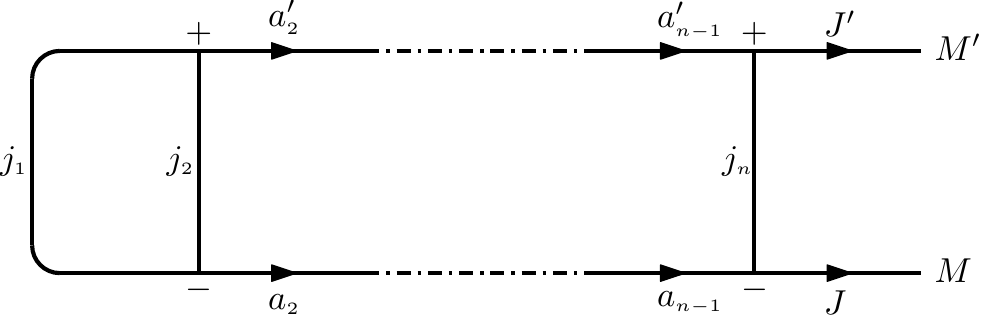}}\notag\\
=&\prod_{i=2}^{n-1}\delta_{a_i,a'_i}\delta_{J,J'}\delta_{M,M'}\notag\\
=&\delta_{\vec{a},\vec{a}'}\delta_{J,J'}\delta_{M,M'}\,,
\end{align}
where, in the third step, we have used Eqs. \eqref{loop-id} and \eqref{two-arrow-cancel}. Hence, the inner product of the spin network functions reads
\begin{align}\label{inner-product-snf-result}
\left(T^{\rm norm}_{\gamma,\vec{j},\vec{i}}, T^{\rm norm}_{\gamma',\vec{j}',\vec{i}'}\right)_{{\cal H}_{\rm kin}}
&=\delta_{\gamma,\gamma'}\prod_{e\in E(\gamma)}\delta_{j_e,j'_e}\prod_{v\in V(\gamma)}\prod_{i=2}^{n-1}\delta_{a_i,a'_i}\delta_{J,J'}\delta_{M,M'}\notag\\
&\equiv\delta_{\gamma,\gamma'}\delta_{\vec{j},\vec{j}'}\delta_{\vec{i},\vec{i}'}\,.
\end{align}
Note that the orthonormal result we obtained in Eq. \eqref{inner-product-snf-result} depends on the premise that the intertwiners at the same vertex $v$ involve the same coupling scheme. If different coupling schemes at the same vertex were chosen, certain additional multiplication of $6j$-symbols would appear in the result. If the spin network functions are gauge invariant, corresponding to $J=0$ and $M=0$, the two Kronecker delta functions become $\delta_{J,J'}=\delta_{M,M'}=1$, which will not appear in the expression \eqref{inner-product-snf-result}.

\section{Proofs for some identities}\label{appendix-B}
\subsection{Proofs of algebraic identities in Eqs. \eqref{spher-rep} and \eqref{two-tau}}\label{appendix-B-1}
By the definition of the matrix element
\begin{align}
{[\pi_j(\tau_i)]^{m'}}_{m}:=\left.\frac{{\rm d}}{{\rm d}t}\right|_{t=0}{[\pi_j(e^{t\tau_i})]^{m'}}_{m}\,,
\end{align}
we get
\begin{align}
{[\pi_j(\tau_1)]^{m'}}_m&=-\frac{i}{2}\sqrt{j(j+1)-m'(m'+1)}\;\delta_{m',m-1}-\frac{i}{2}\sqrt{j(j+1)-m'(m'-1)}\;\delta_{m',m+1}\,,\\
{[\pi_j(\tau_2)]^{m'}}_m&=\frac12\sqrt{j(j+1)-m'(m'+1)}\;\delta_{m',m-1}-\frac12\sqrt{j(j+1)-m'(m'-1)}\;\delta_{m',m+1}\,,\\
{[\pi_j(\tau_3)]^{m'}}_m&=-i\,m'\,\delta_{m',m}\,.
\end{align}
Hence the matrix elements of $\tau_\mu$ ($\mu=0,\pm1$) defined in Eq. \eqref{tau-mu} read
\begin{align}
{[\pi_j(\tau_0)]^{m'}}_m&={[\pi_j(\tau_3)]^{m'}}_m=-i\,m'\,\delta_{m',m}\,,\\
{[\pi_j(\tau_{+1})]^{m'}}_m&=-\frac{1}{\sqrt{2}}\left({[\pi_j(\tau_1)]^{m'}}_m+i\,{[\pi_j(\tau_2)]^{m'}}_m\right)=+i\frac{\sqrt{2}}{2}\sqrt{j(j+1)-m'(m'-1)}\;\delta_{m',m+1}\,,\\
{[\pi_j(\tau_{-1})]^{m'}}_m&=+\frac{1}{\sqrt{2}}\left({[\pi_j(\tau_1)]^{m'}}_m-i\,{[\pi_j(\tau_2)]^{m'}}_m\right)=-i\frac{\sqrt{2}}{2}\sqrt{j(j+1)-m'(m'+1)}\;\delta_{m',m-1}\,.
\end{align}
Taking account of the specialized formulae for the CGCs,
\begin{align}
\langle jm10|jm'\rangle&=\frac{m'}{\sqrt{j(j+1)}}\delta_{m',m}=-\sqrt{2j+1}\,\begin{pmatrix}
j & 1 & j\\
m & 0 & m''
\end{pmatrix}C^{m''m'}_{(j)}\,,\\
\langle jm11|jm'\rangle&=-\sqrt{\frac{j(j+1)-m'(m'-1)}{2j(j+1)}}\delta_{m',m+1}=-\sqrt{2j+1}\,\begin{pmatrix}
j & 1 & j\\
m & 1 & m''
\end{pmatrix}C^{m''m'}_{(j)}\,,\\
\langle jm1-1|jm'\rangle&=+\sqrt{\frac{j(j+1)-m'(m'+1)}{2j(j+1)}}\delta_{m',m-1}=-\sqrt{2j+1}\,\begin{pmatrix}
j & 1 & j\\
m & -1 & m''
\end{pmatrix}C^{m''m'}_{(j)}\,,
\end{align}
we get
\begin{align}
{[\pi_j(\tau_\mu)]^{m'}}_{\,m}&=\frac{i}{2}\sqrt{2j(2j+1)(2j+2)}\,\begin{pmatrix}
  j & 1 & j  \\
  m & \mu & m'' \\
\end{pmatrix}C^{m''m'}_{(j)}=\frac{i}{2}\sqrt{2j(2j+1)(2j+2)}\,\begin{pmatrix}
  1 & j & j \\
  \mu & m'' & m\\
\end{pmatrix}C^{m''m'}_{(j)}\,.
\end{align}

By definition \eqref{tau-mu}, we have
\begin{align}\label{tau-i-tau-mu}
\tau_1=-\frac{1}{\sqrt{2}}(\tau_{+1}-\tau_{-1})\,,\quad \tau_2=\frac{i}{\sqrt{2}}(\tau_{+1}+\tau_{-1})\,,\quad
\tau_3=\tau_0\,.
\end{align}
Then
\begin{align}
\tau_i\tau_i&=\tau_1\tau_1+\tau_2\tau_2+\tau_3\tau_3=-\tau_{+1}\tau_{-1}-\tau_{-1}\tau_{+1}+\tau_0\tau_0\,.
\end{align}
Thus
\begin{align}
{[\pi_{j_I}(\tau_i)]^{m'_I}}_{\,m_I}{[\pi_{j_J}(\tau_i)]^{m'_J}}_{\,m_J}&=-{[\pi_{j_I}(\tau_{+1})]^{m'_I}}_{\,m_I}{[\pi_{j_J}(\tau_{-1})]^{m'_J}}_{\,m_J}-{[\pi_{j_I}(\tau_{-1})]^{m'_I}}_{\,m_I}{[\pi_{j_J}(\tau_{+1})]^{m'_J}}_{\,m_J}+{[\pi_{j_I}(\tau_0)]^{m'_I}}_{\,m_I}{[\pi_{j_J}(\tau_0)]^{m'_J}}_{\,m_J}\notag\\
&=-{[\pi_{j_I}(\tau_{\mu})]^{m'_I}}_{\,m_I}C^{\mu'\mu}_{(1)}\;{[\pi_{j_J}(\tau_{\mu'}
)]^{m'_J}}_{\,m_J}\,.
\end{align}

\subsection{Proofs of graphical identities in Eqs. \eqref{id-1}, \eqref{id-2}, \eqref{id-3} and \eqref{id-4}}\label{appendix-B-2}
Eq. \eqref{id-1} can be proved by
\begin{align}
\makeSymbol{\includegraphics[width=1.5cm]{graph/thm/volumeoperator-I-1}}
&=(-1)^{-(a_I+a_{I-1}+j_I)}\makeSymbol{\includegraphics[width=1.5cm]{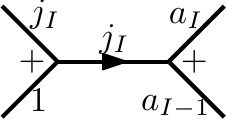}}
=(-1)^{-(a_I+a_{I-1}+j_I)}\sum_{a'_I}(2a'_I+1)(-1)^{a_{I-1}+1+j_I+a'_I}\begin{Bmatrix}  j_I & a_{I-1} & a'_I \\
 a_I & 1 & j_I
\end{Bmatrix}
\makeSymbol{\includegraphics[width=1.5cm]{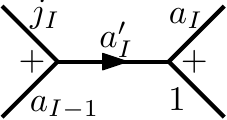}}\notag\\
&=(-1)^{-(a_I+a_{I-1}+j_I)}\sum_{a'_I}(2a'_I+1)(-1)^{a_{I-1}+1+j_I+a'_I}(-1)^{a_I+a'_I+1}(-1)^{j_I+a_{I-1}+a'_I}\begin{Bmatrix}  j_I & a_{I-1} & a'_I \\
 a_I & 1 & j_I
\end{Bmatrix}
\makeSymbol{\includegraphics[width=1.5cm]{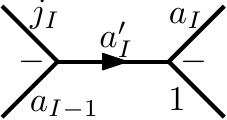}}\notag\\
&=\sum_{a'_I}(2a'_I+1)(-1)^{a_{I-1}-a'_I+j_I}\begin{Bmatrix} a_{I-1} & j_I &  a_I \\
 1 & a'_I & j_I
\end{Bmatrix}\makeSymbol{\includegraphics[width=2cm]{graph/thm/volumeoperator-I-5}}\,,
\end{align}
where Eq. \eqref{3j-orientation-change-graph} was used in the first and third steps, \eqref{6j-interchange} was used in the second step, and we used the fact $(-1)^{-4a'_1}=1$ and the symmetric properties \eqref{6j-symmetry-properties} of the $6j$-symbol in the last step.

Eq. \eqref{id-2} can be shown by
\begin{align}
\makeSymbol{\includegraphics[width=1.5cm]{graph/thm/volumeoperator-K-1}}
&=\makeSymbol{\includegraphics[width=1.5cm]{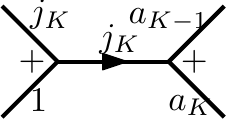}}
=\sum_{b'_{K-1}}(2b'_{K-1}+1)(-1)^{a_K+1+j_K+b'_{K-1}}
\begin{Bmatrix}  j_K & a_K & b'_{K-1} \\
  a_{K-1} & 1 & j_K
\end{Bmatrix}\makeSymbol{\includegraphics[width=1.5cm]{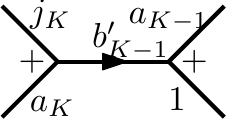}}\notag\\
&=\sum_{b'_{K-1}}(2b'_{K-1}+1)(-1)^{a_K+1+j_K+b'_{K-1}}(-1)^{2b'_{K-1}}
\begin{Bmatrix}  j_K & a_K & b'_{K-1} \\
  a_{K-1} & 1 & j_K
\end{Bmatrix}\makeSymbol{\includegraphics[width=1.5cm]{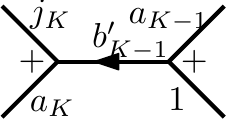}}\notag\\
&=\sum_{b'_{K-1}}(2b'_{K-1}+1)(-1)^{a_K+1+j_K+3b'_{K-1}}
\begin{Bmatrix}  j_K & a_K & b'_{K-1} \\
  a_{K-1} & 1 & j_K
\end{Bmatrix}\makeSymbol{\includegraphics[width=2cm]{graph/thm/volumeoperator-K-5}}\notag\\
&=\sum_{b'_{K-1}}(2b'_{K-1}+1)(-1)^{a_K-b'_{K-1}+j_K+1}
\begin{Bmatrix} a_K & j_K & a_{K-1}  \\
  1 & b'_{K-1} &  j_K
\end{Bmatrix}\makeSymbol{\includegraphics[width=2cm]{graph/thm/volumeoperator-K-5}}\,,
\end{align}
where Eqs. \eqref{6j-interchange} and \eqref{arrow-flip} were used in the second and the third steps, and in the last step we used the fact $(-1)^{-4b'_{K-1}}=1$ and the symmetric properties \eqref{6j-symmetry-properties} of the $6j$-symbol.

Eq. \eqref{id-3} can be proved by
\begin{align}
\makeSymbol{\includegraphics[width=2cm]{graph/thm/volumeoperator-I-n-1}}&=\makeSymbol{\includegraphics[width=1.5cm]{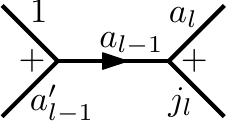}}
=(-1)^{a_{l-1}+a'_{l-1}+1}\makeSymbol{\includegraphics[width=1.5cm]{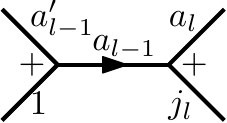}}\notag\\
&=(-1)^{a_{l-1}+a'_{l-1}+1}\sum_{a'_l}(2a'_l+1)(-1)^{j_l+1+a_{l-1}+a'_l}\begin{Bmatrix} a'_{l-1} & j_l & a'_l \\
  a_l & 1 & a_{l-1}
\end{Bmatrix}\makeSymbol{\includegraphics[width=1.5cm]{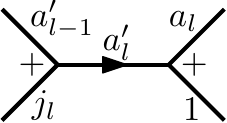}}\notag\\
&=(-1)^{a_{l-1}+a'_{l-1}+1}\sum_{a'_l}(2a'_l+1)(-1)^{j_l+1+a_{l-1}+a'_l}
(-1)^{-(a_l+a'_l+1)}\begin{Bmatrix} a'_{l-1} & j_l & a'_l \\
  a_l & 1 & a_{l-1}
\end{Bmatrix}\makeSymbol{\includegraphics[width=2cm]{graph/thm/volumeoperator-I-n-5}}\notag\\
&=\sum_{a'_l}(2a'_l+1)(-1)^{a'_{l-1}+a_{l-1}+1}(-1)^{a_{l-1}-a_l+j_l}
\begin{Bmatrix}  j_l & a'_{l-1} & a'_l \\
  1 & a_l & a_{l-1}
\end{Bmatrix}\makeSymbol{\includegraphics[width=2cm]{graph/thm/volumeoperator-I-n-5}}\,,
\end{align}
where Eq. \eqref{3j-orientation-change-graph} was used in the second and fourth steps, \eqref{6j-interchange} was used in the third step, and in the last step we used the symmetric properties \eqref{6j-symmetry-properties} of the $6j$-symbol and the exponents were simplified.

Eq. \eqref{id-4} can be proved by
\begin{align}
\makeSymbol{\includegraphics[width=2cm]{graph/thm/volumeoperator-K-n-1}}
&=\makeSymbol{\includegraphics[width=1.5cm]{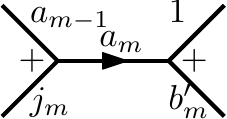}}
=(-1)^{a_m+b'_m+1}\makeSymbol{\includegraphics[width=1.5cm]{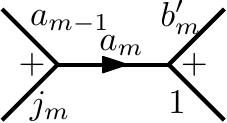}}\notag\\
&=(-1)^{a_m+b'_m+1}\sum_{b'_{m-1}}(2b'_{m-1}+1)(-1)^{1+j_m+a_m+b'_{m-1}}
\begin{Bmatrix}  a_{m-1} & 1 & b'_{m-1}\\
  b'_m & j_m & a_m
\end{Bmatrix}
\makeSymbol{\includegraphics[width=1.5cm]{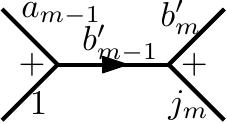}}\notag\\
&=\sum_{b'_{m-1}}(2b'_{m-1}+1)(-1)^{a_m+b'_m+1}(-1)^{1+j_m+a_m+b'_{m-1}}(-1)^{a_{m-1}+b'_{m-1}+1}
\begin{Bmatrix}  a_{m-1} & 1 & b'_{m-1}\\
  b'_m & j_m & a_m
\end{Bmatrix}
\makeSymbol{\includegraphics[width=2cm]{graph/thm/volumeoperator-K-n-5}}\notag\\
&=\sum_{b'_{m-1}}(2b'_{m-1}+1)(-1)^{a_m+b'_m+1}(-1)^{-1-j_m-a_m-b'_{m-1}}(-1)^{a_{m-1}+b'_{m-1}+1}
\begin{Bmatrix}  a_{m-1} & 1 & b'_{m-1}\\
  b'_m & j_m & a_m
\end{Bmatrix}
\makeSymbol{\includegraphics[width=2cm]{graph/thm/volumeoperator-K-n-5}}\notag\\
&=\sum_{b'_{m-1}}(2b'_{m-1}+1)(-1)^{b'_{m-1}+a_{m-1}+1}(-1)^{b'_m-b'_{m-1}-j_m}
\begin{Bmatrix}  j_m & b'_{m-1} & b'_m\\
  1 & a_m & a_{m-1}
\end{Bmatrix}\makeSymbol{\includegraphics[width=2cm]{graph/thm/volumeoperator-K-n-5}}\,,
\end{align}
where Eq. \eqref{3j-orientation-change-graph} was used in the second and fourth steps, \eqref{6j-interchange} was used in the third step, and in the last step we used the symmetric properties \eqref{6j-symmetry-properties} of the $6j$-symbol and the exponents were simplified.

\subsection{Proofs of graphical identities in Eqs. \eqref{id-circle} and \eqref{line-to-3lines}}\label{appendix-B-3}
Eq. \eqref{id-circle} can be proved by
\begin{align}
&\makeSymbol{\includegraphics[width=2.4cm]{graph/thm/volumeoperator-IJK-identity-1}}=(-1)^{a_J+a'_J+1}\,\makeSymbol{\includegraphics[width=2.4cm]{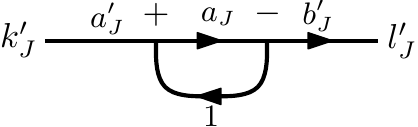}}
=(-1)^{a_J+a'_J+1}\,\makeSymbol{\includegraphics[width=2.4cm]{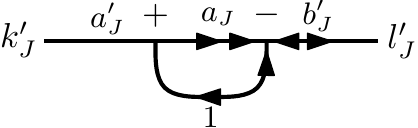}}
=(-1)^{a_J+a'_J+1}\,\makeSymbol{\includegraphics[width=2.4cm]{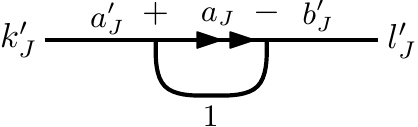}}\notag\\
&=(-1)^{a_J+a'_J+1}(-1)^{2a_J}\,\makeSymbol{\includegraphics[width=2.4cm]{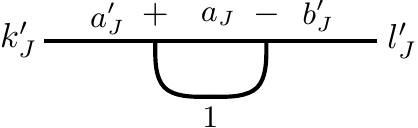}}
=(-1)^{a'_J-a_J+1}\frac{\delta_{a'_J,b'_J}}{2a'_J+1}\,\makeSymbol{\includegraphics[width=1.6cm]{graph/thm/volumeoperator-IJK-identity-6}}\,,
\end{align}
where identities in Eqs. \eqref{3j-orientation-change-graph}, \eqref{three-arrow-adding}, \eqref{two-arrow-cancel}, \eqref{two-arrow-result} and \eqref{loop-id} were used from the first to last steps.

Eq. \eqref{line-to-3lines} can be proved by
\begin{align}
\makeSymbol{\includegraphics[width=1.1cm]{graph/thm/volumeoperator-extending-1}}&=\makeSymbol{\includegraphics[width=1.1cm]{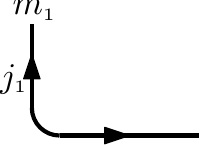}}=\sqrt{2j_1+1}\;\makeSymbol{\includegraphics[width=2cm]{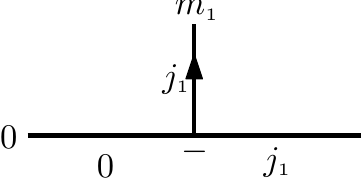}}=\sqrt{2j_1+1}\;\makeSymbol{\includegraphics[width=2cm]{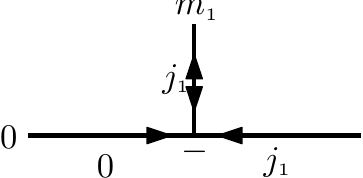}}=\sqrt{2j_1+1}\;\makeSymbol{\includegraphics[width=2cm]{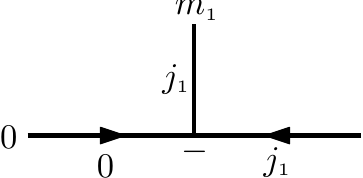}}\notag\\
&=(1)^{2j_1}\sqrt{2j_1+1}\;\makeSymbol{\includegraphics[width=2cm]{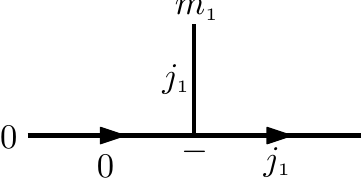}}=(1)^{2j_1}\sqrt{2a_0+1}\sqrt{2a_1+1}\;\;\makeSymbol{\includegraphics[width=2cm]{graph/thm/volumeoperator-extending-7}}\,,
\end{align}
where the rules \eqref{two-arrow-cancel}, \eqref{arrow-3j}, \eqref{three-arrow-adding}, \eqref{two-arrow-cancel} and \eqref{arrow-flip} were used from the first to fifth steps, and in last step we denoted $a_0\equiv 0,a_1\equiv j_1$.

\subsection{Proofs of algebraic identities in Eqs. \eqref{q-IJK-munurho} and \eqref{two-epsilon}}\label{appendix-B-4}
To prove Eq. \eqref{q-IJK-munurho}, we denote ${[\pi_{j_I}(\tau_i)]^{n_I}}_{m_I}{[\pi_{j_J}(\tau_j)]^{
n_J}}_{m_J}{[\pi_{j_K}(\tau_k)]^{n_K}}_{m_K}$ by $[\tau_i\tau_j\tau_k]_{n_Im_In_Jm_Jn_Km_K}$. Then we have
\begin{align}
\epsilon_{ijk}[\tau_i\tau_j\tau_k]_{n_Im_In_Jm_Jn_Km_K}
=&+\epsilon_{123}[\tau_1\tau_2\tau_3]_{n_Im_In_Jm_Jn_Km_K}
+\epsilon_{132}[\tau_1\tau_3\tau_2]_{n_Im_In_Jm_Jn_Km_K}+\epsilon_{213}[\tau_2\tau_1\tau_3]_{n_Im_In_Jm_Jn_Km_K}
+\epsilon_{312}[\tau_3\tau_1\tau_2]_{n_Im_In_Jm_Jn_Km_K}\notag\\
&+\epsilon_{231}[\tau_2\tau_3\tau_1]_{n_Im_In_Jm_Jn_Km_K}
+\epsilon_{321}[\tau_3\tau_2\tau_1]_{n_Im_In_Jm_Jn_Km_K}\notag\\
=&[\tau_1\tau_2-\tau_2\tau_1]_{n_Im_In_Jm_J}[\tau_3]_{n_Km_K}
+[\tau_3]_{n_Im_I}[\tau_1\tau_2-\tau_2\tau_1]_{n_Jm_Jn_Km_K}
-[\tau_1\tau_3\tau_2-\tau_2\tau_3\tau_1]_{n_Im_In_Jm_Jn_Km_K}\,.
\end{align}
Taking account of \eqref{tau-i-tau-mu}, we have
\begin{align}
\tau_1\tau_2-\tau_2\tau_1&=-\frac{1}{\sqrt{2}}\frac{i}{\sqrt{2}}\left[(\tau_{+1}-\tau_{-1})(\tau_{+1}
+\tau_{-1})-(\tau_{+1}
+\tau_{-1})(\tau_{+1}-\tau_{-1})\right]=-i(\tau_{+1}\tau_{-1}-\tau_{-1}\tau_{+1})\,.
\end{align}
Hence, we get
\begin{align}
&\epsilon_{ijk}{[\pi_{j_I}(\tau_i)]^{n_I}}_{m_I}{[\pi_{j_J}(\tau_j)]^{
n_J}}_{m_J}{[\pi_{j_K}(\tau_k)]^{n_K}}_{m_K}\notag\\
=&[\tau_1\tau_2-\tau_2\tau_1]_{n_Im_In_Jm_J}[\tau_3]_{n_Km_K}
+[\tau_3]_{n_Im_I}[\tau_1\tau_2-\tau_2\tau_1]_{n_Jm_Jn_Km_K}
-[\tau_1\tau_3\tau_2-\tau_2\tau_3\tau_1]_{n_Im_In_Jm_Jn_Km_K}\notag\\
=&-i[\tau_{+1}\tau_{-1}-\tau_{-1}\tau_{+1}]_{n_Im_In_Jm_J}[\tau_0]_{n_Km_K}
-i[\tau_0]_{n_Im_I}[\tau_{+1}\tau_{-1}-\tau_{-1}\tau_{+1}]_{n_Jm_Jn_Km_K}
-(-i)[\tau_{+1}\tau_0\tau_{-1}-\tau_{-1}\tau_0\tau_{+1}]_{n_Im_In_Jm_Jn_Km_K}\notag\\
=&-i[\tau_{+1}\tau_{-1}\tau_0-\tau_{-1}\tau_{+1}\tau_0
+\tau_0\tau_{+1}\tau_{-1}-\tau_0\tau_{-1}\tau_{+1}
+\tau_{-1}\tau_0\tau_{+1}-\tau_{+1}\tau_0\tau_{-1}]_{n_Im_In_Jm_Jn_Km_K}\notag\\
=&-i[\epsilon_{+1-1\,0}\tau_{+1}\tau_{-1}\tau_0+\epsilon_{-1+1\,0}\tau_{-1}\tau_{+1}\tau_0
+\epsilon_{0\,+1-1}\tau_0\tau_{+1}\tau_{-1}+\epsilon_{0\,-1+1}\tau_0\tau_{-1}\tau_{+1}
+\epsilon_{-1\,0\,+1}\tau_{-1}\tau_0\tau_{+1}+\epsilon_{+1\,0\,-1}\tau_{+1}\tau_0\tau_{-1}]_{n_Im_In_Jm_Jn_Km_K}\notag\\
=&-i\epsilon_{\mu\nu\rho}[\tau_\mu\tau_\nu\tau_\rho]_{n_Im_In_Jm_Jn_Km_K}\notag\\
=&-i\epsilon_{\mu\nu\rho}
{[\pi_{j_I}(\tau_\mu)]^{n_I}}_{m_I}{[\pi_{j_J}(\tau_\nu)]^{
n_J}}_{m_J}{[\pi_{j_K}(\tau_\rho)]^{n_K}}_{m_K}\,,
\end{align}
where $\epsilon_{\mu\nu\rho}$ is the Levi-Civita symbol defined by $\epsilon_{-1\,0\,+1}=1$.

To show Eq. \eqref{two-epsilon}, notice that
\begin{align}
\begin{pmatrix} 1 & 1&  1 \\
 -1 & 0 & 1
\end{pmatrix}=\frac{1}{\sqrt{6}}\,.
\end{align}
Then one has
\begin{align}
\begin{pmatrix} 1 & 1&  1 \\
 -1 & 0 & 1
\end{pmatrix}=\frac{1}{\sqrt{6}}\epsilon_{-1\,0\,+1}\,.
\end{align}
Recalling the symmetric property of the $3j$-symbol in Appendix \ref{appendix-A-2}
, an even permutation of the columns leaves the numerical value unchanged, while an odd permutation will lead to a factor $(-1)^{1+1+1}=-1$ for the $3j$-symbol in Eq. \eqref{3j-orientation-change-graph}. These symmetries of the $3j$-symbol are the same as those of $\epsilon_{\mu\nu\rho}$. Hence we have $\epsilon_{\mu\nu\rho}=\sqrt{6}\begin{pmatrix}
 1  & 1 & 1 \\
 \mu  & \nu & \rho \\
\end{pmatrix}$.

\section{Complete calculations for the action of the second term in the parenthesis of Eq. \eqref{q-IJK} }\label{appendix-C}
The action of the second term in parenthesis of Eq. \eqref{q-IJK} can be explicitly calculated as
\begin{align}\label{3J-2}
&\quad\hat{q}^{<IJ;JK>}_{IJK}\cdot {\left(i^{\,J;\,\vec{a}}_v\right)_{\,m_1\cdots m_I\cdots m_J\cdots m_K\cdots m_n}}^M\notag\\
&=X(j_I,j_J)^{\frac12}X(j_J,j_K)^{\frac12}\prod_{i=2}^{n-1}\sqrt{2a_i+1}\sqrt{2J+1}\makeSymbol{\includegraphics[width=9.5cm]{graph/volume/volumeoperator-IJK-m-m-1}}\,.
\end{align}
Similar to the calculations in section \ref{section-IV} for the first term in the parenthesis of \eqref{q-IJK}, we perform the calculation of \eqref{3J-2} in four steps. In particular, from the second step we first consider the case of $J>I+1$ and $K>J+1$. In the first and second steps, the calculation are completely similar to those for the first term. Thus we can write down the result directly as
\begin{align}
&\sum_{(a'_I,a'_{I+1}\cdots,a'_{J-1})}\sum_{(b'_{K-1},\cdots,b'_{J+1},b'_J)}(2a'_I+1)(-1)^{a_{I-1}-a'_I+j_I}\begin{Bmatrix} a_{I-1} & j_I &  a_I \\
 1 & a'_I & j_I
\end{Bmatrix}\times(2b'_{K-1}+1)(-1)^{a_K-b'_{K-1}+j_K+1}
\begin{Bmatrix}  a_K & j_K & a_{K-1} \\
  1 & b'_{K-1} & j_K
\end{Bmatrix}\notag\\
&\hspace{3.1cm}\times\prod_{l=I+1}^{J-1}(2a'_l+1)(-1)^{a'_{l-1}+a_{l-1}+1}(-1)^{a_{l-1}-a_l+j_l}
\begin{Bmatrix}  j_l & a'_{l-1} & a'_l \\
  1 & a_l & a_{l-1}
\end{Bmatrix}\notag\\
&\hspace{3.1cm}\times\prod_{m=J+1}^{K-1}(2b'_{m-1}+1)(-1)^{b'_{m-1}+a_{m-1}+1}(-1)^{b'_m-b'_{m-1}-j_m}
\begin{Bmatrix}  j_m & b'_{m-1} & b'_m\\
  1 & a_m & a_{m-1}
\end{Bmatrix}\notag\\
&\quad\times X(j_I,j_J)^{\frac12}X(j_J,j_K)^{\frac12}\prod_{i=2}^{n-1}\sqrt{2a_i+1}\sqrt{2J+1}\makeSymbol{\includegraphics[width=9.5cm]{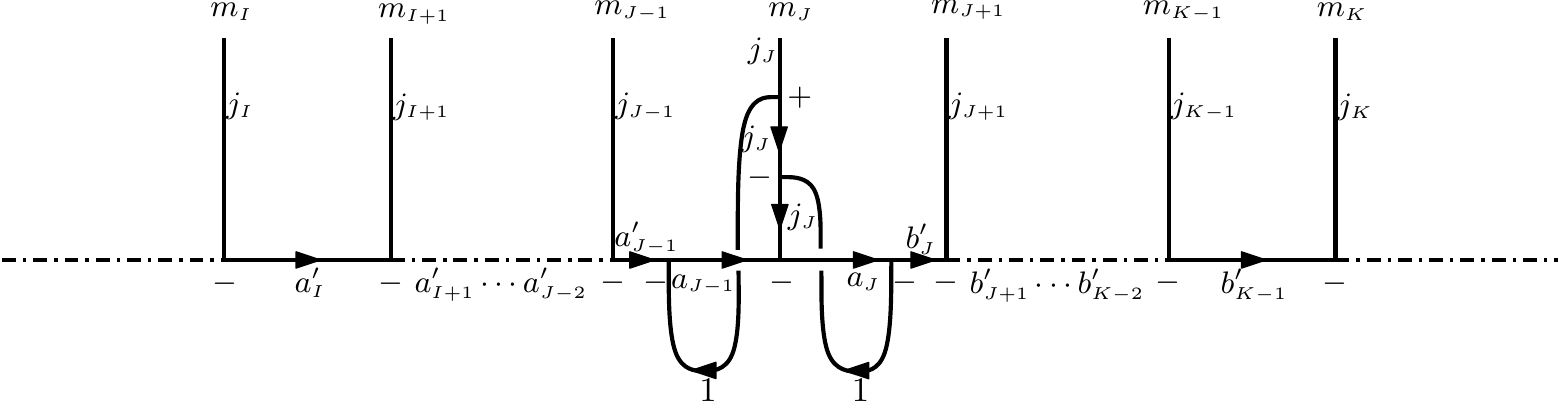}}\,.
\end{align}
In the third step, after dragging two nodes attached to the $j_J$ line down to the horizontal lines by using firstly \eqref{id-1} and then \eqref{id-2}, we have
\begin{align}
&\sum_{(a'_I,a'_{I+1}\cdots,a'_{J-1})}\sum_{(b'_{K-1},\cdots,b'_{J+1},b'_J)}(2a'_I+1)(-1)^{a_{I-1}-a'_I+j_I}\begin{Bmatrix} a_{I-1} & j_I &  a_I \\
 1 & a'_I & j_I
\end{Bmatrix}\times(2b'_{K-1}+1)(-1)^{a_K-b'_{K-1}+j_K+1}
\begin{Bmatrix}  a_K & j_K & a_{K-1} \\
  1 & b'_{K-1} & j_K
\end{Bmatrix}\notag\\
&\hspace{3.1cm}\times(-1)^{a_I-a_{J-1}+\sum_{l=I+1}^{J-1}j_l}\prod_{l=I+1}^{J-1}(2a'_l+1)(-1)^{a'_{l-1}+a_{l-1}+1}
\begin{Bmatrix}  j_l & a'_{l-1} & a'_l \\
  1 & a_l & a_{l-1}
\end{Bmatrix}\notag\\
&\hspace{3.1cm}\times(-1)^{b'_{K-1}-b'_J-\sum_{m=J+1}^{K-1}j_m}\prod_{m=J+1}^{K-1}(2b'_{m-1}+1)(-1)^{b'_{m-1}+a_{m-1}+1}
\begin{Bmatrix}  j_m & b'_{m-1} & b'_m\\
  1 & a_m & a_{m-1}
\end{Bmatrix}\notag\\
&\hspace{3.1cm}\times\sum_{a'_J}(2a'_J+1)(-1)^{a_{J-1}-a'_J+j_J}
\begin{Bmatrix}
a_{J-1} & j_J &  a_J \\
 1 & a'_J & j_J
\end{Bmatrix}
\times\sum_{b'_{J-1}}(2b'_{J-1}+1)(-1)^{a'_J-b'_{J-1}+j_J+1}
\begin{Bmatrix}
  a'_J & j_J & a_{J-1} \\
  1 & b'_{J-1} & j_J
\end{Bmatrix}\notag\\
&\quad\times X(j_I,j_J)^{\frac12}X(j_J,j_K)^{\frac12}\prod_{i=2}^{n-1}\sqrt{2a_i+1}\sqrt{2J+1}\makeSymbol{\includegraphics[width=9.5cm]{graph/volume/volumeoperator-IJK-m-5}}\,.
\end{align}
Now we turn to the forth (last) step. By using identity \eqref{id-circle} and summing over $b'_{J-1}$ and $a'_J$, we obtain
\begin{align}
&\sum_{(a'_I,a'_{I+1}\cdots,a'_{J-1})}\sum_{(b'_{K-1},\cdots,b'_{J+1},b'_J)}(2a'_I+1)(-1)^{a_{I-1}-a'_I+j_I}\begin{Bmatrix} a_{I-1} & j_I &  a_I \\
 1 & a'_I & j_I
\end{Bmatrix}
\times(2b'_{K-1}+1)(-1)^{a_K-b'_{K-1}+j_K+1}
\begin{Bmatrix}  a_K & j_K & a_{K-1} \\
  1 & b'_{K-1} & j_K
\end{Bmatrix}\notag\\
&\hspace{3.1cm}\times(-1)^{a_I-a_{J-1}+\sum_{l=I+1}^{J-1}j_l}\prod_{l=I+1}^{J-1}(2a'_l+1)(-1)^{a'_{l-1}+a_{l-1}+1}
\begin{Bmatrix}  j_l & a'_{l-1} & a'_l \\
  1 & a_l & a_{l-1}
\end{Bmatrix}\notag\\
&\hspace{3.1cm}\times(-1)^{b'_{K-1}-b'_J-\sum_{m=J+1}^{K-1}j_m}\prod_{m=J+1}^{K-1}(2b'_{m-1}+1)(-1)^{b'_{m-1}+a_{m-1}+1}
\begin{Bmatrix}  j_m & b'_{m-1} & b'_m\\
  1 & a_m & a_{m-1}
\end{Bmatrix}\notag\\
&\hspace{3.1cm}\times
\begin{Bmatrix}
a_{J-1} & j_J &  a_J \\
 1 & b'_J & j_J
\end{Bmatrix}(-1)^{a_{J-1}-b'_J+j_J}(-1)^{-b'_J+a_J-1}
\times
\begin{Bmatrix}
  a'_J & j_J & a_{J-1} \\
  1 & a'_{J-1} & j_J
\end{Bmatrix}(-1)^{b'_J-a'_{J-1}+j_J+1}(-1)^{-a'_{J-1}+a_{J-1}-1}\notag\\
&\quad\times X(j_I,j_J)^{\frac12}X(j_J,j_K)^{\frac12}\prod_{i=2}^{n-1}\sqrt{2a_i+1}\sqrt{2J+1}\makeSymbol{\includegraphics[width=9.5cm]{graph/volume/volumeoperator-IJK-m-7}}\,.
\end{align}
Relabeling indices $b'$ by $a'$ yields
\begin{align}\label{general-2}
&\sum_{(a'_I,\cdots,a'_{K-1})}(2a'_I+1)(-1)^{a_{I-1}-a'_I+j_I}\begin{Bmatrix} a_{I-1} & j_I &  a_I \\
 1 & a'_I & j_I
\end{Bmatrix}
\times(2a'_{K-1}+1)(-1)^{a_K-a'_{K-1}+j_K+1}
\begin{Bmatrix}  a_K & j_K & a_{K-1} \\
  1 & a'_{K-1} & j_K
\end{Bmatrix}\notag\\
&\hspace{0.95cm}\times(-1)^{a_I-a_{J-1}+\sum_{l=I+1}^{J-1}j_l}\prod_{l=I+1}^{J-1}(2a'_l+1)(-1)^{a'_{l-1}+a_{l-1}+1}
\begin{Bmatrix}  j_l & a'_{l-1} & a'_l \\
  1 & a_l & a_{l-1}
\end{Bmatrix}\notag\\
&\hspace{0.95cm}\times(-1)^{a'_{K-1}-a'_J-\sum_{m=J+1}^{K-1}j_m}\prod_{m=J+1}^{K-1}(2a'_{m-1}+1)(-1)^{a'_{m-1}+a_{m-1}+1}
\begin{Bmatrix}  j_m & a'_{m-1} & a'_m\\
  1 & a_m & a_{m-1}
\end{Bmatrix}\notag\\
&\hspace{0.95cm}\times
\begin{Bmatrix}
a_{J-1} & j_J &  a_J \\
 1 & a'_J & j_J
\end{Bmatrix}(-1)^{a_{J-1}-a'_J+j_J}(-1)^{-a'_J+a_J-1}
\times
\begin{Bmatrix}
  a'_J & j_J & a_{J-1} \\
  1 & a'_{J-1} & j_J
\end{Bmatrix}(-1)^{a'_J-a'_{J-1}+j_J+1}(-1)^{-a'_{J-1}+a_{J-1}-1}\notag\\
&\quad\times X(j_I,j_J)^{\frac12}X(j_J,j_K)^{\frac12}\prod_{i=2}^{n-1}\sqrt{2a_i+1}\sqrt{2J+1}\makeSymbol{\includegraphics[width=9.5cm]{graph/volume/volumeoperator-IJK-m-8}}\,.
\end{align}
The exponents in \eqref{general-2} can be simplified as
\begin{align}
&(-1)^{a_{I-1}-a'_I+j_I}(-1)^{a_K\bcancel{-a'_{K-1}}+j_K+1}(-1)^{a_I\cancel{-a_{J-1}}+\sum_{l=I+1}^{J-1}j_l}(-1)^{\bcancel{a'_{K-1}}-a'_J-\sum_{m=J+1}^{K-1}j_m}(-1)^{\cancel{a_{J-1}}-a'_J+j_J}(-1)^{\xcancel{-a'_J}+a_J-1}(-1)^{\xcancel{a'_J}-a'_{J-1}+j_J+1}(-1)^{-a'_{J-1}+a_{J-1}-1}\notag\\
&=(-1)^{a_{I-1}+j_I+a_K+j_K}(-1)^{a_I-a'_I}(-1)^{\sum_{l=I+1}^{J-1}j_l}(-1)^{-\sum_{m=J+1}^{K-1}}(-1)^{a_{J-1}+a_j}(-1)^{-2(a'_J+a'_{J-1}-j_J)}\notag\\
&=(-1)^{a_{I-1}+j_I+a_K+j_K}(-1)^{a_I-a'_I}(-1)^{\sum_{l=I+1}^{J-1}j_l}(-1)^{-\sum_{m=J+1}^{K-1}}(-1)^{a_{J-1}+a_j}\,.
\end{align}
By \eqref{interchange} we obtain the result
\begin{align}\label{2-general-final}
&\sum_{(a'_I,\cdots,a'_{K-1})}(-1)^{a_{I-1}+j_I+a_K+j_K}(-1)^{a_I-a'_I}(-1)^{\sum_{l=I+1}^{J-1}j_l}(-1)^{-\sum_{m=J+1}^{K-1}}X(j_I,j_J)^{\frac12}X(j_J,j_K)^{\frac12}\notag\\
&\hspace{0.95cm}\times\sqrt{(2a'_I+1)(2a_I+1)}\sqrt{(2a'_J+1)(2a_J+1)}\begin{Bmatrix} a_{I-1} & j_I &  a_I \\
 1 & a'_I & j_I
\end{Bmatrix}
\begin{Bmatrix}  a_K & j_K & a_{K-1} \\
  1 & a'_{K-1} & j_K
\end{Bmatrix}\notag\\
&\hspace{0.95cm}\times\prod_{l=I+1}^{J-1}\sqrt{(2a'_l+1)(2a_l+1)}(-1)^{a'_{l-1}+a_{l-1}+1}
\begin{Bmatrix}  j_l & a'_{l-1} & a'_l \\
  1 & a_l & a_{l-1}
\end{Bmatrix}\notag\\
&\hspace{0.95cm}\times\prod_{m=J+1}^{K-1}\sqrt{(2a'_m+1)(2a_m+1)}(-1)^{a'_{m-1}+a_{m-1}+1}
\begin{Bmatrix}  j_m & a'_{m-1} & a'_m\\
  1 & a_m & a_{m-1}
\end{Bmatrix}\notag\\
&\hspace{0.95cm}\times
(-1)^{a_{J-1}+a_j}\begin{Bmatrix}
a_{J-1} & j_J &  a_J \\
 1 & a'_J & j_J
\end{Bmatrix}
\begin{Bmatrix}
  a'_J & j_J & a_{J-1} \\
  1 & a'_{J-1} & j_J
\end{Bmatrix}\notag\\
&\quad\times\prod_{i=2}^{I-1}\sqrt{2a_i+1}\prod_{s=I}^{K-1}\sqrt{2a'_s+1}\prod_{k=K}^{n-1}\sqrt{2a_k+1}\sqrt{2J+1}\makeSymbol{\includegraphics[width=9.5cm]{graph/volume/volumeoperator-IJK-m-8}}\,.
\end{align}
In what follows we now consider the remaining cases.\\\\
(I). The first case of $J=I+1$ and $K=J+1$\\
The result can be directly obtained from that in the case of $J>I+1$ and $K>J+1$ in Eq. \eqref{general-2} by omitting those terms appeared in second step as\\
\begin{align}\label{2-case-1-1}
&\sum_{(a'_I,\cdots,a'_{K-1})}(2a'_I+1)(-1)^{a_{I-1}-a'_I+j_I}\begin{Bmatrix} a_{I-1} & j_I &  a_I \\
 1 & a'_I & j_I
\end{Bmatrix}\times(2a'_{K-1}+1)(-1)^{a_K-a'_{K-1}+j_K+1}
\begin{Bmatrix}  a_K & j_K & a_{K-1} \\
  1 & a'_{K-1} & j_K
\end{Bmatrix}\notag\\
&\hspace{0.95cm}\times
\begin{Bmatrix}
a_{J-1} & j_J &  a_J \\
 1 & a'_J & j_J
\end{Bmatrix}(-1)^{a_{J-1}-a'_J+j_J}(-1)^{-a'_J+a_J-1}
\times
\begin{Bmatrix}
  a'_J & j_J & a_{J-1} \\
  1 & a'_{J-1} & j_J
\end{Bmatrix}(-1)^{a'_J-a'_{J-1}+j_J+1}(-1)^{-a'_{J-1}+a_{J-1}-1}\notag\\
&\quad\times X(j_I,j_J)^{\frac12}X(j_J,j_K)^{\frac12}\prod_{i=2}^{n-1}\sqrt{2a_i+1}\sqrt{2J+1}\makeSymbol{\includegraphics[width=6.5cm]{graph/volume/volumeoperator-IJK-case-I}}\,.
\end{align}
The exponents of \eqref{2-case-1-1} can be simplified as
\begin{align}
&(-1)^{a_{I-1}-a'_I+j_I}(-1)^{a_K-a'_{K-1}+j_K+1}(-1)^{a_{J-1}-a'_J+j_J}(-1)^{\bcancel{-a'_J}+a_J-1}(-1)^{\bcancel{a'_J}-a'_{J-1}+j_J+1}(-1)^{-a'_{J-1}+a_{J-1}-1}\notag\\
&=(-1)^{a_{I-1}+j_I+a_K+j_K}(-1)^{a_{J-1}-a'_I}(-1)^{a_{J-1}+a_J}(-1)^{-a'_{K-1}-a'_J-2a'_{J-1}+2j_J}\notag\\
&=(-1)^{a_{I-1}+j_I+a_K+j_K}(-1)^{a_I-a'_I}(-1)^{a_{J-1}+a_J}(-1)^{-2(a'_J+a'_{J-1}-j_J)}\notag\\
&=(-1)^{a_{I-1}+j_I+a_K+j_K}(-1)^{a_I-a'_I}(-1)^{a_{J-1}+a_J}\,,
\end{align}
where the identities $a_{J-1}=a_I$ and $a'_{K-1}=a'_J$ were used in the second step. After replacing $(2a'_{K-1}+1)$ by $(2a'_J+1)$ and properly adjusting the ordering of multi-products of $\sqrt{2a+1}$, Eq. \eqref{2-case-1-1} reduces to
\begin{align}\label{2-case-1-final}
&\sum_{(a'_I,\cdots,a'_{K-1})}(-1)^{a_{I-1}+j_I+a_K+j_K}(-1)^{a_I-a'_I}X(j_I,j_J)^{\frac12}X(j_J,j_K)^{\frac12}\notag\\
&\hspace{0.95cm}\times\sqrt{(2a'_I+1)(2a_I+1)}\sqrt{(2a'_J+1)(2a_J+1)}\begin{Bmatrix} a_{I-1} & j_I &  a_I \\
 1 & a'_I & j_I
\end{Bmatrix}
\begin{Bmatrix}  a_K & j_K & a_{K-1} \\
  1 & a'_{K-1} & j_K
\end{Bmatrix}\notag\\
&\hspace{0.95cm}\times
(-1)^{a_{J-1}+a_J}\begin{Bmatrix}
a_{J-1} & j_J &  a_J \\
 1 & a'_J & j_J
\end{Bmatrix}
\begin{Bmatrix}
  a'_J & j_J & a_{J-1} \\
  1 & a'_{J-1} & j_J
\end{Bmatrix}\notag\\
&\quad\times\prod_{i=2}^{I-1}\sqrt{2a_i+1}\prod_{s=I}^{K-1}\sqrt{2a'_s+1}\prod_{k=K}^{n-1}\sqrt{2a_k+1}\sqrt{2J+1}\makeSymbol{\includegraphics[width=6.5cm]{graph/volume/volumeoperator-IJK-case-I}}\,.
\end{align}
In fact, \eqref{2-case-1-final} can also be regarded as a special case of Eq. \eqref{2-general-final}. In \eqref{2-general-final}, the multi-products $\prod_{l=I+1}^{J-1}$ and $\prod_{m=J+1}^{K-1}$ and summations $\sum_{l=I+1}^{J-1}$ and $\sum_{m=J+1}^{K-1}$ do not exist for the case of $J=I+1$ and $K=J+1$. Hence the terms involving these multi-products and summations can be omitted, and this yields the result \eqref{2-case-1-final}.\\\\\
(II). The second case of $J=I+1$ and $K>J+1$\\
The result can be directly obtained from that in the case of $J>I+1$ and $K>J+1$ by omitting certain terms appeared in the first line of the second step in Eq. \eqref{general-2} as
\begin{align}\label{2-case-2-1}
&\sum_{(a'_I,\cdots,a'_{K-1})}(2a'_I+1)(-1)^{a_{I-1}-a'_I+j_I}\begin{Bmatrix} a_{I-1} & j_I &  a_I \\
 1 & a'_I & j_I
\end{Bmatrix}
\times(2a'_{K-1}+1)(-1)^{a_K-a'_{K-1}+j_K+1}
\begin{Bmatrix}  a_K & j_K & a_{K-1} \\
  1 & a'_{K-1} & j_K
\end{Bmatrix}\notag\\
&\hspace{0.95cm}\times(-1)^{a'_{K-1}-a'_J-\sum_{m=J+1}^{K-1}j_m}\prod_{m=J+1}^{K-1}(2a'_{m-1}+1)(-1)^{a'_{m-1}+a_{m-1}+1}
\begin{Bmatrix}  j_m & a'_{m-1} & a'_m\\
  1 & a_m & a_{m-1}
\end{Bmatrix}\notag\\
&\hspace{0.95cm}\times
\begin{Bmatrix}
a_{J-1} & j_J &  a_J \\
 1 & a'_J & j_J
\end{Bmatrix}(-1)^{a_{J-1}-a'_J+j_J}(-1)^{-a'_J+a_J-1}
\times
\begin{Bmatrix}
  a'_J & j_J & a_{J-1} \\
  1 & a'_{J-1} & j_J
\end{Bmatrix}(-1)^{a'_J-a'_{J-1}+j_J+1}(-1)^{-a'_{J-1}+a_{J-1}-1}\notag\\
&\quad\times X(j_I,j_J)^{\frac12}X(j_J,j_K)^{\frac12}\prod_{i=2}^{n-1}\sqrt{2a_i+1}\sqrt{2J+1}\makeSymbol{\includegraphics[width=8.5cm]{graph/volume/volumeoperator-IJK-case-II}}\,.
\end{align}
The exponents in \eqref{2-case-2-1} can be simplified as
\begin{align}
&(-1)^{a_{I-1}-a'_I+j_I}(-1)^{a_K\bcancel{-a'_{K-1}}+j_K+1}(-1)^{\bcancel{a'_{K-1}}-a'_J-\sum_{m=J+1}^{K-1}j_m}(-1)^{a_{J-1}-a'_J+j_J}(-1)^{\cancel{-a'_J}+a_J-1}(-1)^{\cancel{a'_J}-a'_{J-1}+j_J+1}(-1)^{-a'_{J-1}+a_{J-1}-1}\notag\\
&=(-1)^{a_{I-1}+j_I+a_K+j_K}(-1)^{a_{J-1}-a'_I}(-1)^{-\sum_{m=J+1}^{K-1}j_m}(-1)^{a_{J-1}+a_J}(-1)^{-2(a'_J+a'_{J-1}-j_J)}\notag\\
&=(-1)^{a_{I-1}+j_I+a_K+j_K}(-1)^{a_I-a'_I}(-1)^{-\sum_{m=J+1}^{K-1}j_m}(-1)^{a_{J-1}+a_J}(-1)^{-2(a'_J+a'_{J-1}-j_J)}\notag\\
&=(-1)^{a_{I-1}+j_I+a_K+j_K}(-1)^{a_I-a'_I}(-1)^{-\sum_{m=J+1}^{K-1}j_m}(-1)^{a_{J-1}+a_J}\,,
\end{align}
where in the second step we replaced $a_{J-1}$ by $a_I$. By using \eqref{interchange} and
 properly adjusting the ordering of multi-products of $\sqrt{2a+1}$, we have
\begin{align}\label{2-case-2-final}
&\sum_{(a'_I,\cdots,a'_{K-1})}(-1)^{a_{I-1}+j_I+a_K+j_K}(-1)^{a_I-a'_I}(-1)^{-\sum_{m=J+1}^{K-1}}X(j_I,j_J)^{\frac12}X(j_J,j_K)^{\frac12}\notag\\
&\hspace{0.95cm}\times\sqrt{(2a'_I+1)(2a_I+1)}\sqrt{(2a'_J+1)(2a_J+1)}\begin{Bmatrix} a_{I-1} & j_I &  a_I \\
 1 & a'_I & j_I
\end{Bmatrix}
\begin{Bmatrix}  a_K & j_K & a_{K-1} \\
  1 & a'_{K-1} & j_K
\end{Bmatrix}\notag\\
&\hspace{0.95cm}\times\prod_{m=J+1}^{K-1}\sqrt{(2a'_m+1)(2a_m+1)}(-1)^{a'_{m-1}+a_{m-1}+1}
\begin{Bmatrix}  j_m & a'_{m-1} & a'_m\\
  1 & a_m & a_{m-1}
\end{Bmatrix}\notag\\
&\hspace{0.95cm}\times
(-1)^{a_{J-1}+a_J}\begin{Bmatrix}
a_{J-1} & j_J &  a_J \\
 1 & a'_J & j_J
\end{Bmatrix}
\begin{Bmatrix}
  a'_J & j_J & a_{J-1} \\
  1 & a'_{J-1} & j_J
\end{Bmatrix}\notag\\
&\quad\times\prod_{i=2}^{I-1}\sqrt{2a_i+1}\prod_{s=I}^{K-1}\sqrt{2a'_s+1}\prod_{k=K}^{n-1}\sqrt{2a_k+1}\sqrt{2J+1}\makeSymbol{\includegraphics[width=8.5cm]{graph/volume/volumeoperator-IJK-case-II}}\,.
\end{align}
Again, \eqref{2-case-2-final} can also be directly written down from Eq. \eqref{2-general-final}. In \eqref{2-general-final}, the multi-products $\prod_{l=I+1}^{J-1}$ and summations $\sum_{l=I+1}^{J-1}$ do not exist for the case of $J=I+1$. Hence the terms involving these multi-products and summations can be omitted, and this yields the result \eqref{2-case-2-final}.\\\\
(III). The third case of $J>I+1$ and $K=J+1$\\
The result can be directly obtained from that in the case of $J>I+1$ and $K>J+1$ by omitting some terms appeared in the second line of the second step in Eq. \eqref{general-2} as
\begin{align}
&\sum_{(a'_I,\cdots,a'_{K-1})}(2a'_I+1)(-1)^{a_{I-1}-a'_I+j_I}\begin{Bmatrix} a_{I-1} & j_I &  a_I \\
 1 & a'_I & j_I
\end{Bmatrix}\label{2-case-3-1}
\times(2a'_{K-1}+1)(-1)^{a_K-a'_{K-1}+j_K+1}
\begin{Bmatrix}  a_K & j_K & a_{K-1} \\
  1 & a'_{K-1} & j_K
\end{Bmatrix}\notag\\
&\hspace{0.95cm}\times(-1)^{a_I-a_{J-1}+\sum_{l=I+1}^{J-1}j_l}\prod_{l=I+1}^{J-1}(2a'_l+1)(-1)^{a'_{l-1}+a_{l-1}+1}
\begin{Bmatrix}  j_l & a'_{l-1} & a'_l \\
  1 & a_l & a_{l-1}
\end{Bmatrix}\notag\\
&\hspace{0.95cm}\times
\begin{Bmatrix}
a_{J-1} & j_J &  a_J \\
 1 & a'_J & j_J
\end{Bmatrix}(-1)^{a_{J-1}-a'_J+j_J}(-1)^{-a'_J+a_J-1}
\times
\begin{Bmatrix}
  a'_J & j_J & a_{J-1} \\
  1 & a'_{J-1} & j_J
\end{Bmatrix}(-1)^{a'_J-a'_{J-1}+j_J+1}(-1)^{-a'_{J-1}+a_{J-1}-1}\notag\\
&\quad\times X(j_I,j_J)^{\frac12}X(j_J,j_K)^{\frac12}\prod_{i=2}^{n-1}\sqrt{2a_i+1}\sqrt{2J+1}\makeSymbol{\includegraphics[width=8.5cm]{graph/volume/volumeoperator-IJK-case-III}}\,.
\end{align}
The exponents in \eqref{2-case-3-1} can be simplified as
\begin{align}
&(-1)^{a_{I-1}-a'_I+j_I}(-1)^{a_K-a'_{K-1}+j_K+1}(-1)^{a_I\bcancel{-a_{J-1}}+\sum_{l=I+1}^{J-1}j_l}(-1)^{\bcancel{a_{J-1}}-a'_J+j_J}(-1)^{\cancel{-a'_J}+a_J-1}(-1)^{\cancel{a'_J}-a'_{J-1}+j_J+1}(-1)^{-a'_{J-1}+a_{J-1}-1}\notag\\
&=(-1)^{a_{I-1}+j_I+a_K+j_K}(-1)^{a_I-a'_I}(-1)^{\sum_{l=I+1}^{J-1}j_l}(-1)^{a_{J-1}+a_J}(-1)^{-a'_{K-1}-a'_J-2a'_{J-1}+2j_J}\notag\\
&=(-1)^{a_{I-1}+j_I+a_K+j_K}(-1)^{a_I-a'_I}(-1)^{\sum_{l=I+1}^{J-1}j_l}(-1)^{a_{J-1}+a_J}(-1)^{-2(a'_J+a'_{J-1}-j_J)}\notag\\
&=(-1)^{a_{I-1}+j_I+a_K+j_K}(-1)^{a_I-a'_I}(-1)^{\sum_{l=I+1}^{J-1}j_I}(-1)^{a_{J-1}+a_J}\,,
\end{align}
where in the second step we used $a'_{K-1}=a'_J$. By replacing $(2a'_{K-1}+1)$ by $(2a'_J+1)$ and properly adjusting the ordering of multi-products of $\sqrt{2a+1}$, we have
\begin{align}\label{2-case-3-final}
&\sum_{(a'_I,\cdots,a'_{K-1})}(-1)^{a_{I-1}+j_I+a_K+j_K}(-1)^{a_I-a'_I}(-1)^{\sum_{l=I+1}^{J-1}j_l}X(j_I,j_J)^{\frac12}X(j_J,j_K)^{\frac12}\notag\\
&\hspace{0.95cm}\times\sqrt{(2a'_I+1)(2a_I+1)}\sqrt{(2a'_J+1)(2a_J+1)}\begin{Bmatrix} a_{I-1} & j_I &  a_I \\
 1 & a'_I & j_I
\end{Bmatrix}
\begin{Bmatrix}  a_K & j_K & a_{K-1} \\
  1 & a'_{K-1} & j_K
\end{Bmatrix}\notag\\
&\hspace{0.95cm}\times\prod_{l=I+1}^{J-1}\sqrt{(2a'_l+1)(2a_l+1)}(-1)^{a'_{l-1}+a_{l-1}+1}
\begin{Bmatrix}  j_l & a'_{l-1} & a'_l \\
  1 & a_l & a_{l-1}
\end{Bmatrix}\notag\\
&\hspace{0.95cm}\times
(-1)^{a_{J-1}+a_J}\begin{Bmatrix}
a_{J-1} & j_J &  a_J \\
 1 & a'_J & j_J
\end{Bmatrix}
\begin{Bmatrix}
  a'_J & j_J & a_{J-1} \\
  1 & a'_{J-1} & j_J
\end{Bmatrix}\notag\\
&\quad\times\prod_{i=2}^{I-1}\sqrt{2a_i+1}\prod_{s=I}^{K-1}\sqrt{2a'_s+1}\prod_{k=K}^{n-1}\sqrt{2a_k+1}\sqrt{2J+1}\makeSymbol{\includegraphics[width=8.5cm]{graph/volume/volumeoperator-IJK-case-III}}\,.
\end{align}
The above result can also be directly written down from Eq. \eqref{2-general-final}. In \eqref{2-general-final}, the multi-products $\prod_{m=J+1}^{K-1}$ and summations $\sum_{m=J+1}^{K-1}$ do not exist for the case of $K=J+1$. Hence the terms involving these multi-products and summations can be omitted, and this yields the result \eqref{2-case-3-final}.

The above discussions in case by case indicate that the general form of the action can be written as \eqref{2-general-final}, i.e.,
\begin{align}\label{2-general-final-re-appendix}
\hat{q}^{<IJ;JK>}_{IJK}\cdot {\left(i^{\,J;\,\vec{a}}_v\right)_{\,m_1\cdots m_I\cdots m_J\cdots m_K\cdots m_n}}^M
&=\sum_{(a'_I,\cdots,a'_{K-1})}(-1)^{a_{I-1}+j_I+a_K+j_K}(-1)^{a_I-a'_I}(-1)^{\sum_{l=I+1}^{J-1}j_l}(-1)^{-\sum_{m=J+1}^{K-1}}X(j_I,j_J)^{\frac12}X(j_J,j_K)^{\frac12}\notag\\
&\hspace{0.95cm}\times\sqrt{(2a'_I+1)(2a_I+1)}\sqrt{(2a'_J+1)(2a_J+1)}\begin{Bmatrix} a_{I-1} & j_I &  a_I \\
 1 & a'_I & j_I
\end{Bmatrix}
\begin{Bmatrix}  a_K & j_K & a_{K-1} \\
  1 & a'_{K-1} & j_K
\end{Bmatrix}\notag\\
&\hspace{0.95cm}\times\prod_{l=I+1}^{J-1}\sqrt{(2a'_l+1)(2a_l+1)}(-1)^{a'_{l-1}+a_{l-1}+1}
\begin{Bmatrix}  j_l & a'_{l-1} & a'_l \\
  1 & a_l & a_{l-1}
\end{Bmatrix}\notag\\
&\hspace{0.95cm}\times\prod_{m=J+1}^{K-1}\sqrt{(2a'_m+1)(2a_m+1)}(-1)^{a'_{m-1}+a_{m-1}+1}
\begin{Bmatrix}  j_m & a'_{m-1} & a'_m\\
  1 & a_m & a_{m-1}
\end{Bmatrix}\notag\\
&\hspace{0.95cm}\times
(-1)^{a_{J-1}+a_J}\begin{Bmatrix}
a_{J-1} & j_J &  a_J \\
 1 & a'_J & j_J
\end{Bmatrix}
\begin{Bmatrix}
  a'_J & j_J & a_{J-1} \\
  1 & a'_{J-1} & j_J
\end{Bmatrix}\notag\\
&\quad\times{\left(i^{\,J;\,\vec{\tilde{a}}}_v\right)_{\,m_1\cdots m_I\cdots m_J\cdots m_K\cdots m_n}}^M\,.
\end{align}
For special cases of $J-1<I+1$ and $K-1<J+1$, the results can be obtained from the form \eqref{2-general-final-re-appendix} by omitting the corresponding multi-products $\prod_{l=I+1}^{J-1}$ and $\prod_{m=J+1}^{K-1}$ and summations $\sum_{l=I+1}^{J-1}$ and $\sum_{m=J+1}^{K-1}$.

\providecommand{\href}[2]{#2}\begingroup\raggedright\endgroup



\begin{thebibliography}{10}

\bibitem{Ashtekar:2004eh}
A.~Ashtekar and J.~Lewandowski, {Background independent quantum gravity: A
  status report}, \href{http://dx.doi.org/10.1088/0264-9381/21/15/R01}{{\em
  Class. Quant. Grav.} {\bfseries 21} (2004) R53},
  \href{http://arxiv.org/abs/gr-qc/0404018}{{\ttfamily arXiv:gr-qc/0404018
  [gr-qc]}}.

\bibitem{Han:2005km}
M.~Han, Y.~Ma, and W.~Huang, {Fundamental structure of loop quantum gravity},
  \href{http://dx.doi.org/10.1142/S0218271807010894}{{\em Int. J. Mod. Phys. D}
  {\bfseries 16} (2007) 1397--1474},
\href{http://arxiv.org/abs/gr-qc/0509064}{{\ttfamily arXiv:gr-qc/0509064
  [gr-qc]}}.

\bibitem{Rovelli:2004tv}
C.~Rovelli, {\em Quantum Gravity}.
\newblock Cambridge University Press, Cambridge, 2004.

\bibitem{Thiemann:2007bk}
T.~Thiemann, {\em Modern Canonical Quantum General Relativity}.
\newblock Cambridge University Press, Cambridge, 2007.

\bibitem{Thiemann:1996at}
T.~Thiemann, {A length operator for canonical quantum gravity},
  \href{http://dx.doi.org/10.1063/1.532445}{{\em J. Math. Phys.} {\bfseries 39}
  (1998) 3372--3392}, \href{http://arxiv.org/abs/gr-qc/9606092}{{\ttfamily
  arXiv:gr-qc/9606092 [gr-qc]}}.

\bibitem{Bianchi:2008es}
E.~Bianchi, {The length operator in loop quantum gravity},
  \href{http://dx.doi.org/10.1016/j.nuclphysb.2008.08.013}{{\em Nucl. Phys. B}
  {\bfseries 807} (2009) 591--624},
\href{http://arxiv.org/abs/0806.4710}{{\ttfamily arXiv:0806.4710 [gr-qc]}}.

\bibitem{Ma:2010fy}
Y.~Ma, C.~Soo, and J.~Yang, {New length operator for loop quantum gravity},
  \href{http://dx.doi.org/10.1103/PhysRevD.81.124026}{{\em Phys. Rev. D}
  {\bfseries 81} (2010) 124026},
\href{http://arxiv.org/abs/1004.1063}{{\ttfamily arXiv:1004.1063 [gr-qc]}}.

\bibitem{Rovelli:1994ge}
C.~Rovelli and L.~Smolin, {Discreteness of area and volume in quantum gravity},
  \href{http://dx.doi.org/10.1016/0550-3213(95)00150-Q,
  10.1016/0550-3213(95)00150-Q}{{\em Nucl. Phys. B} {\bfseries 442} (1995)
  593--622},
\href{http://arxiv.org/abs/gr-qc/9411005}{{\ttfamily arXiv:gr-qc/9411005
  [gr-qc]}}.

\bibitem{Ashtekar:1996eg}
A.~Ashtekar and J.~Lewandowski, {Quantum theory of geometry: I. Area
  operators}, \href{http://dx.doi.org/10.1088/0264-9381/14/1A/006}{{\em Class.
  Quant. Grav.} {\bfseries 14} (1997) A55--A82},
  \href{http://arxiv.org/abs/gr-qc/9602046}{{\ttfamily arXiv:gr-qc/9602046
  [gr-qc]}}.

\bibitem{Ashtekar:1997fb}
A.~Ashtekar and J.~Lewandowski, {Quantum theory of geometry: II. Volume
  operators}, {\em Adv. Theor. Math. Phys.} {\bfseries 1} (1998) 388--429,
  \href{http://arxiv.org/abs/gr-qc/9711031}{{\ttfamily arXiv:gr-qc/9711031
  [gr-qc]}}.

\bibitem{Thiemann:1996au}
T.~Thiemann, {Closed formula for the matrix elements of the volume operator in
  canonical quantum gravity}, \href{http://dx.doi.org/10.1063/1.532259}{{\em J.
  Math. Phys.} {\bfseries 39} (1998) 3347--3371},
  \href{http://arxiv.org/abs/gr-qc/9606091}{{\ttfamily arXiv:gr-qc/9606091
  [gr-qc]}}.

\bibitem{DePietri:1996pja}
R.~De~Pietri and C.~Rovelli, {Geometry eigenvalues and scalar product from
  recoupling theory in loop quantum gravity},
  \href{http://dx.doi.org/10.1103/PhysRevD.54.2664}{{\em Phys. Rev. D}
  {\bfseries 54} (1996) 2664--2690},
\href{http://arxiv.org/abs/gr-qc/9602023}{{\ttfamily arXiv:gr-qc/9602023
  [gr-qc]}}.

\bibitem{Brunnemann:2004xi}
J.~Brunnemann and T.~Thiemann, {Simplification of the spectral analysis of the
  volume operator in loop quantum gravity},
  \href{http://dx.doi.org/10.1088/0264-9381/23/4/014}{{\em Class. Quant. Grav.}
  {\bfseries 23} (2006) 1289--1346},
  \href{http://arxiv.org/abs/gr-qc/0405060}{{\ttfamily arXiv:gr-qc/0405060
  [gr-qc]}}.

\bibitem{Dass:2006sp}
N.~H. Dass and M.~Mathur, {On loop states in loop quantum gravity},
  \href{http://dx.doi.org/10.1088/0264-9381/24/9/002}{{\em Class. Quant. Grav.}
  {\bfseries 24} (2007) 2179--2192},
\href{http://arxiv.org/abs/gr-qc/0611156}{{\ttfamily arXiv:gr-qc/0611156
  [gr-qc]}}.

\bibitem{Borissov:1997ji}
R.~Borissov, R.~De~Pietri, and C.~Rovelli, {Matrix elements of Thiemann's
  Hamiltonian constraint in loop quantum gravity},
  \href{http://dx.doi.org/10.1088/0264-9381/14/10/008}{{\em Class. Quant.
  Grav.} {\bfseries 14} (1997) 2793--2823},
\href{http://arxiv.org/abs/gr-qc/9703090}{{\ttfamily arXiv:gr-qc/9703090
  [gr-qc]}}.

\bibitem{Alesci:2007tx}
E.~Alesci and C.~Rovelli, {The complete LQG propagator. I. Difficulties with
  the Barrett-Crane vertex},
  \href{http://dx.doi.org/10.1103/PhysRevD.76.104012}{{\em Phys. Rev. D}
  {\bfseries 76} (2007) 104012},
\href{http://arxiv.org/abs/0708.0883}{{\ttfamily arXiv:0708.0883 [gr-qc]}}.

\bibitem{Alesci:2010gb}
E.~Alesci and C.~Rovelli, {A regularization of the Hamiltonian constraint
  compatible with the spinfoam dynamics},
  \href{http://dx.doi.org/10.1103/PhysRevD.82.044007}{{\em Phys. Rev. D}
  {\bfseries 82} (2010) 044007},
\href{http://arxiv.org/abs/1005.0817}{{\ttfamily arXiv:1005.0817 [gr-qc]}}.

\bibitem{Alesci:2011ia}
E.~Alesci, T.~Thiemann, and A.~Zipfel, {Linking covariant and canonical LQG:
  New solutions to the Euclidean scalar constraint},
  \href{http://dx.doi.org/10.1103/PhysRevD.86.024017}{{\em Phys. Rev. D}
  {\bfseries 86} (2012) 024017},
\href{http://arxiv.org/abs/1109.1290}{{\ttfamily arXiv:1109.1290 [gr-qc]}}.

\bibitem{Alesci:2013kpa}
E.~Alesci, K.~Liegener, and A.~Zipfel, {Matrix elements of Lorentzian
  Hamiltonian constraint in loop quantum gravity},
  \href{http://dx.doi.org/10.1103/PhysRevD.88.084043}{{\em Phys. Rev. D}
  {\bfseries 88} (2013) 084043},
\href{http://arxiv.org/abs/1306.0861}{{\ttfamily arXiv:1306.0861 [gr-qc]}}.

\bibitem{brink1968angular}
D.~M. Brink and G.~R. Satchler, {\em Angular Momentum}.
\newblock Oxford Library of the Physical Sciences. Clarendon Press, 1968.

\bibitem{Alesci:2015wla}
E.~Alesci, M.~Assanioussi, J.~Lewandowski, and I.~M{\"a}kinen, {Hamiltonian
  operator for loop quantum gravity coupled to a scalar field},
\href{http://arxiv.org/abs/1504.02068}{{\ttfamily arXiv:1504.02068 [gr-qc]}}.

\bibitem{graph-II}
J.~Yang and Y.~Ma, {Graphical method in loop quantum gravity: II. The
  Hamiltonian constraint and inverse volume operators},
\href{http://arxiv.org/abs/1505.00225}{{\ttfamily arXiv:1505.00225 [gr-qc]}}.

\bibitem{Yang:2015zda}
J.~Yang and Y.~Ma, {New Hamiltonian constraint operator for loop quantum
  gravity}, \href{http://dx.doi.org/10.1016/j.physletb.2015.10.062}{{\em Phys.
  Lett. B} {\bfseries 751} (2015) 343--347},
\href{http://arxiv.org/abs/1507.00986}{{\ttfamily arXiv:1507.00986 [gr-qc]}}.

\bibitem{Ashtekar:1987gu}
A.~Ashtekar, {New Hamiltonian formulation of general relativity},
\href{http://dx.doi.org/10.1103/PhysRevD.36.1587}{{\em Phys. Rev. D} {\bfseries
  36} (1987) 1587--1602}.

\bibitem{Barbero:1994ap}
J.~F. Barbero~G., {Real Ashtekar variables for Lorentzian signature space-times}, 
\href{http://dx.doi.org/10.1103/PhysRevD.51.5507}{{\em Phys. Rev. D}
  {\bfseries 51} (1995) 5507--5510},
\href{http://arxiv.org/abs/gr-qc/9410014}{{\ttfamily arXiv:gr-qc/9410014
  [gr-qc]}}.

\bibitem{Ashtekar:1995zh}
A.~Ashtekar, J.~Lewandowski, D.~Marolf, J.~Mourao, and T.~Thiemann,
  {Quantization of diffeomorphism invariant theories of connections with local
  degrees of freedom}, \href{http://dx.doi.org/10.1063/1.531252}{{\em J. Math.
  Phys.} {\bfseries 36} (1995) 6456--6493},
  \href{http://arxiv.org/abs/gr-qc/9504018}{{\ttfamily arXiv:gr-qc/9504018
  [gr-qc]}}.

\bibitem{Edmonds}
A.~R. Edmonds, {\em Angular Momentum in Quantum Mechanics}.
\newblock Princeton University Press, Princeton, New Jersey, 1974.

\bibitem{Wigner-bk}
E.~P. Wigner, {\em Group Theory and Its Application to the Quantum Mechanics of
  Atomic Spectra}.
\newblock Academic Press Inc., New York, 1959.

\bibitem{Yutsis:1962bk}
A.~P. Yutsis, I.~B. Levinson, and V.~V. Vanagas, {\em Mathematical Apparatus of
  the Theory of Angular Momentum}.
\newblock Israel Program for Scientific Translation, Jerusalem, 1962.
\newblock Translated from the Russian by A. Sen and R. N. Sen.

\bibitem{Giesel:2005bk}
K.~Giesel and T.~Thiemann, {Consistency check on volume and triad operator
  quantization in loop quantum gravity: I},
  \href{http://dx.doi.org/10.1088/0264-9381/23/18/011}{{\em Class. Quant.
  Grav.} {\bfseries 23} (2006) 5667--5692},
  \href{http://arxiv.org/abs/gr-qc/0507036}{{\ttfamily arXiv:gr-qc/0507036
  [gr-qc]}}.

\bibitem{Giesel:2005bm}
K.~Giesel and T.~Thiemann, {Consistency check on volume and triad operator
  quantization in loop quantum gravity: II},
  \href{http://dx.doi.org/10.1088/0264-9381/23/18/012}{{\em Class. Quant.
  Grav.} {\bfseries 23} (2006) 5693--5772},
  \href{http://arxiv.org/abs/gr-qc/0507037}{{\ttfamily arXiv:gr-qc/0507037
  [gr-qc]}}.

\bibitem{Thiemann:1996aw}
T.~Thiemann, {Quantum spin dynamics (QSD)},
  \href{http://dx.doi.org/10.1088/0264-9381/15/4/011}{{\em Class. Quant. Grav.}
  {\bfseries 15} (1998) 839--873},
  \href{http://arxiv.org/abs/gr-qc/9606089}{{\ttfamily arXiv:gr-qc/9606089
  [gr-qc]}}.

\bibitem{Fecko}
M.~Fecko, {\em Differential Geometry and Lie Groups for Physicists}.
\newblock Cambridge university press, 2006.

\bibitem{Wigner}
E.~P. Wigner, On the matrices which reduce the Kronecker products of
  representation of simply reducible groups, {\em Princeton (unpublished
  manuscript), 1937} .

\end{thebibliography}
\end{document}